\documentclass[12pt,english,1p]{elsarticle}
\usepackage[T1]{fontenc}
\usepackage[latin9]{inputenc}
\usepackage{color}
\usepackage{babel}
\usepackage{url}
\usepackage{amsmath}
\usepackage{amsthm}
\usepackage{amssymb}
\usepackage{graphicx}
\usepackage[unicode=true,
 bookmarks=true,bookmarksnumbered=false,bookmarksopen=false,
 breaklinks=false,pdfborder={0 0 0},pdfborderstyle={},backref=false,colorlinks=true]
 {hyperref}

\makeatletter

\providecommand{\tabularnewline}{\\}

\theoremstyle{plain}
\newtheorem{thm}{\protect\theoremname}

\usepackage[T1]{fontenc}
\usepackage{newtxtext}
\usepackage[libertine]{newtxmath}
\usepackage{bm}

\DeclareMathOperator*{\argmin}{arg\,min}

\makeatother

\providecommand{\theoremname}{Theorem}

\addto\captionsenglish{}

\begin{document}

\begin{frontmatter}{}

\title{Probabilistic Recalibration of Forecasts}

\author[carlo]{Carlo Graziani\corref{c1}}
\ead{cgraziani@anl.gov}

\author[bob]{Robert Rosner}

\author[jennifer]{Jennifer M. Adams}

\author[reason]{Reason L. Machete}

\address[carlo]{Argonne National Laboratory, Lemont, IL, USA}

\address[bob]{Department of Astronomy \& Astrophysics, University of Chicago, Chicago,
IL, USA}

\address[jennifer]{Center for Ocean-Land-Atmosphere Studies, Fairfax, VA,
USA}

\address[reason]{Climate Change Division, Botswana Institute for Technology Research and Innovation, Gaborone, Botswana}

\cortext[c1]{Corresponding Author}
\begin{abstract}
We present a scheme by which a probabilistic
forecasting system whose predictions have poor probabilistic calibration may be recalibrated by
incorporating past performance information to produce a new forecasting
system that is demonstrably superior to the original, in that
one may use it to consistently win wagers against someone using the original system. The scheme
utilizes Gaussian process (GP) modeling to estimate a probability distribution over the Probability
Integral Transform (PIT) of a scalar predictand. The GP density estimate
gives closed-form access to information entropy measures associated with the estimated distribution,
which allows prediction of winnings in wagers against the base forecasting
system. A separate consequence of the procedure is that the recalibrated forecast has a uniform expected PIT distribution. 
A distinguishing feature of the procedure is that it is appropriate even if the PIT values are not i.i.d.
The recalibration scheme is formulated in a framework that
exploits the deep connections between information theory, forecasting,
and betting. We demonstrate the effectiveness of the scheme in two
case studies: a laboratory experiment with a nonlinear circuit and
seasonal forecasts of the intensity of the El Ni\~no-Southern Oscillation
phenomenon.
\end{abstract}

\end{frontmatter}{}

\section{Introduction}

A forecast, being an expression of uncertainty about the future, is
necessarily a probabilistic affair. Probabilistic forecasts of events
falling along a continuum---such as short-term weather
forecasts \citep{Gneiting_etal_2014,brier1950verification}, medium-term
seasonal rainfall \citep{KRZYSZTOFOWICZ2014643,2001JHyd..249....2K,SHARMA2000232},
fluctuations of financial asset prices \citep{diebold1998,kamstra2001combining}
or electrical demand \citep{MACIEJOWSKA2016957}, rates of spread
of infectious disease \citep{held2017probabilistic,moran2016epidemic},
macroeconomic indicators \citep{casillas2006probability,greenwood2012probabilistic},
wind power availability \citep{pinson2012very,zhang2014review}, species
endangerment and extinction \citep{araujo2007ensemble}, human population
growth \citep{lutz2001end}, and seismic activity \citep{kagan2000probabilistic,marzocchi2007probabilistic}---are of urgent interest to many kinds of decision-makers,
and have occasioned much scientific literature across a broad range
of fields. 

Weather forecasting through numerical weather prediction (NWP) has
substantially improved its performance over the past few decades in
consequence of improvements in observational data, computational models,
and computational power and currently is capable of providing, on
average, reasonably robust weather forecasts for periods on the order of
10 days \citep{QJ:QJ2559}. Unfortunately, compared to empirical forecasting schemes, these physics-based forecasting
schemes are known to lose skill for periods longer than $\approx$10
days, and the obvious question arises whether it is possible to achieve
skillful forecasting for periods longer than 10 days, and if so, whether
there is an upper bound on such forecasting.

Given the difficulty of achieving the computing power and model fidelity
required to abate model errors of NWP, the best way forward for the
present may well be to attempt to achieve some kind of synthetic hybrid
between NWP and empirical forecast methods in an effort to leverage
the information content of the former to enhance the predictive power
of the latter. Important approaches that have been attempted include applying some kind of statistical
recalibration to the NWP simulation output such as Model Output Statistics
\citep{glahn1972use} and then infer probability distributions
for predictands from the corrected simulations by some smoothing procedure
\citep{coelho2004forecast,2005MWRv..133.1098G}; leaving the simulations
as they are and adapt the smoothing procedure itself to validation
data \citep{brocker2008ensemble}; or doing a bit of both \citep{raftery2005using,Fraley_et_al-2010,dutton2013calibration}.
A difficulty of such programs is that the smoothing procedure itself
has statistical properties that are usually not under very good control,
since they frequently take the form of highly simplified models such
as Gaussian mixtures, which are not generally in well-motivated
correspondence with the processes that relate the simulation output
to the random predictand.

One common feature of the continuous forecast probability density functions (PDFs) produced from NWP output is that they are more often than not \textit{probabilistically miscalibrated}---that is, the long-term frequencies of observations do not match stated probabilities of predictions (see \cite{glahn2009}, for example). In such cases, the interpretation of the PDFs requires caution, and the value of having a probabilistic forecast rather than a point forecast can be questionable, particularly given the risk of underestimating frequencies of extremes.

As discussed by Diebold, Hahn, and Tay (\cite{diebold1999multivariate}, hereafter ``DHT''), the phenomenon of probabilistic miscalibration creates another opportunity for recalibration: direct recalibration of the forecast probability distributions. This possibility arises because
whatever the methodology adopted to produce a forecast system,
long enough use of that system leads to additional performance information---through comparison of a series of forecasts with their
predictands---that can be incorporated into current forecasts
to produce improved forecasts. Such information, which is commonly used
to assess forecast system quality, was shown by DHT \cite{diebold1999multivariate} to permit \textit{correction} of future forecasts, assuming an i.i.d. restriction on predictands. More recently, a similar approach was devised in the context of deep learning by Kuleshov, Fenner, and Ermon (\cite{pmlr-v80-kuleshov18a}, hereafter ``KFE''), who recast the usual machine learning activities of regression and classification as forecasting problems, and regarded artificial neural network outputs as discrete or continuous predictands, respectively. Under an i.i.d. restriction on outputs, KFE  \cite{pmlr-v80-kuleshov18a} obtain recalibration procedures that are equivalent to those of DHT \cite{diebold1999multivariate}, now cast as calibration procedures for classification and regression. 

In this work we generalize the work described in \cite{diebold1999multivariate,pmlr-v80-kuleshov18a} in two respects. In the first place, we establish a mathematical framework for treating predictands without i.i.d. restrictions. In the process of doing so, we strongly emphasize the role of conditional information in forecast distributions, and make use of ideas from information theory to characterize the effect of miscalibration. Additionally, we replace the PDF estimation schemes suggested in \cite{diebold1999multivariate} and the isotonic regression adopted by \cite{pmlr-v80-kuleshov18a}  with a Gaussian-process (GP) density estimation scheme, which allows us to also estimate -- with quantified uncertainties -- information entropy measures associated with the estimated pdfs. Using this technique
we demonstrate that if a series of forecasts shows evidence of poor probabilistic calibration then we may use past forecast performance information to produce new current forecasts that have well-calibrated expected distributions, and that have greater expected logarithmic forecast skill score than the original forecasts, irrespective of whether the predictands are i.i.d. Furthermore, the expected performance improvement in logarithmic skill score is computable in advance, together with an uncertainty estimate.

We demonstrate the method using two case studies: a laboratory experiment with a nonlinear circuit and seasonal forecasts of the intensity of the El Ni\~no-Southern Oscillation (ENSO) phenomenon.

\section{Probabilistic Forecasts}

A probabilistic forecast of a continuous scalar random variable $X$
is simply a probability distribution $P(X|I)$ over the value of the
predictand $X$, which is to be observed at a later date. The distribution
is conditioned on prior information $I$ such as current and past
conditions, (often approximate) deterministic and probabilistic model
structure, empirically determined training parameters, and simulation
output. Such forecasts are often generated
as a time series $P(X_n|I_{n},C)$, with $n\in\mathbb{Z}$ an index
that labels time $t_{n}$, so that $n_{2}>n_{1}\implies t_{n_{2}}>t_{n_{1}}$.
Here, $X_n$ is the random predictand at time $t_n$, $I_{n}$ represents information that varies with $n$, while
$C$ represents static conditioning information that is constant for
a particular forecasting system. Typically, the information $I_{n}$
is stochastic, and fluctuates randomly with $n$. Consequently,
the distribution $P(X_n|I_{n},C)$ is itself a distribution-valued
random variable \citep{gneiting2013combining,Gneiting_etal_2014}.
Note that implicit in the notation $P(X_n|I_n,C)$ is the assumption that the particular
realization $I_n=i$ completely determines the distribution of $X_n$ irrespective
of $n$, so that $P(X_n|I_n=i,C)=P(X_m|I_m=i,C)$ for $m\neq n$. This will allow us to consistently drop time subscripts from expressions such as $P(X|I,C)$ in what
follows.

As an example, in the case of weather prediction, $I_{n}$ might represent
a discrete vector of weather observations at a finite number of weather
stations over the course of several previous days, while $C$ might
represent climatological data. Another example is provided by the
empirical time-series modeling that underlies many analyses of financial
and economic data, where the $I_{n}$ could be the last
$M$ values of the time series $X_n$ and $C$ the parameters of an autoregressive
moving-average (ARMA) time-series model \citep{cressie2015statistics}.

In our methodological development we assume the system is approximately
stationary, so that secular drifts due to external forcings are ignored.
We also overlook annual-type periodicities, which are in principle
tractable by adding cycle phase information to $I$. One may easily show that a unique distribution
$P(X|I,C)$ always exists in principle. This follows simply from the existence of a unique joint
distribution $P(X,I|C)$, which is ascertainable empirically from
a sufficiently large archive of $(X_n,I_n)$ values. The forecast distribution
$P(X|I,C)$ is then just $P(X,I|C)/P(I|C)$. This unique distribution
is called the \emph{ideal} forecast with respect to the information
$I$ \citep{gneiting2013combining,Gneiting_etal_2014}.

Above and beyond empirical observation, often
some kind of dynamical law exists from which $P(X|I,C)$ could in principle be inferred.
In such cases, however, accurate inference of $P(X|I,C)$
from first principles is often impractical, because either the dynamical law
is not known (as in the case of most time series in economics) or
it is known imperfectly (as is the case with weather forecasting),
or it is not feasibly computable even where it is well understood. 

Weather forecasting furnishes an instructive example. The dynamical
origin of the distribution $P(X|I,C)$ is intelligible in terms of the nonlinear
physics of weather systems.
However, while describable, this forecast distribution is in no way feasibly
computable, because of limitations in model fidelity and in computational
resources. Instead, limited-fidelity computational models \citep{fritsch2004improving,knutti2013robustness}
are used to filter the information in $I$, incorporating techniques
of data assimilation \citep{bengtsson1981dynamic},
evolving ensembles of states not chosen by a fair sampling of the
distribution on the observation-constrained submanifold of the chaotic attractor, and in any case with too
few ensemble members to be sufficiently informative about the distribution's
structure. Postprocessing of a training set of ensembles and corresponding
validation values of X must be used to construct an approximation
to $P\left(X|I,C\right)$ \citep{raftery2005using,Roulston_Smith-2002}. 

Clearly, by the time this approximation has been
constructed, it is no longer necessarily conditioned directly on $I$,
but rather on some highly processed information $J[I]$. In ensemble
NWP, $J[I]$ has both a deterministic aspect (the NWP simulations)
and a stochastic aspect (the selection of the random ensemble of initial
conditions to evolve). Quite generally, we can assume that
some probabilistic model $J\sim P(J|I)$ describes the dependence
of $J$ on $I$. If that mapping should happen to be deterministic,
the probability distribution $P(J|I)$ would degenerate to a product
of Dirac $\delta$-distributions. In general,
the dimensionality of $J$ is not necessarily inferior to that of 
$I$---in the case of NWP, the simulations generate data over grids whose
data mass far exceeds that of the input information. Invariably, however,
the \emph{information content} of $J$ is degraded
in comparison with that of $I$, by the very approximations described
above. This is merely the observation that $P(X|J[I],C)$
is expected to be---and generally is---inferior
to the computationally infeasible $P(X|I,C)$ in quality measures
such as calibration and sharpness (discussed below).

Another practical concern in obtaining $P(X|I,C)$ is the fact that
the static conditioning information $C$ may not be exactly known
and must be estimated from data. For example, even if a financial
time series were known to be well approximated by a stationary autoregressive
process, the process parameters would not in general be known,
and would have to be fit from data. Similarly,
in weather, climatological information would have to be fit from noisy
data.

We will assume that forecasts are appropriately modeled by absolutely continuous
distributions, which may therefore be represented by probability densities
over $X$. The forecasting system converts the information $J_{n}=J[I_{n}]$
and $C$ into a \emph{published forecast} $p(X_n;J_{n},C)$, a
density over $X_n$, at each time $n$. The data-generating process
that is being forecast then generates a realization $X_n=x_{n}$.
Suppose we generate $N$ forecasts and $N$ corresponding observations.
The series of pairs $\left\{ {\cal P}_{n}=\left(x_{n},p(\cdot;J_{n},C)\right),n=1,2,\ldots N\right\} $
is called the \emph{Forecast-Observation Archive }(FOA) \citep{suckling2013evaluation,smith2015towards}.
The pairs ${\cal P}_{n}$ may be viewed as elements of a set called
a \emph{prediction space}, whose mathematical properties were analyzed
in \citep{Gneiting_etal_2014}.

One of the most important tools for assessing the validity of published
forecasts is the \emph{Probability Integral Transform}, or PIT \citep{10.2307/2981683,diebold1998,gneiting2007probabilistic}.
This is defined in terms of the cumulative probability distribution
function $\tilde{F}(\cdot;J,C)$ associated with $p(\cdot;J,C)$, 
\begin{equation}
\tilde{F}(x;J,C)=\int_{-\infty}^{x}dx^\prime\,p(x^\prime;J,C).\label{eq:PIT_def}
\end{equation}
The PIT associated with the pair ${\cal P}_{n}=(X_n=x_{n},p(\cdot;J_{n},C))$
is simply the value $f_{n}\equiv\tilde{F}(x_{n};J_{n},C)$. The reason
for the usefulness of the PIT is that if the density $p(\cdot;J_{n},C)$
correctly models the stochastic behavior of $X_n|J_{n},C$, then the
random variable $F_{n}=\tilde{F}(X_n;J_{n},C)$ must be uniformly
distributed over the interval $\left[0,1\right]$ irrespective of
$J_{n}$. If this is the case, we say that $p(\cdot;J,C)$ is \emph{probabilistically
calibrated }\citep{gneiting2013combining,Gneiting_etal_2014}. Probabilistic
calibration is a desirable feature in a published forecast because
it means that the forecast is ``honest'' about the probabilities
of its quantiles, since those probabilities correspond to long-term
average frequencies. The property of being probabilistically calibrated
may be checked, given a sufficiently large FOA, by histogramming the
values of $F_{n}$ and inspecting the histogram for evidence of nonuniformity
\citep{10.2307/2981683,diebold1998,gneiting2007probabilistic}. All
ideal forecasts are probabilistically calibrated, although the reverse
is not true---many different calibrated
forecasts can easily be constructed, but only one is ideal with respect to the input information.

It is perhaps surprising to realize that calibration, while a desirable
feature of a forecast system, is not sufficient to prefer one forecast
system to another. As discussed in \citep{gneiting2007probabilistic,2001MWRv..129..550H},
it is quite possibly for forecast systems yielding distributions
that vary widely in precision to all be equally probabilistically
calibrated. For example, a ``climatological'' forecast system that
uses only long-term historical averages to make predictions and an
idealized perfect NWP model that makes approximation-free use of current
information $I_{n}$ to make ideal forecasts are both equally probabilistically
calibrated from the point of view of PIT histogram uniformity. Clearly,
the former provides forecasts that are vague compared with those of
the latter, which are more informative and precise. The term \emph{sharpness}
was introduced by Bross and Bross \citep{bross1953design} to characterize this
distinction. It refers to the degree of concentration on small outcome
sets of the published forecast density, and is sometimes expressed
as distributional variance or as width of a central fixed-probability
(e.g., 90\%) interval \citep{gneiting2007probabilistic}. Thus a sharper
forecast is less vague in its predictions than is a
less-sharp one, independently of the relative degree to which their respective PIT histograms
are close to uniform.

Clearly, the difference between probabilistically calibrated
forecasts of different sharpness ---the difference between
the climatological and the idealized NWP forecaster, for example---is 
\emph{purely in the information on which the forecasts are conditioned.}
States of more specific information lead to sharper forecasts. In the case
of the idealized NWP forecaster, for example, a large increase in
the number of available weather stations necessarily leads to sharper
forecasts, while an increase in the measurement uncertainty of current
weather conditions necessarily leads to less-sharp forecasts. The
explicit highlighting of the relevant conditioning information is
therefore essential to the discussion of probabilistic forecasting.

In the case of poorly calibrated forecasts, the source of the misspecification
of the forecast distributions is necessarily to be sought in erroneous
conditioning information, such as model errors that
distort the information borne by the processed input data $J[I]$,
model errors in constructing the published forecast distribution,
or poor approximations encoded in the static conditioning information
$C$. Forecast interpretation in the presence of misinformation is
an important subject in decision support \citep{smith2016integrating}.
One may be confronted with cases of sharp but uncalibrated
forecasts, that are (for example) biased, but which have smaller mean-square error than climatology.
In these cases the incorrect
conditioning information may not entirely be condemned, because such
forecasts can have better predictive skill than climatology. One naturally
wonders about the extent to which this partially correct information
can be exploited to produce probabilistically
calibrated forecast distributions.

In the next section, we will show that knowledge of the PIT histogram
of a sufficiently large FOA can be used to correct a current published
forecast $p(\cdot;J_{n},C)$ prior to the observation of the predictand
$X_n$, to produce a new, updated forecast $p_{1}(\cdot;J_{n},C)$ that
outperforms $p(\cdot;J_{n},C)$, in that it has a better expected logarithmic (``ignorance'')
score (the ignorance score is defined in \citep{Roulston_Smith-2002}, and is further discussed below in \S\ref{subsec:scoring}). This \emph{probabilistic recalibration}
procedure allows us to better exploit the correct part of the conditioning
information.

\section{Probabilistic Recalibration\label{sec:recalibration}}

The essence of the probabilistic recalibration procedure is that the
PIT histogram of a sufficiently large FOA can be subjected to an empirical
fit, so that the underlying distribution may be inferred by regression.
Assuming that current forecasts suffer from the same miscalibration as
those in the FOA, the fit distribution may then be used to correct
a current forecast distribution to produce a new forecast that outperforms
the original by various objective measures, including the ignorance score.
We now set out the procedure. 

Since we have raised the issue of misinformation in connection with
miscalibration, we adopt notation that distinguishes
between distributions that are ideal---that correctly reflect their conditioning
information, that is---and distributions that
may be misinformed or poorly calibrated. In what follows, therefore,
we will reserve the symbol $\pi$ and the notation $\pi\left(A|B\right)$
or $\pi(A=a|B)da$ for the probability density function of a random
variable $A$ correctly conditioned on information $B$, so that $\pi(A|B)$
is ideal. Published forecast densities, which may be ``misinformed'' and
hence incorrectly reflect the dependence on conditioning information,
we denote by simple function notation such as $p(x;J)$.

We denote the input information to the $t=t_n$ published forecast by the random
variable ${\cal J}_n$, whose realization is $J_n$. We assume the
existence of a unique ideal distribution relative to $J,C$ with density
$\pi\left(X|{\cal J}=J,C\right)$. Again, such a
unique forecast clearly exists, by its relation to the empirically ascertainable
joint distribution $\pi(X,{\cal J}|C)$.

A published forecast $p(\cdot;J_{n},C)$ and the corresponding
observations $x_{n}$ give rise to a PIT value
$f_{n}=\tilde{F}(x_{n},J_{n},C)$, which is a realization of a random
variable $F_n\equiv\tilde{F}(X_{n},J_{n},C)$. The variable $F_n$ is
simply a change of random variables from $X_n$, which implies 
an ideal density $\pi(F_n|{\cal J}=J_{n},C)$ for
$F_n$ satisfying
\begin{eqnarray}
\pi\left(X_n=x_n|{\cal J}_n=J_{n},C\right) & = & \pi\left(F_n=\tilde{F}(x_n,J_{n},C)|{\cal J}_n=J_{n},C\right)\frac{d\tilde{F}}{dx}\nonumber \\
 & = & \pi\left(F_n=\tilde{F}(x_n,J_{n},C)|{\cal J}_n=J_{n},C\right)p(x_n;J_{n},C).\label{eq:ChOfVar}
\end{eqnarray}

Equation (\ref{eq:ChOfVar}) connects the published forecast $p(\cdot;J_{n},C)$
to $\pi\left(X_n|{\cal J}_n=J_{n},C\right)$, the unknown ideal forecast
distribution density relative to $J_{n}$, by a pointwise multiplication
with another unknown density $\pi\left(F_n|{\cal J}_n=J_{n},C\right)$. \footnote{For the sake
of simplicity, we assume that published forecasts $p(x;J,C)$ are always non-zero for 
all $x$. If this were not the case, an interval $[x_1,x_2]$ over which $p(x;J,C)$
is zero would be mapped to a single point $F_1$ by the $x\rightarrow F$ change of variables.
Should the ideal distribution density $\pi\left(X|\mathcal{J},C\right)$ happen to have nonzero
probability mass over such an interval, that finite probability would be mapped to the
single point $F_1$. It would then be necessary to represent this effect by an additive
Dirac-$\delta$ distributional component in $\pi(F|\mathcal{J},C)$. Such
a generalization would be cumbersome, and we avoid having to address it by specifying $p(x;J_{n},C)>0$ for all $x$.}

If the observables $F_n$
are i.i.d.,  we may drop the dependence of $\pi\left(F_n|\mathcal{J}_n,C\right)$ on $\mathcal{J}_n$ in Equation (\ref{eq:ChOfVar}). In this case one may proceed
straightforwardly estimating the time-independent distribution $\pi(F|C)$ (where $F$ may be any of the identically-distributed $F_n$)
by regression on the FOA PIT data $\mathcal{F}\equiv\left\{F_{n}:\,n=1,2,\ldots,N\right\}$ as described in DHT \cite{diebold1999multivariate}. Denoting this estimate by
$\pi\left(F|{\cal F},C\right)$ and replacing
$\pi\left(F_n|\mathcal{J}_n,C\right)$ by $\pi\left(F_n|{\cal F},C\right)$ in Equation (\ref{eq:ChOfVar})  results in forecasts with improved calibration
properties. Equivalently, KFE \cite{pmlr-v80-kuleshov18a} perform isotonic regression on what is, in effect,
the i.i.d CDF of the $F_n$ (as opposed to their i.i.d. PDF), to obtain improved probabilistic calibration of
deep learning classifiers and regressors. The theory developed in \cite{diebold1999multivariate,pmlr-v80-kuleshov18a}
does not address the important general case of non-i.i.d. $F_n$, however, because
the regression estimate $\pi\left(F|\mathcal{F},C\right)$  from the FOA averages over all conditioning data $\mathcal{J}$,
and in this sense is a ``climatological'' distribution that is ignorant of current conditionining information $\mathcal{J}_n=J_n$.\footnote{Note that the i.i.d assumption in \cite{diebold1999multivariate,pmlr-v80-kuleshov18a} applies to the distribution of the $F_n$, and not to the forecast distribution of the $X_n$. The latter are not ``climatological'' in that they are conditioned on their individual $\mathcal{J}_n$.}
DFE \cite{diebold1999multivariate} address the i.i.d. restriction by considering published and ideal forecasts belonging to different ``scale-location'' families with the same ``scale-location'' parameters, showing that this case does give rise to i.i.d. $F_n$. The generality of this restriction is problematic, however. As we discuss below in \S\ref{subsec:thinning}, and explicitly demonstrate in \S\ref{subsec:elnino}, it is not uncommon for time-series of $F_n=f_n$ realizations to be exhibit strong correlations. In such cases the i.i.d. assumption on the $F_n$ is simply not tenable.

Equation (\ref{eq:ChOfVar}) is nonetheless the starting point of our probabilistic recalibration procedure:
we will show that if in Equation (\ref{eq:ChOfVar}) we replace the
density $\pi\left(F|{\cal J},C\right)$ with the predictive distribution
density $\pi\left(F|{\cal F},C\right)$ estimated by a Bayesian regression
fit to the FOA PIT data ${\cal F}$, we will obtain a new forecast
distribution which is not ideal, but which nonetheless improves on the logarithmic (``ignorance'') skill score of the published forecast
irrespective of whether the $F$ data are i.i.d.

\subsection{Bayesian PIT-Fit}

We now perform the regression fit to the data ${\cal F}$. Diebold et al. \cite{diebold1999multivariate}
recommended either using a kernel estimator, or simply and directly the empirical PIT distribution, whereas
Kuleshov et al. \cite{pmlr-v80-kuleshov18a} recommended isotonic regression on the PIT CDF. Here we adopt a non-parametric procedure: 
Gaussian process measure estimation
(GPME), described in detail in \ref{sec:AppendixA}. This is a more complicated procedure than previously used for
this task, but it has benefits that will be described presently. As used here, GPME effectively fits functions from an infinite-dimensional
function space to estimate the predictive density $\pi\left(F|{\cal F},C\right)$. 

Our stationarity assumption implies that the FOA PIT data ${\cal F}$ may be viewed as a realization of a process that repeatedly samples a climatologically averaged
distribution, corresponding to many different random realizations
$J$ of the random variable ${\cal J}$ representing the input
information. The climatology gives rise to a distribution $\pi\left({\cal J}|C\right)$,
which we use to average $\pi\left(F|J,C\right)$, obtaining
\begin{eqnarray}
\pi\left(F|C\right) & = & \int dJ\,\pi\left({\cal J}=J|C\right)\times\pi\left(F|{\cal J}=J,C\right)\nonumber \\
 & = & E_{{\cal J}}\left\{ \pi\left(F|{\cal J},C\right)\right\} .\label{eq:pi_F_Lambda}
\end{eqnarray}

The distribution $\pi(F|C)$ is unknown and must be estimated by
regression from the noisy FOA PIT data in ${\cal F}$. This density
function estimate bears uncertainty represented by the Gaussian process
posterior distribution over the density function $\pi(F|C)$ given
${\cal F}$. This uncertainty is purely epistemic, in contrast to
the uncertainty consequent on the stochastic nature of ${\cal J}$,
represented by the climatological distribution $\pi({\cal J}|C)$. 

We will notationally represent the uncertainty in the determination
of the ideal density function $\pi(X|{\cal J},C)$ using a density
function-valued random variable $\Pi(X|{\cal J},C)$, whose realizations
are possible densities $\pi(X|{\cal J},C)$. We refer to density function-valued
random variables such as $\Pi(X|{\cal J},C)$ as \emph{imperfectly known
distributions}. The distribution $\Pi(X|{\cal J},C)$ is imperfectly
known because our knowledge of it comes from a database of time series
of pairs (${\cal J}$,$X$), from which the joint distribution $\pi(X,{\cal J}|C)$,
and hence $\pi(X|{\cal J},C)$, could be estimated empirically, with
considerable uncertainty.

One may use Equation (\ref{eq:ChOfVar}) to define the imperfectly known
distribution $\Pi(F=f|{\cal J}=J,C)=\Pi\left(X=x|{\cal J}=J,C\right)/p\left(x;J,C\right)$,
where $\tilde{F}(x;J,C)=f$, whose realizations are possible densities
$\pi(F|{\cal J},C)$.

The uncertainty in $\pi(F|C)$ is then represented by an imperfectly known
distribution $\Pi(F|C)\equiv E_{{\cal J}}\left[\Pi(F|{\cal J},C)\right]$,
whose realizations are possible density functions $\pi(F|C)$. The
prior distribution over $\Pi(F|C)$ is described in GPME by a Gaussian
process over $\log\Pi(F|C)$ with a chosen kernel $K(f_{1},f_{2})$
(here squared-exponential) and a constant mean function. The posterior
distribution over $\Pi(F|C)$ given ${\cal F}$ is described in GPME by
an updated Gaussian process over $\log\Pi(F|C)$, with a mean function
$\lambda(f)$ given by Equation (\ref{eq:lambda_pred_fn}), and a
covariance $C(f_{1},f_{2})$ given by Equation (\ref{eq:C_pred_fn}).

The \emph{predictive distribution} $\pi(F|{\cal F},C)$ is the expectation of $\Pi(F|C)$ under this posterior distribution $\Pi(F|C)|{\cal F}$, that is
\begin{eqnarray}
\pi\left(F|{\cal F},C\right) & = & E_{\Pi(F|C)|{\cal F}}\left\{ \Pi(F|C)\right\},\label{eq:Pred_dist} \\
 & = & E_{\Pi(F|C)|{\cal F}}\left\{ E_{{\cal J}}\left[\Pi(F|{\cal J},C)\right]\right\} .\label{eq:Pred_dist2}
\end{eqnarray}
Equation (\ref{eq:Pred_dist}) expresses the operation by which $\pi\left(F|{\cal F},C\right)$ is obtained from the GPME posterior distribution.
Equation (\ref{eq:Pred_dist2}) illustrates the fact that the predictive distribution incorporates both epistemic
fit uncertainties and climatological averaging. This distribution, which
is the key quantity enabling the probabilistic recalibration procedure, is
given explicitly in Equation (\ref{eq:pi_E}) and is the
principal output of the GPME procedure.

\subsection{The Recalibration Procedure\label{subsec:The-Recalibration-Procedure}}

As adumbrated above, the recalibration procedure consists in replacing
the published forecast density $p(x;J,C)$ with the recalibrated forecast
density $p_{1}(x;J,C)$ given by the \emph{recalibration equation}
\begin{equation}
p_{1}(x;J,C)=\pi\left(F=\tilde{F}(x;J,C)|{\cal F},C\right)\times p(x;J,C),\label{eq:Recalibration}
\end{equation}
which is obtained from Equation (\ref{eq:ChOfVar}) simply through
the replacement of $\pi(F|{\cal J},C)$ by the predictive distribution
${\cal \pi}(F|{\cal F},C)$, estimated by the GPME procedure.

We may gauge how much has been gained by the recalibration procedure,
by introducing the Kullback-Leibler divergence or relative entropy
\citep{kullback1951} of a density $f_{1}(x)$ relative to another
density $f_{2}(x)$,
\begin{equation}
KL[f_{2}\,||\,f_{1}]=\int dx\,f_{2}(x)\,\log_{2}\frac{f_{2}(x)}{f_{1}(x)},\label{eq:KL_Div}
\end{equation}
which may be usefully viewed as a measure of the information embodied
by $f_{2}$ relative to a prior state of information that is embodied
by $f_{1}$. The divergence $KL[f_{2}\,||\,f_{1}]$ has the well-known
property of being non-negative definite, and of being zero only if
$f_{1}=f_{2}$ almost everywhere. 

By setting $f_{2}$ to the imperfectly known ideal distribution $\Pi(X|{\cal J}=J,C)$
and setting $f_{1}$ alternatively to the published forecast $p(x;J,C)$
and to the recalibrated forecast $p_{1}(x;J,C)$, we may, by taking
expectations over the climatology ${\cal J}$ and over the posterior
GPME model of the density $\Pi(F|C)\,|\,{\cal F}$, assess whether
the ideal distribution is closer in information to the recalibrated
forecast than the original published forecast.

This leads to the following theorem:
\begin{thm}
\label{thm:Thm1}(i) The distribution $\Pi\left(X=x|{\cal J}=J,C\right)$
is closer to the recalibrated forecast $p_{1}(x;J,C)$ in expected
(over ${\cal J}$ and $\Pi(F|C)\,|\,{\cal F}$) relative entropy than
it is to the published forecast $p(x;J,C)$ unless $p_{1}=p$ almost
everywhere. (ii) The recalibrated forecast $p_{1}(x;J,C)$ is on average
(over $\Pi(F|C)\,|\,{\cal F}$) probabilistically calibrated.
\end{thm}
To prove (i), we first define the difference in the relative entropies,
\begin{eqnarray}
\Delta s & \equiv & KL\left[\Pi(X|{\cal J}=J,C)\,||\,p(\cdot;J,C)\right]-KL\left[\Pi(X|{\cal J}=J,C)\,||\,p_{1}(\cdot;J,C)\right]\nonumber \\
 & = & \int dx\,\Pi(X=x|{\cal J}=J,C)\,\log_{2}\left(\frac{p_{1}(x;J,C)}{p(x;J,C)}\right)\nonumber \\
 & = & \int_{0}^{1}df\,\Pi(F=f|{\cal J}=J,C)\,\log_{2}\left[\pi(F=f|{\cal F},C)\right].\label{eq:DeltaI}
\end{eqnarray}
This quantity is a random variable in consequence of the stochastic
nature of ${\cal J}$ and the epistemic uncertainty in $\Pi(F|{\cal J},C)\,|\,{\cal F}$.
Taking the required expectations, we obtain
\begin{eqnarray}
E_{\Pi(F|C)\,|\,{\cal F}}\left\{ E_{{\cal J}}\left[\Delta s\right]\right\}  & = & E_{\Pi\left(F|C\right)\,|\,{\cal F}}\left\{ \int_{0}^{1}df\,\Pi\left(F=f|C\right)\,\log_{2}\left[\pi(F=f|{\cal F},C\right]\right\} \nonumber \\
 & = & \int_{0}^{1}df\,\pi\left(F=f|{\cal F},C\right)\,\log_{2}\left[\pi\left(F=f|{\cal F},C\right)\right]\nonumber \\
 & \ge & 0,\label{eq:Thm}
\end{eqnarray}
with equality holding only when $\pi(F=f|{\cal F})=1$ almost everywhere,
which is to say, when the original distribution was probabilistically
calibrated. In this case, $p_{1}=p$.

To see (ii), observe that the cumulative distribution of $p_{1}(x;J,C)$
defines a change of variables from the random variable $F$ to a new
random variable $G$ through the function $\tilde{G}(F;C)$ given
by
\begin{eqnarray}
\tilde{G}(f;C) & \equiv & \int_{-\infty}^{\tilde{F}^{-1}(f;J,C)}dx^{\prime}\,p(x^{\prime};J,C)\,\pi(F=\tilde{F}(x^{\prime};J,C)|{\cal F},C)\nonumber \\
 & = & \int_{0}^{f}df^{\prime}\,\pi(F=f^{\prime}|{\cal F},C).\label{eq:GTilde}
\end{eqnarray}
The function $\tilde{G}(\tilde{F}(x;J,C);C)$ is the PIT function
of the recalibrated forecast distribution $p_{1}(x;J,C)$. In terms
of the imperfectly known distribution $\Pi(F=f|{\cal J},C)$, the
imperfectly known distribution $\Pi(G=g|{\cal J},C)$ is
\begin{eqnarray}
\Pi(G=g|{\cal J},C) & = & \frac{\Pi(F=f|{\cal J},C)}{|d\tilde{G}/df|}\nonumber \\
 & = & \frac{\Pi(F=f|{\cal J},C)}{\pi(F=f|{\cal F},C)},\label{eq:PiG1}
\end{eqnarray}
where $g=\tilde{G}(f;C)$. The imperfectly known PIT distribution
for $p_{1}$ is obtained by averaging over ${\cal J}$:
\begin{eqnarray}
\Pi(G=g|C) & = & E_{{\cal J}}\left\{ \Pi(G=g|{\cal J},C)\right\} \nonumber \\
 & = & \frac{\Pi(F=f|C)}{\pi(F=f|{\cal F},C)}.\label{eq:PiG2}
\end{eqnarray}
Averaging over the imperfectly known distribution $\Pi(F|C)\,|\,{\cal F}$
and using Equation (\ref{eq:Pred_dist}), we find
\begin{eqnarray}
E_{\Pi(F|C)\,|\,{\cal F}}\left\{ \Pi(G=g|C)\right\}  & = & \frac{\pi(F=f|{\cal F},C)}{\pi(F=f|{\cal F},C)}\nonumber \\
 & = & 1.\label{eq:PiG_is_calibrated}
\end{eqnarray}

Hence, on average over $\Pi(F|C)\,|\,{\cal F}$, the PIT of the recalibrated
forecast $p_{1}$ is uniform. $\square$.

Note a curious feature of the theorem: it does not use any fact about
the GPME procedure, other than that it allows averaging over the posterior
distribution $\Pi(F|C)\,|\,{\cal F}$. The statements of the theorem
would be true given \emph{any} such distribution, even one that is
wildly incorrect about the likely shape of the true $\pi(F|C)$. If the posterior distribution over $\Pi(F|C)\,|\,{\cal F}$ did happen
to be wildly wrong, however, the theorem, while still true, would no longer furnish
the basis for a working recalibration procedure. The reason is that the
actual data-generating process produces an observable value of $E_{{\cal J}}\left\{ \Delta s_{True}\right\} $
given by 
\begin{equation}
\Delta S_{True}\equiv E_{{\cal J}}\left\{ \Delta s_{True}\right\} =\int_{0}^{1}df\,\pi(F=f|C)\log_{2}\pi(F=f|{\cal F},C),\label{eq:RealKL}
\end{equation}
where the subscript indicates that the true
(unknown) distribution $\pi(F|C)$ is used in the average, and where the distinction
between $\Delta s$ and $\Delta S$ is that the latter is averaged over $\mathcal{J}$.
This expression differs from the expression in Equation (\ref{eq:Thm}) and
is under no obligation to be non-negative definite. This expression
can be expected to be strongly positive only when $\pi(F|{\cal F},C)$
approaches $\pi(F|C)$, which is to say when the fitting procedure
results in a posterior distribution that is well concentrated near
density functions that look a lot like $\pi(F|C)$. The success of the density-fitting
regression procedure is therefore essential to the success of the
recalibration procedure.

This observation applies equally to other styles of regression estimates for $\pi(F|{\cal F},C)$, including the kernel density and histogram estimators of DHT \cite{diebold1999multivariate}, and the isotonic regression of KFE \cite{pmlr-v80-kuleshov18a}. So long as those procedures succeed in furnishing an estimate of $\pi(F|{\cal F},C)$ that is reasonably close to $\pi(F|C)$ and not too uncertain they too should produce recalibrated forecasts that are on average (over the uncertainty in those estimates) informationally closer to the ideal forecast than is the published forecast. The novel thing here is that we can now see that this ought to happen \textit{irrespective of whether the} $F_n$ \textit{are i.i.d.}. This is an important observation, since it was not previously clear to what extent correlations among the $F_n$ could be expected to damage or even vitiate recalibration. As a result, the work of KFE and DHT can be seen to have broader applicability than might otherwise have been believed to be the case.
 
The recalibrated forecast $p_{1}(x;J,C)$, while on-average probabilistically
calibrated, is not in general the same distribution as the unique
ideal distribution $\pi(X=x|{\cal J}=J,C)$, even when the size of
the FOA is large and the uncertainty in $\Pi(F|C)$ is very small.
With a sufficiently large database of pairs $(x,J)$, one could empirically
estimate $\pi(X=x|{\cal J}=J,C)$ and discern its differences from
$p_{1}(x;J,C)$. What the theorem establishes is that on average $p_{1}$
is a better approximation to the ideal forecast than $p$, in the
sense that it is closer in information to the ideal forecast.

\subsection{Scoring, Betting, and Information\label{subsec:scoring}}

A natural connection exists between the relative entropy difference
$\Delta S$ and the ignorance score \citep{Roulston_Smith-2002}.
For continuous predictands $X$, the ignorance score $\textrm{Ign}[p]$
of a forecast distribution $p(x;J,C)$ is defined, using the climatological
density $\pi(X|C)$ as a reference distribution, by the expression
\begin{eqnarray}
\textrm{Ign}[p] & = & -E_{{\cal J}}\left\{ E_{X|{\cal J},C}\left[\log_{2}\frac{p(X;{\cal J},C)}{\pi(X|C)}\right]\right\} \nonumber \\
 & = & -\int dJdx\,\pi(X=x,{\cal J}=J|C)\,\log_{2}\frac{p(x;J,C)}{\pi(X=x|C)}.\label{eq:IGN}
\end{eqnarray}
Because this final average is over the joint distribution on $X,{\cal J}$,
the ignorance score may be estimated empirically from an FOA simply
by the average
\begin{equation}
\textrm{Ign[p]}\approx-\frac{1}{N}\sum_{n=1}^{N}\log_{2}\frac{p(x_{n};J_{n},C)}{\pi(X=x_{n}|C)},\label{eq:IGN_Empirical}
\end{equation}
whose expected value is the expression in Equation (\ref{eq:IGN}).
Note that for i.i.d. predictands $x$, the expression on the RHS of Equation (\ref{eq:IGN_Empirical}) is proportional to the log-likelihood for the model represented by $p(x;J,C)$,
up to an additive constant. The resemblance is purely formal for non-i.i.d. predictands, however.

The difference in the ignorance scores of the published and recalibrated
forecasts is
\begin{eqnarray}
\Delta\textrm{Ign}[p_{1},p] & \equiv & -E_{{\cal J}}\left\{ \int dx\,\pi(X=x|{\cal J},C)\,\log_{2}\frac{p_{1}(x;{\cal J},C)}{p(x;{\cal J},C)}\right\} \nonumber \\
 & = & -\int_{0}^{1}df\,\pi(F=f|C)\,\log_{2}\pi(F=f|{\cal F},C)\nonumber \\
 & \approx & -E_{\Pi(F|C)\,|\,{\cal F}}\left\{ E_{{\cal J}}\left[\Delta s\right]\right\} \nonumber \\
 & \le & 0,\label{eq:Delta_IGN}
\end{eqnarray}
where in the third line we have approximated $\pi(F|C)$ by $\pi(F|{\cal F},C)$
and in the last line we appealed to Theorem \ref{thm:Thm1}. We therefore
expect that the recalibrated forecast $p_{1}$ will have a lower (i.e.,
better) ignorance score than the original published forecast. By the
standards of the ignorance score, then, $p_{1}$
is an improvement on $p$.

It was pointed out in \citep{Roulston_Smith-2002,hagedorn2009communicating}
that the Ignorance Score has an interpretation as a tool for practical
decision-making under uncertainty. The interpretation is couched in
terms of a horse race, in which a bettor and an oddsmaker make optimal
decisions about their choices with respect to the discrete possible
outcomes of a horse-race. Kelly \citep{6771227} described a strategy
for a player allocating wealth to bets on $N$ outcomes $i=1,\ldots,N$
offering wealth multiplier odds $o_{i}$, assuming the player works
with a forecast distribution $h_{i}$, $\sum_{i=1}^{N}h_{i}=1$. Kelly
showed that the optimal strategy for the player is to allocate wealth
$W_{i}$ to bet on outcome $i$ according to the rule $W_{i}=Wh_{i}$.
By contrast, the optimal strategy for the pari-mutuel bookmaker with
a forecast distribution $g_{i}$ is to set odds $o_{i}=1/g_{i}$.
With these strategies for wagering and odds-setting, supposing the
ideal forecast is $p_{i},$ the player's wealth grows at the expected
rate $2^{\Delta I}$, where
\begin{eqnarray}
\Delta I & = & \sum_{i=1}^{N}p_{i}\log_{2}\frac{h_{i}}{g_{i}}\nonumber \\
 & = & KL\left[p||g\right]-KL\left[p||f\right]\nonumber \\
 & = & -\Delta\textrm{Ign}[f,g]\label{eq:Kelly}
\end{eqnarray}
is the difference between the entropy divergences of the two forecast
distributions relative to the true distribution $p_{i}$, and hence also
the negative Ignorance Score difference. This game can be regarded
as a symmetric game between two forecasters who alternate the roles
of bookmaker and bettor---so long as the current bookmaker
with forecast $g_{i}$ sets odds in reciprocal proportion to $g_{i}$
and the bettor with forecast $h_{i}$ sets bets in proportion to $h_{i}$,
Equation $($\ref{eq:Kelly}) shows that it makes no difference which
is which. A player with a forecast system that is consistently closer
in information to the ideal distribution $p_{i}$ (and hence has a lower
ignorance score) than the other player's system has a positive expected
wealth growth rate in this game. 

This connection is attractive because it furnishes an
example of decision-making under uncertainty that is
improved by using forecasts with lower ignorance scores, and in particular
by using the recalibration procedure described above. The view that
underpins this work is that there ought to be \emph{some} form of
symmetric-rule game in which decisions made according to a higher-scoring
forecasting system should consistently dominate those made under a
lower-scoring one. This view of forecast quality differs from the standard
``proper scoring'' outlook \citep{Gneiting_etal_2014,gneiting2007strictly,Roulston_Smith-2002,good1952rational,brier1950verification},
in which forecasts vie amongst one another for high scores using scoring
rules designed to encourage forecaster honesty and furnish usable
utility functions for estimation. It is in better correspondence with the
Bayesian decision theory outlook on proper scoring \cite{dawid1999coherent,gneiting2007strictly}.

The Kelly horse race is not ideal for furnishing a
fiducial symmetric-rule game for continuous probabilistic forecast
assessment since it concerns itself with discrete outcomes and contains
an appearance of asymmetry in the bettor-odds--maker model that it is
based on. It can be adapted to continuous forecast distributions,
for example by using fine quantiles (such as percentiles) as outcomes
to be wagered on. A more natural symmetric-rule game can be
described, however, by observing that we may recast the second line of Equation
(\ref{eq:Delta_IGN}) as
\begin{eqnarray}
-\Delta\textrm{Ign}[p_{1},p] & = & \Delta S_{True}\nonumber \\
 & = & \int dJdx\,\pi(X=x,{\cal J}=J|C)\,\log_{2}\frac{p_{1}(x;J,C)}{p(x;J,C)}.\label{eq:Egame_Exp}
\end{eqnarray}

Observe that the right-hand side of Equation (\ref{eq:Egame_Exp}) may be interpreted
as the expected winnings in a symmetric-rule game, which we call the
\emph{entropy game}. The rules of the game are as follows: two players
compete, one using the published forecast distribution $p(\cdot;J,C)$,
the other using the recalibrated distribution $p_{1}(\cdot;J,C)$.
At the $n$-th turn of the game, values $J_{n}$,$x_{n}$ are observed
from the data-generating distribution $\pi({\cal J},X|C)$. The players
compute the log-ratio of their respective densities at $x_{n}$, $w=\log_{2}\left[p_{1}\left(x_{n};J_{n},C\right)/p(x_{n};J_{n},C)\right]$.
If $w>0$, then the player using the published forecast $p$ pays
the amount $w$ to the player using the recalibrated distribution
$p_{1}$. Otherwise, the recalibrated player pays $|w|$ to the player
using $p$. Equation (\ref{eq:Egame_Exp}) states that $\Delta S_{True}$
is the expected winnings per turn of the recalibrated player.

The entropy game may be viewed as the logarithm of the Kelly horse race,
since the expected per turn winnings of the entropy game are in fact
equal to the log (base 2) of the expected wealth increase rate per
turn of a bettor at a horse racing track. Kelly-style bets on (say)
percentiles of the original forecast distribution have an expected
wealth growth rate whose logarithm is approximately equal to $\Delta S_{True}$.
The entropy game is a useful alternative to the Kelly horse race because
it is better adapted to continuous distributions and because its
rules present a more symmetric appearance than does the bettor-and-bookie
model of the horse race. It is our fiducial game for assessing the
improvement in recalibrated forecasts, and we will make extensive
use of it in what follows.

\subsection{Predicting Recalibrated Forecast Performance}

The entropy game winnings $\Delta S_{True}$ in Equation (\ref{eq:Egame_Exp})
are expressed in terms of the distribution $\pi(X,J|C)\propto\pi(X|J,C)$,
which is known only approximately through the GPME fit. An important
and useful feature of the GPME procedure is that it allows us to compute
expected entropy game winnings per turn in advance of the game. 

We first simplify $\Delta S_{True}$ by re-expressing it in PIT-space,
using the fact that $\pi(X=x|J,C)dx=\pi(F=f|J,C)df$:
\begin{eqnarray}
\Delta S_{True} & = & \int dJ\,\pi({\cal J=}J|C)\int dx\,\pi(X=x|{\cal J}=J,C)\log_{2}\frac{p_{1}(x;J,C)}{p(x;J,C)}\nonumber \\
 & = & \int_{0}^{1}df\,\pi(F=f|C)\log_{2}\pi\left(F=f|{\cal F},C\right).\label{eq:DS_True}
\end{eqnarray}
Note that unlike the expression in Equation (\ref{eq:Thm}), $\Delta S_{True}$
is not non-negative definite but rather may in principle attain negative
values if $\pi(F|{\cal F},C)$ is a poor estimate of $\pi(F|C)$.

We define the quantity
\begin{equation}
\Delta S\equiv E_{\mathcal{J}}\left\{\Delta s\right\}=\int_{0}^{1}df\,\Pi(F=f|C)\log_{2}\pi\left(F=f|{\cal F},C\right),\label{eq:Egame_RV}
\end{equation}
which is a random variable (unlike $\Delta S_{True}$) in consequence
of the use of the distribution-valued random variable $\Pi(F|C)$
for the average. This randomness expresses our epistemic uncertainty
about the value of $\Delta S_{True}$ consequent on our uncertainty
about the shape of the distribution $\pi(F|C)$. The probability distribution
of $\Delta S$ is not directly computable by using GPME. However, we
can calculate
\begin{equation}
\overline{\Delta S}\equiv E_{\Pi(F|C)\,|\,{\cal F}}\left\{ \Delta S\right\} \label{eq:Egame_PredWin}
\end{equation}
and
\begin{equation}
\textrm{Var}\left(\Delta S\right)\equiv E_{\Pi(F|C)\,|\,{\cal F}}\left\{ \Delta S^{2}\right\} -\left(\overline{\Delta S}\right)^{2}.\label{eq:Egame_PredVar}
\end{equation}

These quantities represent the expectation and variance of $\Delta S$
with respect to the epistemic uncertainty contained in the GPME posterior
distribution over the density $\Pi(F|C)\,|\,{\cal F}$. They constitute
predictions of the outcome of rounds of the entropy game to be conducted
out of sample with respect to the FOA training data that furnishes
${\cal F}$. Thus, not only can we verify the improvement in the recalibrated
forecast using many rounds of the entropy game out of sample, but
we can also predict in advance, with uncertainty bounds, what the
average outcome of those games will be. The degree to which the predictions
match the outcomes can be a useful gauge of model validity, as we
will see below.

The expression for $\overline{\Delta S}$ is readily obtained (even
without appealing to the GPME theory): 
\begin{eqnarray}
\overline{\Delta S} & = & E_{\Pi(F|C)\,|\,{\cal F}}\left\{ \int_{0}^{1}df\,\Pi(F=f|C)\log_{2}\pi\left(F=f|{\cal F},C\right)\right\} \nonumber \\
 & = & \int_{0}^{1}df\,\pi(F=f|{\cal F},C)\log_{2}\pi(F=f|{\cal F},C).\label{eq:DS1}
\end{eqnarray}
That is, $\overline{\Delta S}$ is just $KL[\pi(F|{\cal F},C)\,||\,U(F)]$,
where $U(F)$ is the uniform distribution on $[0,1]$. Unsurprisingly,
we have $\overline{\Delta S}\ge0$. Note that it is not the case in
general that $\overline{\Delta S}=\Delta S_{True}$, since the weighted
averages in Equations $($\ref{eq:DS_True}) and (\ref{eq:DS1}) are
different. The two quantities are ``close'' only to the extent that
the distribution $\pi(F|{\cal F},C)$ approaches $\pi(F|C)$. The
GPME model fit makes certain approximations (such as statistical independence
of the data in ${\cal F}$, and the Laplace approximation for the
log-density) and assumptions (such as kernel and mean function choice)
that can in principle produce model errors that make $\overline{\Delta S}$
a flawed estimator for $\Delta S_{True}$. As we will see in the case
studies below, GPME does a good modeling job in general, and such
modeling errors can be acceptably small in practical cases. 

Note also that despite the fact that $\overline{\Delta S}\ge0$, it
is not the case that the random variable $\Delta S\ge0$, as can be
seen from Equation (\ref{eq:Egame_RV}). The random variable $\Delta S$
itself may certainly attain negative values for some realizations
$\pi(F|C)$ of $\Pi(F|C)$, just as $\Delta S_{True}$ could in principle
turn out to be negative if $\pi(F|{\cal F},C)$ badly mis-estimates
$\pi(F|C)$. It would be reassuring to have some measure of how unlikely
it is for $\Delta S<0$. This can be obtained from the variance of
$\Delta S$.

We reproduce here the GPME expression for $\textrm{Var}(\Delta S)$,
given in Equation (\ref{eq:VarDeltaS}):
\begin{eqnarray}
\textrm{Var}(\Delta S) & = & \int_{0}^{1}df_{1}\int_{0}^{1}df_{2}\left[\pi(F=f_{1}|{\cal F},C)\log_{2}\pi(F=f_{1}|{\cal F},C)\right]\nonumber \\
 &  & \hspace{2.4cm}\times\left[\pi(F=f_{2}|{\cal F},C)\log_{2}\pi(F=f_{2}|{\cal F},C)\right]\nonumber \\
 &  & \hspace{2.4cm}\times\left[e^{C(f_{1},f_{2})}-1\right],\label{eq:PredVar_GPME}
\end{eqnarray}
where $C(f_{1},f_{2})$ is the GPME posterior covariance over $\ln\text{\ensuremath{\Pi}}(F|C)$.
Using this formula, we may compute the uncertainty in the estimate
$\overline{\Delta S}$ in terms of a straightforward two-dimensional
quadrature on $[0,1]\times[0,1]$. In the Appendix, we show that
in the asymptotic limit $N\rightarrow\infty$ of unlimited training
data, $\textrm{Var}(\Delta S)\sim N^{-1}$. Thus the uncertainty in
the estimate $\overline{\Delta S}$ scales asymptotically as ${\cal O}(N^{-1/2})$.

We may use this fact to introduce a \emph{forecast advantage measure}
(FAM):

\begin{equation}
\textrm{FAM}=\frac{\overline{\Delta S}}{\sqrt{\textrm{Var}\left(\Delta S\right)}},\label{eq:FAM}
\end{equation}
which asymptotically scales as ${\cal O}(N^{1/2})$. The FAM gives
us a measure of a-priori confidence in the positivity of the out-of-sample
entropy game winnings of the recalibrated forecast.

To summarize the results so far: Given an FOA and its associated PIT
data ${\cal F}$, we have a recalibration procedure that allows us
to improve current published forecasts $p(X;J,C$), replacing them
with recalibrated forecasts $p_{1}(X;J,C)$. The recalibrated forecasts
are expected to outperform the original published forecasts in ignorance
score, and, equivalently, in entropy game performance. The extent
of this superior performance---the average per round Entropy
Game winnings $\Delta S_{True}$---may be estimated in
advance using only the data in ${\cal F}$, and that estimate is
attended by an uncertainty that is also computable using only the
data in ${\cal F}$. Our confidence in the positivity of 
$\Delta S_{True}$---that is, in the superiority of the recalibrated forecast---can 
be expressed by the expression in Equation (\ref{eq:FAM})
for the FAM, which in the GPME theory grows with training dataset
size $N$ as $N^{1/2}$. Hence we can reassure ourselves of the superior
performance of the recalibrated forecast by accumulating a sufficiently
large training set.

\subsection{Fit Quality}

As discussed at the end of \S\ref{subsec:The-Recalibration-Procedure},
success of the density estimation procedure is crucial to the success
of the recalibration procedure, since the latter success depends on
the estimated distribution $\pi(F|{\cal F},C)$ not departing too
much from the true distribution $\pi(F|C)$. The GPME theory furnishes
a quantitative measure of the departure $\pi(F|{\cal F},C)$ from
$\pi(F|C)$, through the quantity
\begin{equation}
EI\left[\pi(F|{\cal F},C)\right]\equiv E_{\Pi(F|C)\,|\,{\cal F}}\left\{ \int_{0}^{1}df\,\Pi(F=f|C)\log_{2}\frac{\Pi(F=f|C)}{\pi(F=f|{\cal F},C)}\right\} .\label{eq:EI1}
\end{equation}
The quantity inside the expectation is the K-L divergence of the imperfectly known
distribution $\Pi(F|C)$ from the estimated distribution $\pi(F|{\cal F},C)$.
The GPME theory yields a closed-form expression for this fit quality
measure. It is given in Equation (\ref{eq:ES_2}), which we reproduce
here:
\begin{equation}
EI\left[\pi(F|{\cal F},C)\right]=\frac{1}{2\ln2}\int_{0}^{1}df\,\pi(F=f|{\cal F},C)\times C(f,f),\label{eq:EI2}
\end{equation}
where $C(f_{1},f_{2})$ is the GPME posterior covariance over $\ln\text{\ensuremath{\Pi}}(F|C)$. 

Equation (\ref{eq:EI2}) expresses a very sensible result: the expected
divergence between the estimated probability density and the true
density is proportional to the average of the posterior variance weighted
by the effective posterior probability density. As the quality of
the fit improves, the variance decreases and takes $EI[\pi(F|{\cal F},C)]$
down with it. Thus, in some sense, $EI[\pi(F|{\cal F},C)]$ expresses
fit quality. One must be cautious in interpretation, however, since
no probabilistic interpretation (such as a ``P-value'')
attaches to $EI[\cdot]$, so it is difficult to say in an absolute
sense how small a value of $EI[\cdot]$ is adequate. Moreover, relying
on $EI[\cdot]$ for fit quality is, in effect, asking the model to
report on its own success. The result will be conditioned by model
assumptions (such as covariance kernel choice), and cannot directly
detect the effects of model error on fit quality. 

It is more useful to employ the fact, derived in the Appendix, that
asymptotically, $EI[\pi(F|{\cal F},C)]\rightarrow B/2N$, where $B$
is the number of bins used in the GPME fit and $N$ is the number
of datapoint in ${\cal F}$. The $N^{-1}$ scaling is checkable by
varying the size of the training set ${\cal F}$. This behavior can
be a useful diagnostic of model adequacy, as we will see below, since
departures from this scaling at large $N$ may indicate model
errors that are masked by noise at smaller values of $N$.

\subsection{Thinning Data To Remove Correlations\label{subsec:thinning}}

The GPME theory has the considerable benefit that besides providing
an efficient method to obtain the predictive distribution $\pi\left(F|{\cal F}\right)$,
it also provides closed-form expressions for the desired entropy-related
quantities $EI$, $\overline{\Delta S}$, $\textrm{Var}(\Delta S)$.
It has one serious defect for our application, however: it assumes
that a pointlike Poisson process governs the generation of the PIT
values in ${\cal F}$. This assumption is more often false than true.
In the case of weather, forecast cadences are generally more rapid
than the characteristic times on which the dynamical system loses
memory (hence the adage that the best predictor of tomorrow's weather
is today's weather). This means that successive members of an FOA---both 
forecasts and observations---typically
resemble each other more than do well-separated members. This effect
manifests itself in correlations of successive PIT values in ${\cal F}$ --
examples are displayed in the right panels of Figure \ref{fig:ENSO_1} in \S\ref{subsec:elnino}.
These correlations technically invalidate the point-like Poisson process assumption
that underlies the GPME theory. Two bad consequences are that (1)
the fit may be skewed, especially if the training set is not large;
and (2) the a priori estimates of recalibrated forecast improvement
over base forecast are not reliable, since they are based on an inaccurate
statistical model.

To recover the utility of the GPME theory in such cases,
one must \emph{thin} the training dataset by a factor that
may be inferred from the autocorrelation function of the PIT values
in ${\cal F}$. This is a process analogous to the thinning of samples
output by an MCMC chain \citep[p. 149]{gamerman2006markov} and is
necessary for the same reason: the samples obtained after appropriate
thinning have good independence properties. They may therefore be
appropriately modeled by a point Poisson process. The downside is
that if data is not abundant, the thinning of the training set may
be harmful to predictive performance.

One may, with some justice, ask what was the point of emphasizing the non-i.i.d.
nature of the recalibration procedure, if an i.i.d. restriction is then re-introduced
through the GPME fitting procedure. The answer is that GPME only imposes an i.i.d.
restriction on the \textit{model training}. The results in \S\ref{sec:recalibration} on performance improvement
of recalibrated forecasts are still valid for non-i.i.d. \textit{forecasts} of future events,
given an acceptable, statistically consistent regression estimate of $\pi\left(F|\mathcal{F},C\right)$
from the GPME procedure. There may quite possibly exist a generalization of GPME that
takes proper account of non-i.i.d. behavior in $\mathcal{F}$. Locating such a procedure
would be a promising avenue of future research, since the result would be a recalibration
procedure that is entirely free of the i.i.d. restriction.

Note, however, that the GPME i.i.d. restriction on training data is \textit{only} necessary to preserve the predictive performance of our recalibration procedure -- that is, to be able to state in advance the expected improvement in forecast logarithmic skill. The restriction is \textit{not} necessary to improve forecast skill by some (possibly difficult to predict) amount.  As demonstrated by \cite{diebold1999multivariate} and \cite{pmlr-v80-kuleshov18a}, several different styles of regression on the data $\mathcal{F}$ are capable of furnishing estimates of $\pi(F|\mathcal{F},C)$ that improve calibration.
We can say that the present work advances the state of the art from the work of \cite{diebold1999multivariate,pmlr-v80-kuleshov18a} in that it is now clear that recalibration may be expected to work even in the case where the $F_n$ are not i.i.d., so long as some reasonable regression model for $\pi(F|\mathcal{F},C)$ is produced.

\section{Verification}

We now exhibit practical examples of the forecast recalibration procedure
in two separate applications: a laboratory experiment with predictions
of the output from a nonlinear circuit and a seasonal metereology
example using ensemble forecasts of El Ni\~no Southern Oscillations
(ENSO) temperature fluctuations.

\subsection{A Nonlinear Circuit\label{subsec:A-Nonlinear-Circuit}}

Our first application of the forecasting recalibration scheme is a
laboratory experiment with predictions of the output from a nonlinear
circuit. The circuit was first introduced in
\citep{machete2007modelling} and later discussed in more detail in \citep{machete2013}.
References  \citep{machete2013} and \citep{Reason_Smith_2016} discuss different aspects of its predictability properties. The circuit is designed to produce
output voltages that mimic the Moore-Spiegel \citep{moore1966thermally}
three-dimensional system of ordinary differential equations:
\begin{eqnarray}
\dot{x} & = & y\nonumber \\
\dot{y} & = & -y+Rx-\Gamma(x+z)-Rxz^{2}\label{eq:Moore-Spiegel}\\
\dot{z} & = & x.\nonumber
\end{eqnarray}
This system is a simplified model of a parcel of fluid moving vertically
in a stratified fluid, with which it exchanges heat, while tethered
by a harmonic force to a point \citep{moore1966thermally}. The variable $z$ represents the height of the fluid element. 

The circuit is set to operate at parameter values $R=10$, $\Gamma=3.6$, at which
values the system exhibits chaotic behavior. Voltages $(V_1,V_2,V_3)$ corresponding to the variables $(x,y,z)$ are measured at three points on the circuit. The ODE system of Equations (\ref{eq:Moore-Spiegel}) is scaled to endow its variables with the dimensions of voltage. The voltage $V_3$ corresponding to $z$ is our predictand. As
noted in \citep{machete2013}, the system
of Equations (\ref{eq:Moore-Spiegel}) poorly predicts the behavior
of the circuit due to model imperfection. An alternative prediction model is constructed using radial basis functions.

\begin{figure}[tp]
\begin{centering}
\includegraphics[clip,width=7cm]{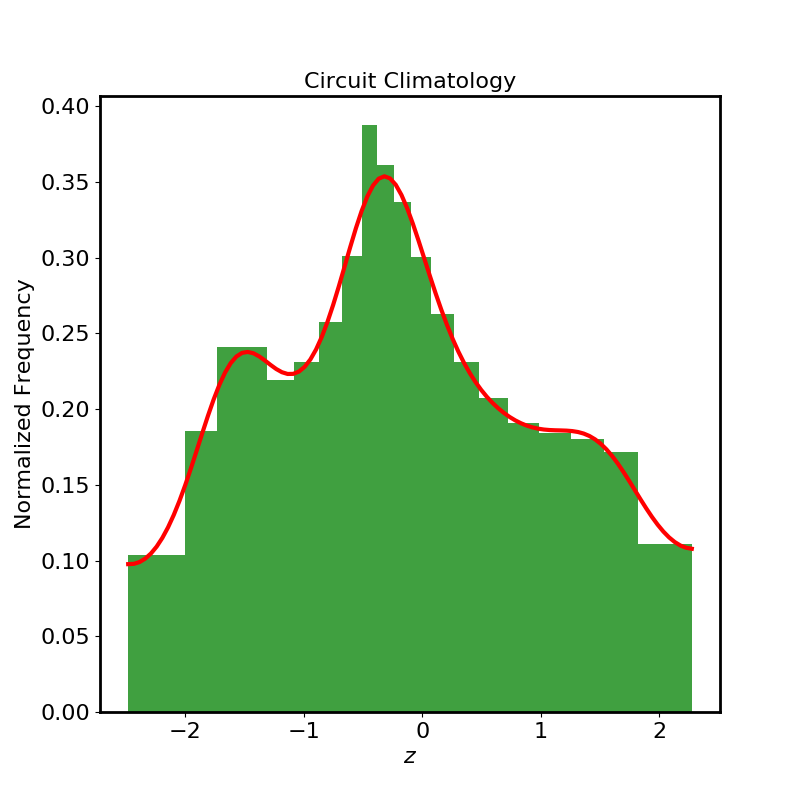}\includegraphics[clip,width=7cm]{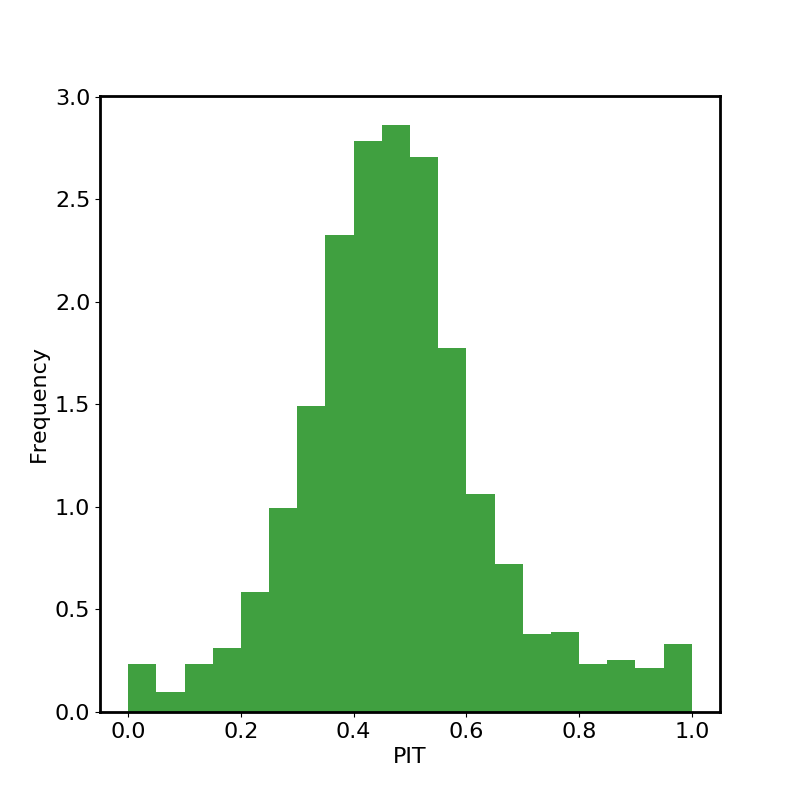}
\par\end{centering}
\centering{}\includegraphics[clip,width=7cm]{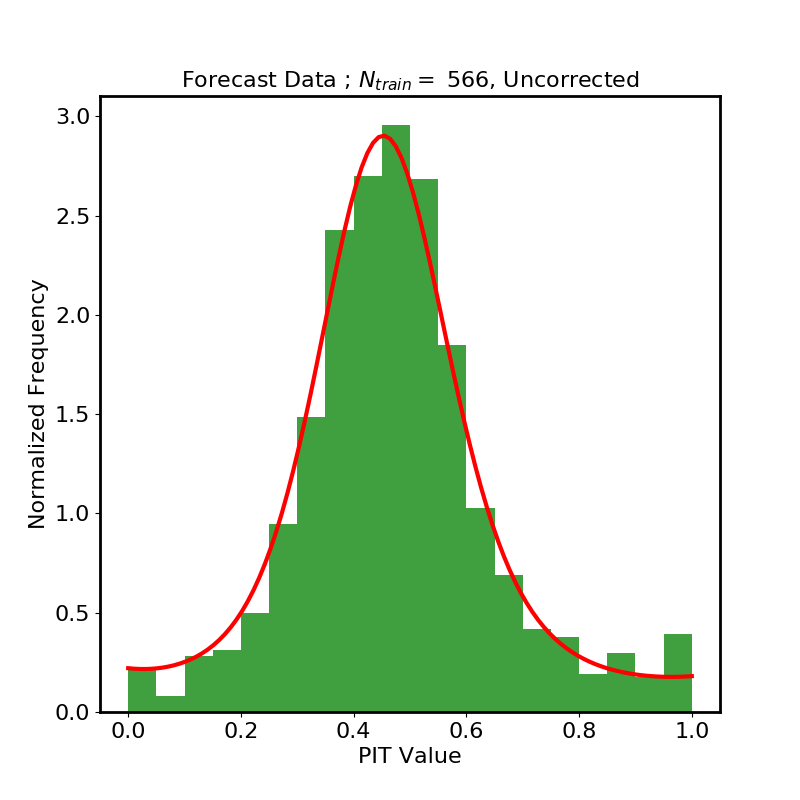}\includegraphics[clip,width=7cm]{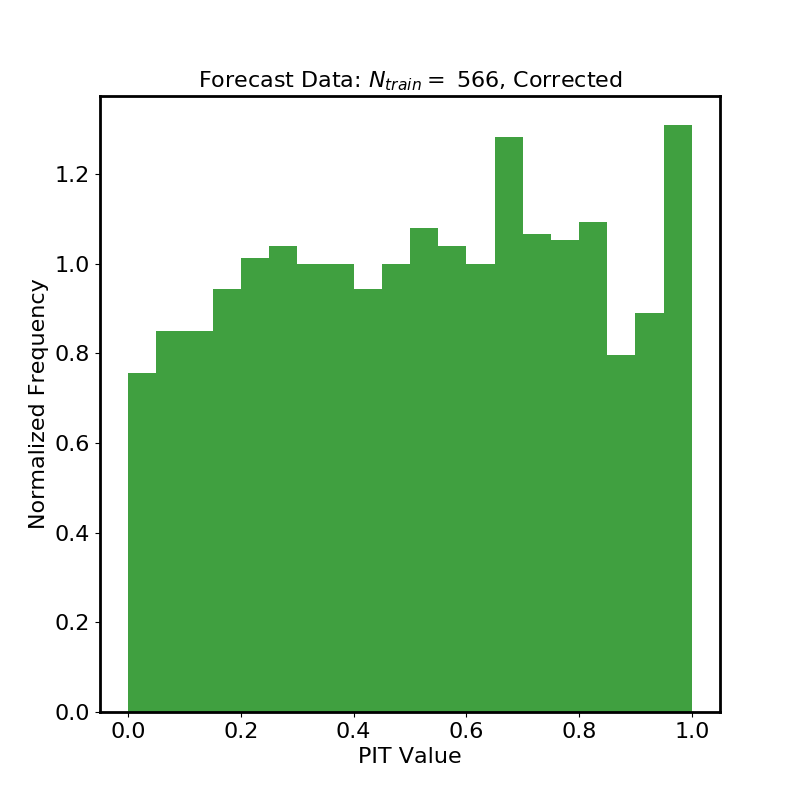}\caption{\label{fig:Circuit_1}Recalibration of nonlinear circuit ensemble
forecasts. Top left: Climatology histogram. The red line shows the
probability density estimate blended into the ensemble forecast. Top
right: PIT histogram of all 2,048 ensemble forecasts. The forecasts
are clearly overdispersed. Bottom left: PIT histogram of forecasts,
with training data excluded, for the case $N_{t}=566$. The red line
shows the density inferred from the training data. Bottom right: PIT
histogram of recalibrated forecasts for the case $N_{t}=566$. The
probabilistic calibration of the recalibrated forecast is excellent,
particularly compared with that of the original forecast.}
\end{figure}

Probe voltages for the three voltage probes were collected over a duration of 14 hours at a sampling rate of 10 kH. A sample of 2000 points corresponding to the $z$ voltage probe was used to empirically estimate the climatological
distribution $\rho(z)$ of $V_3$ (corresponding to $z$) (top-left panel of Figure
\ref{fig:Circuit_1}) . Then 2,048 uncorrelated
  voltage states were sampled to furnish initial conditions from which forecasts could be initialized. Each of these
states was used to create an ensemble of 127 forecasts by small Gaussian additive perturbations about the observed state and evolving
the resulting states using a radial basis function model up to eight time steps ahead, that is, up to a forecast lead time of 0.8 ms. These 0.8 ms lead time forecasts
were then converted to probabilistic forecasts for $V_3$ by kernel
dressing and blending with climatology \citep{brocker2008ensemble},
wherein the forecast distribution density is expressed as a sum of
kernels, each centered at the value of one of the 127 simulation values,
and the result is linearly blended with the climatology $\rho(z)$.
The kernels were chosen to be Gaussians, with equal widths chosen to
minimize the ignorance score, and the linear blending parameter was
also chosen to minimize the ignorance score \citep{Reason_Smith_2016}.

The continuous ensemble forecasts thus generated were compared with
the corresponding observed values of the $z$ voltage to obtain the
PIT distribution shown in the top-right panel of Figure \ref{fig:Circuit_1}.

The 2,048 available observations and forecasts were divided into
training and test sets, with training set sizes $N_{t}\in\{200,283,400,566,800,1131,1600\}$
(each about a factor of $\sqrt{2}$ larger than the previous value).
In each case, the test set comprised all the remaining data.
We carried out the recalibration procedure with each training set
to compute the corresponding PIT posterior predictive density $\pi(F|{\cal F},C)$
and carried out entropy games over the corresponding test sets, recording
the performance predictors ($EI[\pi(F|{\cal F},C)]$, $\overline{\Delta S}$,
$\textrm{Var}(\Delta S)$, FAM) and the game outcomes.

The lower-left panel of Figure \ref{fig:Circuit_1} shows the PIT fit
from the training data (red line) superposed on the PIT histogram
from the test data for the case $N_{t}=566$. The lower-right panel
shows the PIT distribution of the recalibrated forecast for the same
case. Comparing this figure with the one to its
left we can see that the recalibration procedure successfully produced
updated forecasts that are probabilistically calibrated.

\begin{figure}[p]
\begin{centering}
\includegraphics[width=7cm]{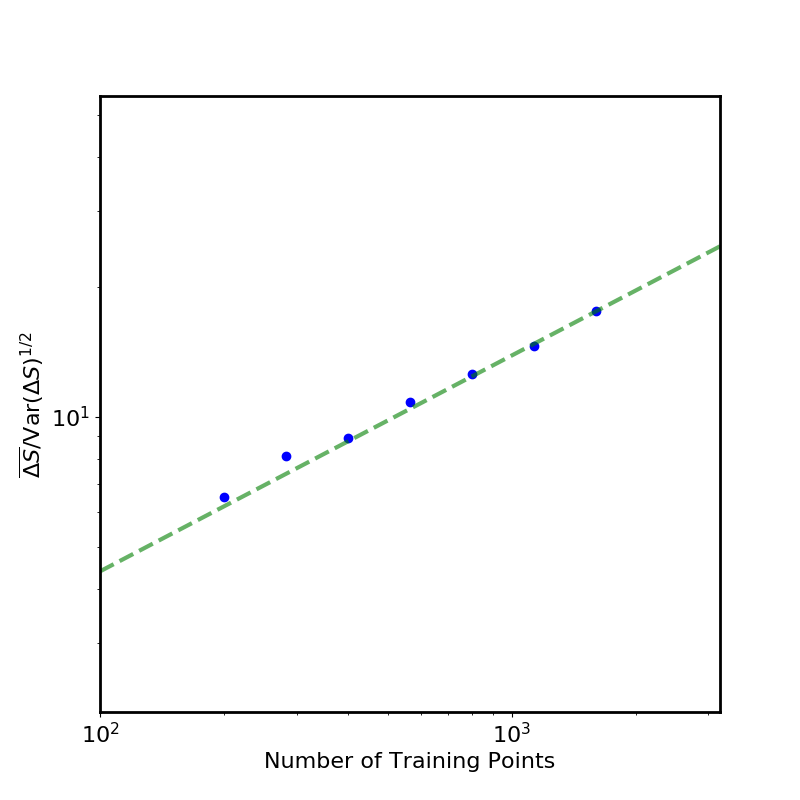}\includegraphics[width=7cm]{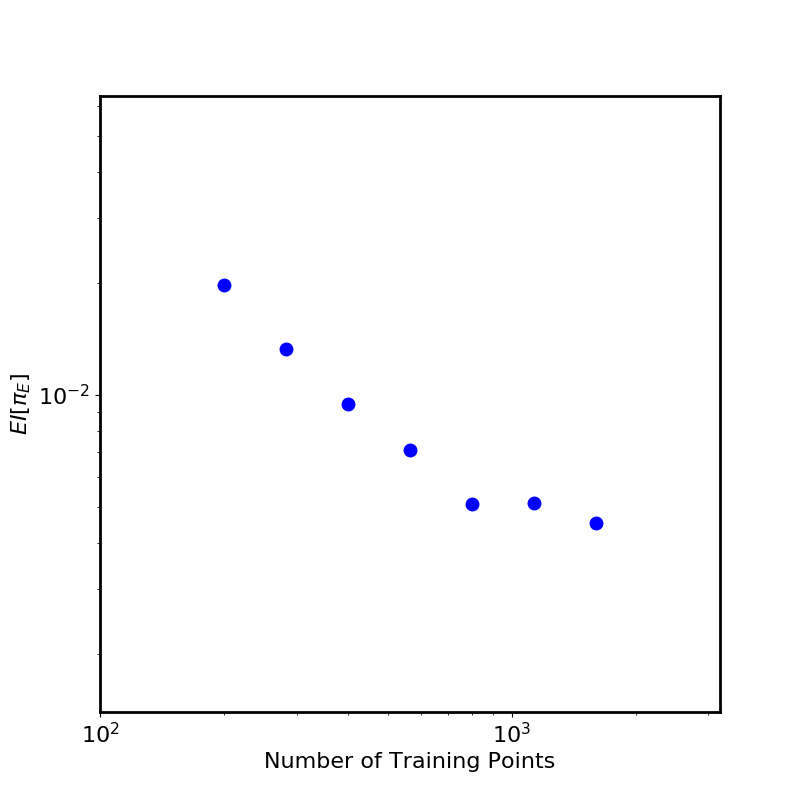}
\par\end{centering}
\centering{}\includegraphics[width=7cm]{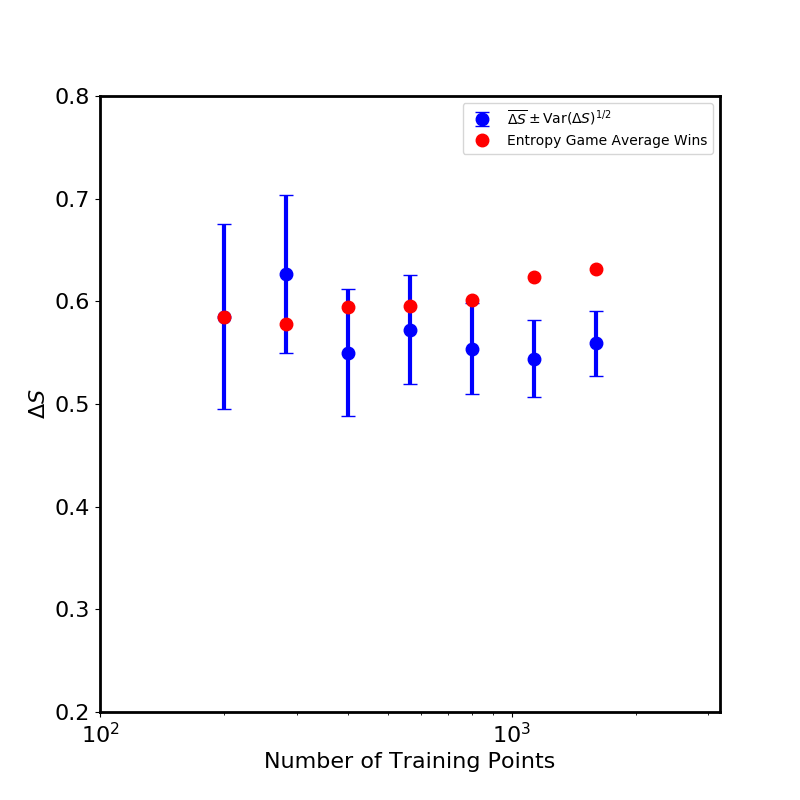}\includegraphics[width=7cm]{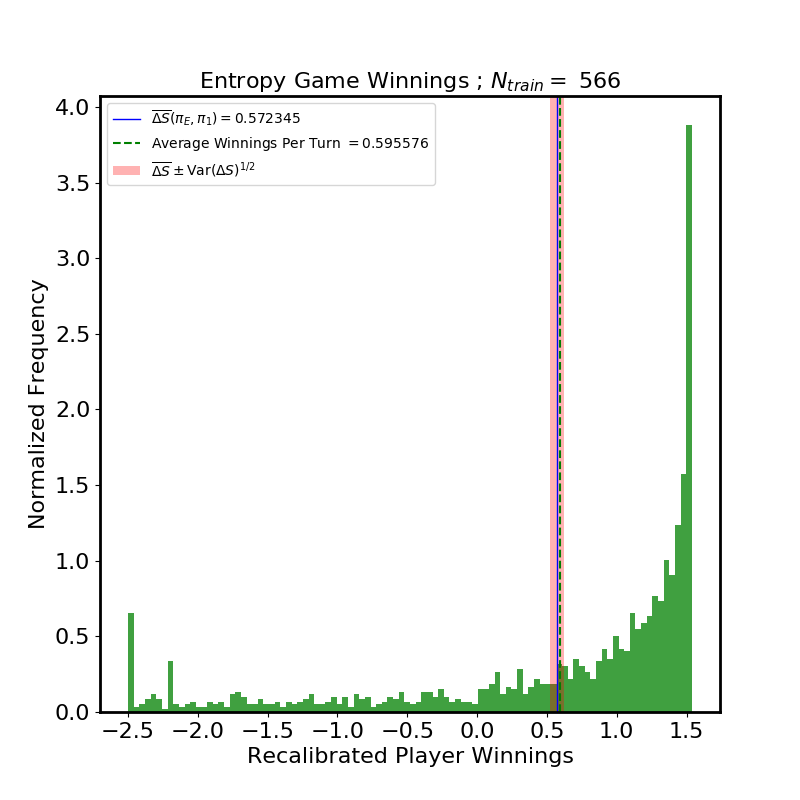}\caption{\label{fig:Circuit_2}Top Left: FAM plot showing expected $N_{t}^{1/2}$
trend. Top Right: Plot of $EI[\pi(F|{\cal F},C)]$, showing possible
evidence of model inadequacy at the largest training set sizes. Lower
Left: entropy game winnings and a-priori predictions. Lower Right:
Histogram of outcomes of 1482 rounds of the entropy game, for the
case of 566 training samples. The blue line is the prediction $\overline{\Delta S}$,
computed from the training data. The pink band that surrounds the
blue line is the predicted 1-$\sigma$ interval, with $\sigma=\sqrt{\mathrm{Var}(\Delta S)}$ computed using Equation~(\ref{eq:PredVar_GPME}). The green dashed
line is the empirical average of the distribution.}
\end{figure}

The top-left panel of Figure \ref{fig:Circuit_2} shows the run
of FAM with $N_{t}$, displaying the expected $N_{t}^{1/2}$ trend.
The top right panel of Figure \ref{fig:Circuit_2} displays the run
of $EI[\pi(F|{\cal F},C)]$ with $N_{t}$. The initial expected drop
appears to level off at the highest values of $N_{t}$, possibly indicating
some model inadequacy (for example, a poor choice of GP kernel) that
reveals itself as the noise in the training histogram is suppressed
by larger values of $N_{t}$. 

The lower-left panel of Figure \ref{fig:Circuit_2} displays Entropy
Game winnings (red dots) together with the predicted winnings $\overline{\Delta S}$
(blue dots) and predicted uncertainty $\textrm{Var}(\Delta S)^{1/2}$
(error bars). Here again we see a tendency at the highest values of
$N_{t}$ of the average winnings to depart from predictions---the 
actual winnings seem somewhat higher than predicted. Again, this
discrepancy is possibly explainable in terms of inadequacies of
the GP model used to estimate $\pi(f|{\cal F},C)$, which are
perceptible only when the histogram noise abates at higher values of $N_{t}$.
Nonetheless, the success of the model in predicting the entropy game
winnings is gratifying, and the recalibrated model is clearly winning
systematically against the ensemble forecasts.

\begin{figure}[tp]
\begin{centering}
\includegraphics[width=4.6cm]{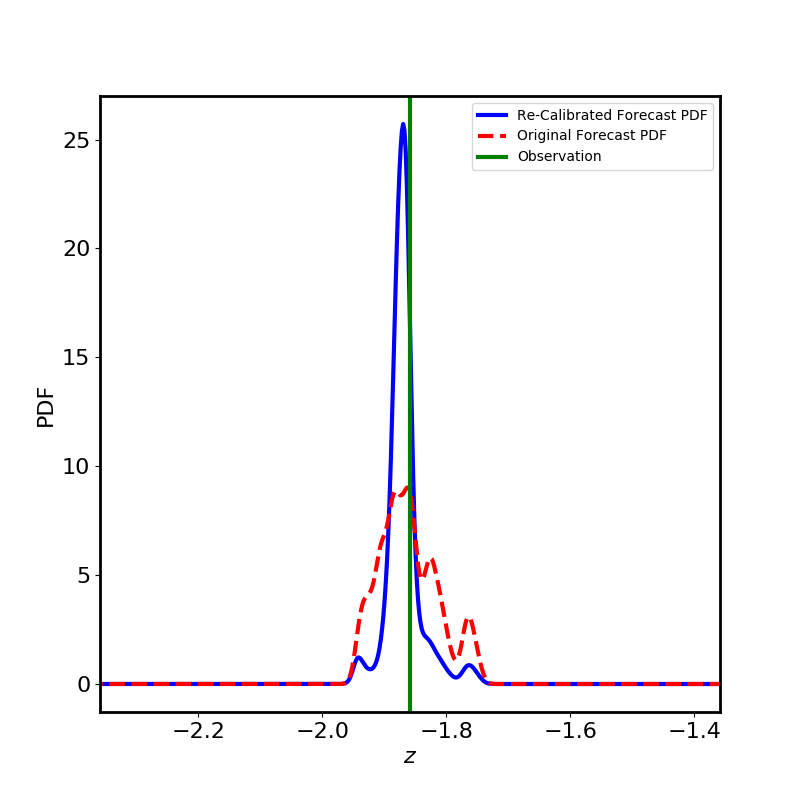}\includegraphics[width=4.6cm]{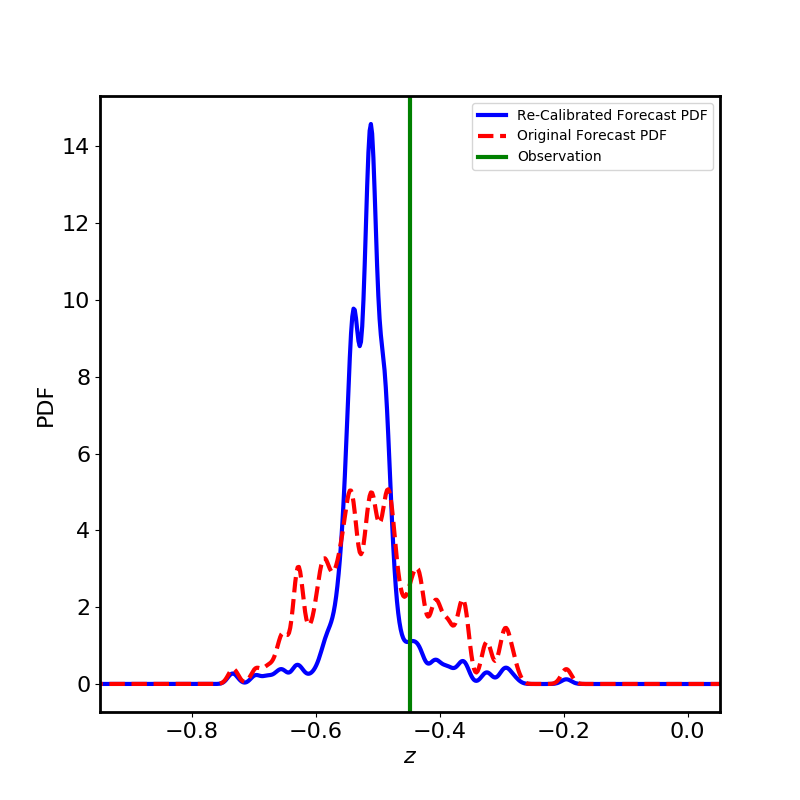}\includegraphics[width=4.6cm]{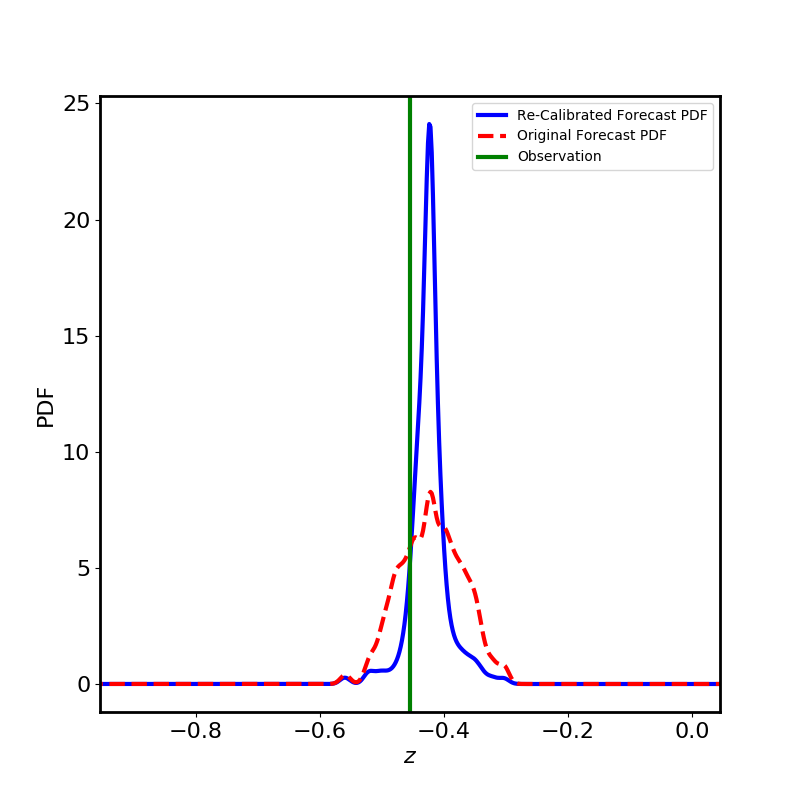}
\par\end{centering}
\begin{centering}
\includegraphics[width=4.6cm]{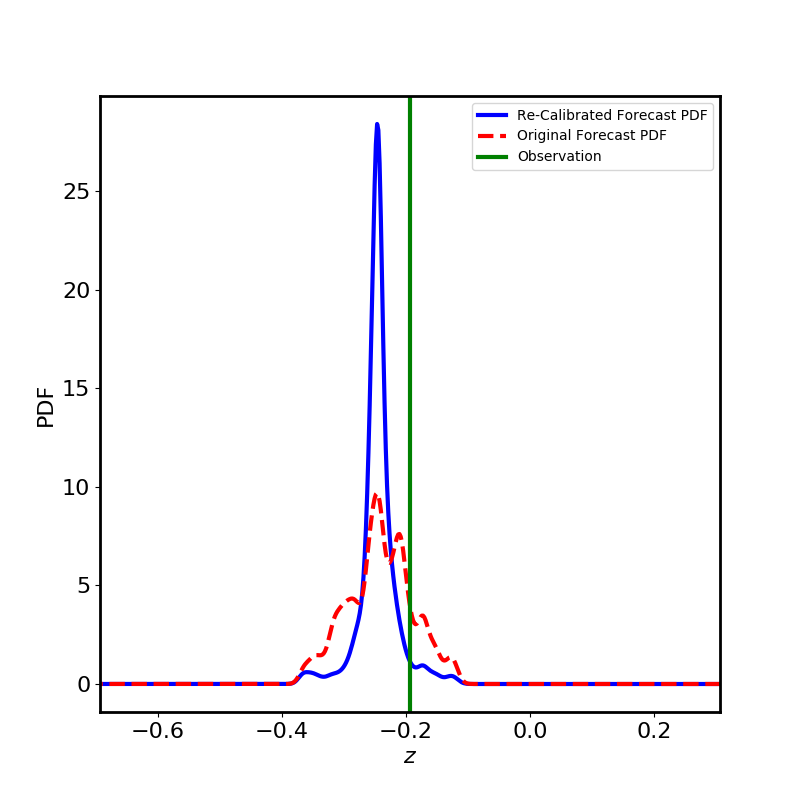}\includegraphics[width=4.6cm]{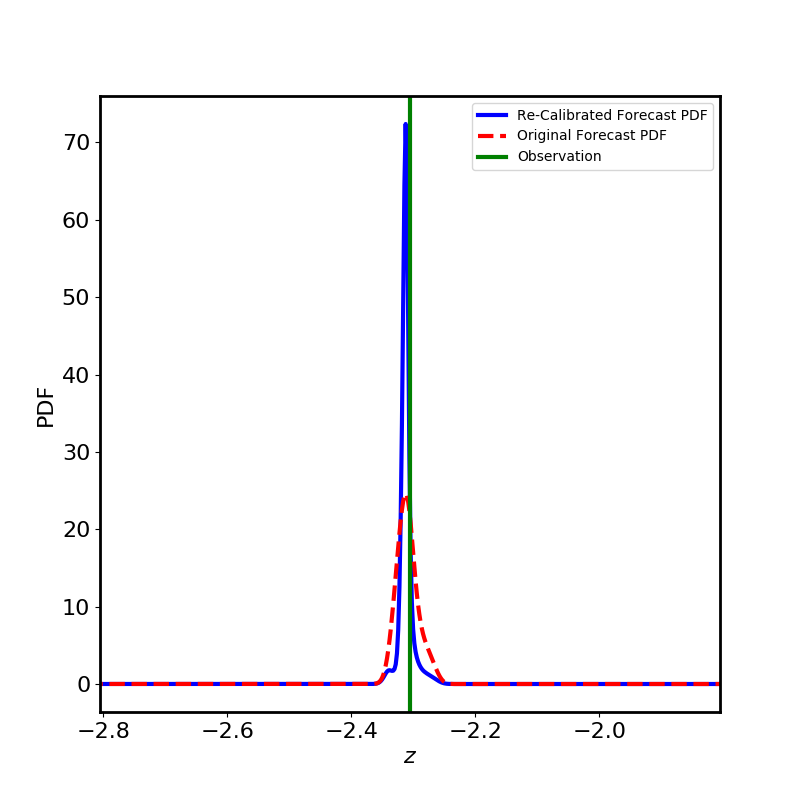}\includegraphics[width=4.6cm]{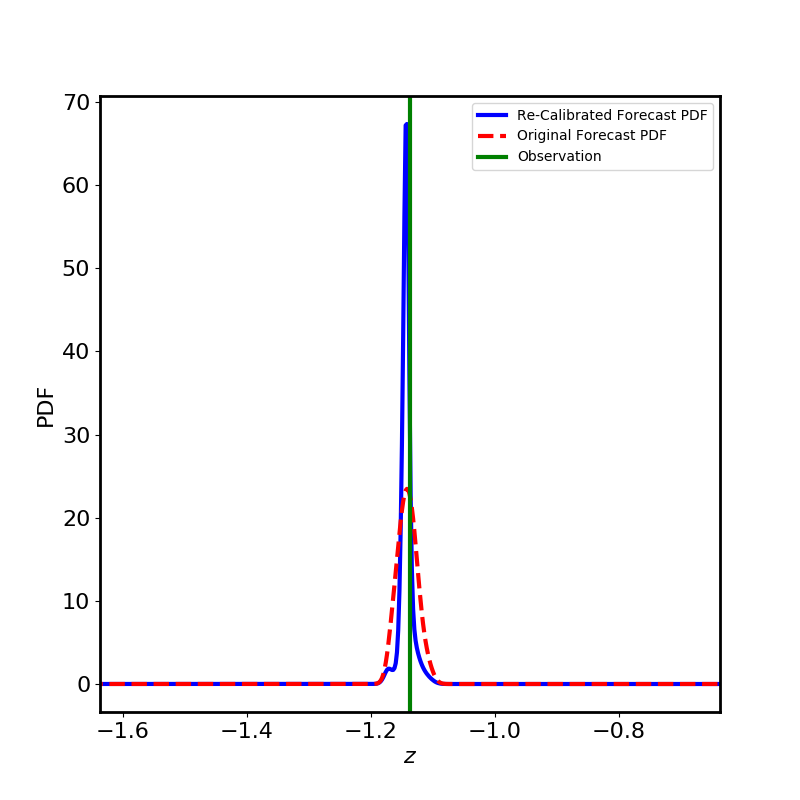}
\par\end{centering}
\centering{}\includegraphics[width=4.6cm]{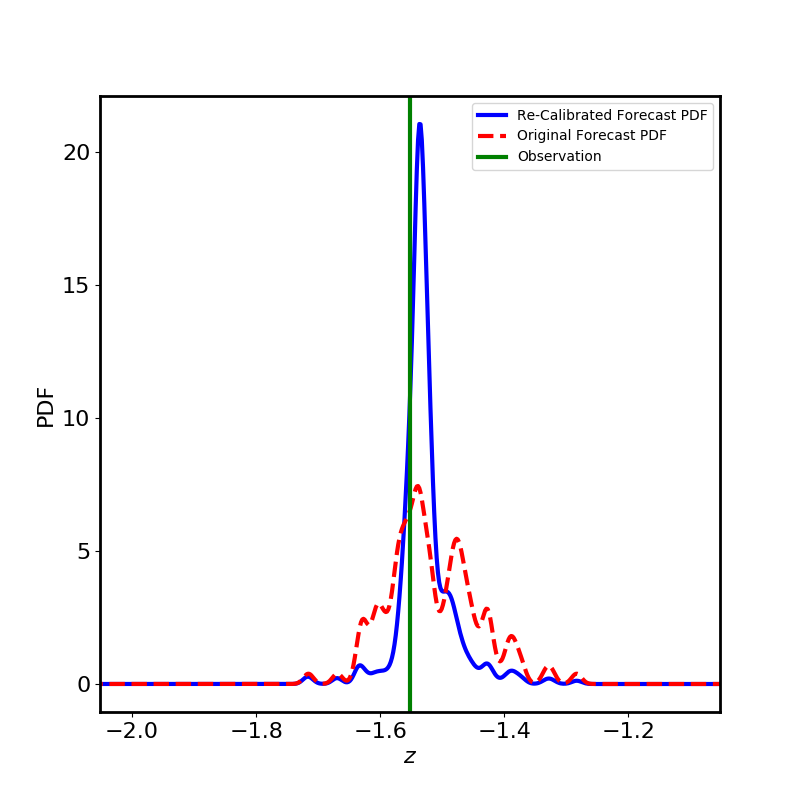}\includegraphics[width=4.6cm]{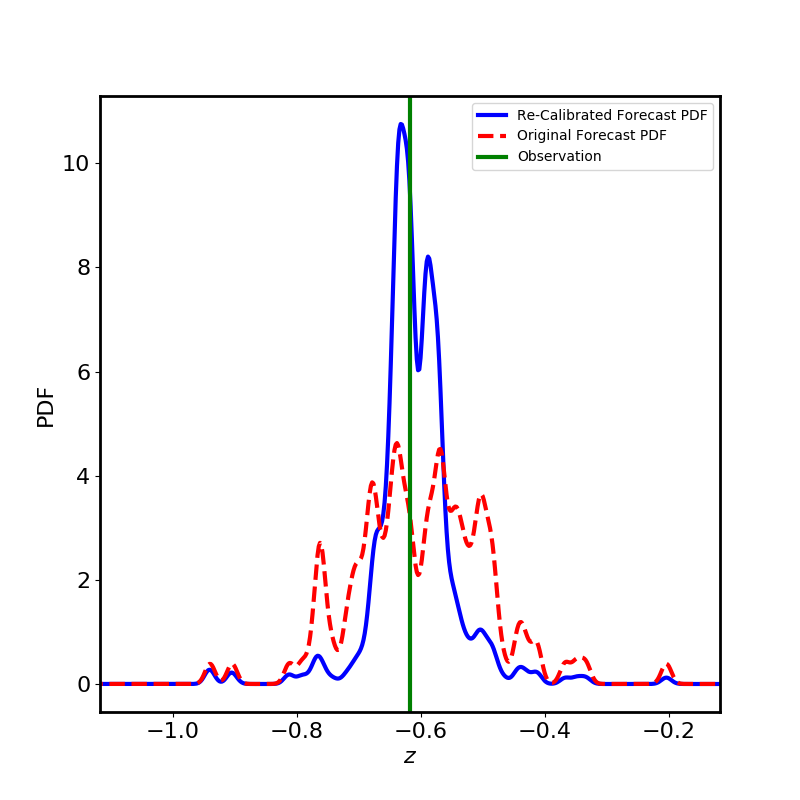}\includegraphics[width=4.6cm]{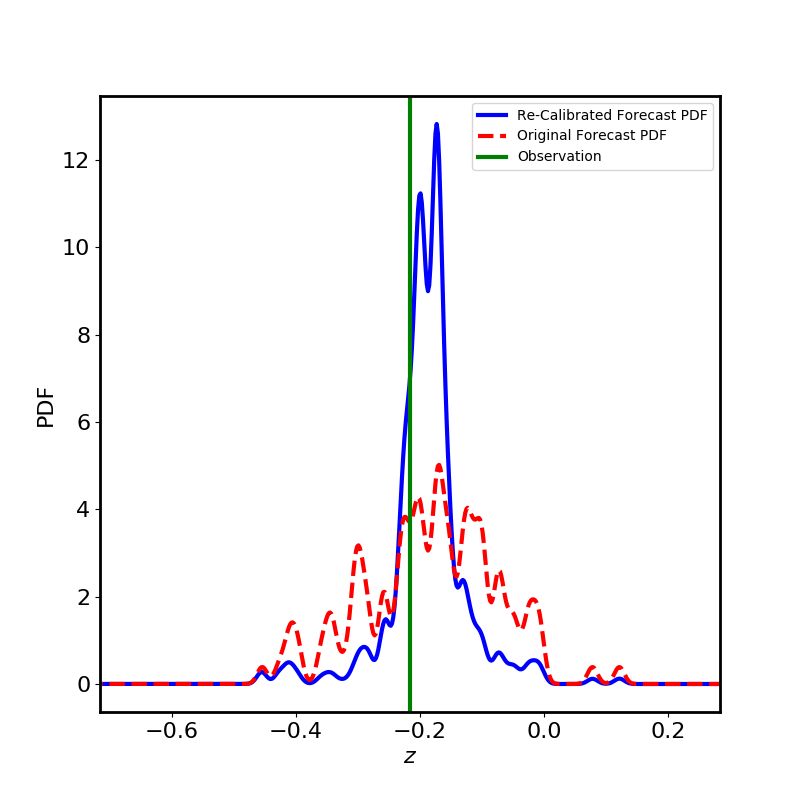}\caption{\label{fig:Circuit_3}Nonlinear circuit. The panels show ensemble
forecasts, recalibrated forecasts, and observation for the first 9
forecasts that follow the training set of 1,600 values. Dashed red
curve is the ensemble forecast, solid blue curve is the recalibrated
forecast, and green vertical line shows the observation.}
\end{figure}

The lower-right panel of Figure \ref{fig:Circuit_2} shows a histogram
of the outcomes of 1,482 rounds of entropy game for the case $N_{t}=566$,
together with the empirical mean (green dashed line), predicted $\overline{\Delta S}$
(blue solid line) and predicted $1-\sigma$ interval (pink band), where $\sigma=\sqrt{\mathrm{Var}(\Delta S)}$ is computed using Equation~(\ref{eq:PredVar_GPME}).
The distribution of outcomes is quite dispersed, comprising a sharp
positive peak with a long tail of negative ``bad busts.'' The shape
of the histogram is easily understood in terms of the GPME fit---the 
red line in the lower-left panel of Figure \ref{fig:Circuit_1}.
In PIT space, this line approximates the actual PIT distribution,
while the original forecast distribution is represented by a horizontal
line at unit normalized frequency. The winnings cutoff at the right
of the winnings histogram corresponds to the log of the maximum ratio
between these two distributions, which coincides with the mode of
the ``actual'' distribution. This is the reason that the mode of
the winnings distribution is at the cutoff. By means of a second-order
Taylor series at the mode of the ``actual'' PIT distribution, we
can also show that the expected behavior in winnings space near the
cutoff should approach $(w_{cutoff}-w)^{-1/2}$, which is (integrably)
divergent. The long tail to the left is also intelligible, as it corresponds
to the two tails near PIT values of 0 and 1 where the ``actual''
distribution is smallest. At these places, the fit distribution density
has value of about $0.2$, so the tail should extend to values of
$w$ near $w_{tail}=\log_{2}\left(0.2/1.0\right)=-2.3$. Note that
these properties are to be expected of overdispersed base forecasts,
because of the hump-shaped PIT distribution, and are not expected
for, say, underdispersed predictions, where the PIT distribution looks
like a pair of peaks near 0 and 1 with a valley in-between.

In Figure \ref{fig:Circuit_3} we have displayed, for the case $N_{t}=1600$,
the first nine ensemble forecasts (red dashed lines), recalibrated forecasts
(blue solid lines), and observed $z$ voltages (green vertical lines).
The plots show the overdispersion of the ensemble forecasts, manifest
in the fact that the observations are too frequently near the median
of the ensemble forecast. The recalibrated forecasts are sharper than
the ensemble forecasts in this case. This would not be expected in
general but is true here because of the overdispersion of
the ensemble forecasts---the relative sharpness of the
recalibrated forecasts restores the missing scatter in the PIT distribution.
A noteworthy feature of this plot is that the recalibrated
forecasts are less noisy than the ensemble forecasts, which are more
prone to show their underlying discrete basis of ensemble-members
that anchor the Gaussian mixture model of the continuous ensemble
forecast. The transition from $p(x)$ to $p_{1}(x)$ appears to smooth
out this noise somewhat.

In summary, the recalibration procedure is highly successful at improving
the performance of the published ensemble forecasts of the nonlinear
circuit. The average winnings of about 0.6 bits corresponds to a wealth
amplification factor of $2^{0.6}=1.5$ per turn in a Kelly-style betting
contest between the two forecasts---offering and wagering
on odds on percentiles of the published ensemble forecast, say---which 
means that the ensemble forecaster would likely meet ruin in
only a few rounds of betting.

\subsection{El Ni\~no Temperature Fluctuations\label{subsec:elnino}}

Our second application of the recalibration technique is to a seasonal
forecasting problem. The seasonal forecast dataset used in this study
is from the North American Multimodel Ensemble (NMME) project \citep{Kirtman_etal-2014}.
The NMME is a collection of global ensemble forecasts from coupled
atmosphere-ocean models produced by operational and research centers
in the United States and Canada. The NMME forecasts are generated in real time
but also include a 30-year set of retropsective monthly forecasts
(hindcasts) for assessing systematic biases in the models.

\begin{table}
\centering{}%
\caption{\label{tab:NMME}NMME models selected for this study, and respective
ensemble sizes.}
\begin{tabular}{|l|c|}
\hline 
Model & Ensemble Size\tabularnewline
\hline 
\hline 
COLA-RSMAS-CCSM3 & 6\tabularnewline
\hline 
COLA-RSMAS-CCSM4 & 10\tabularnewline
\hline 
GFDL-CM2p1-aer04 & 10\tabularnewline
\hline 
GFDL-CM2p5-FLOR-A06 & 12\tabularnewline
\hline 
GFDL-CM2p5-FLOR-B01 & 12\tabularnewline
\hline 
\end{tabular}
\end{table}

\begin{figure}[tp]
\begin{centering}
\includegraphics[width=6cm]{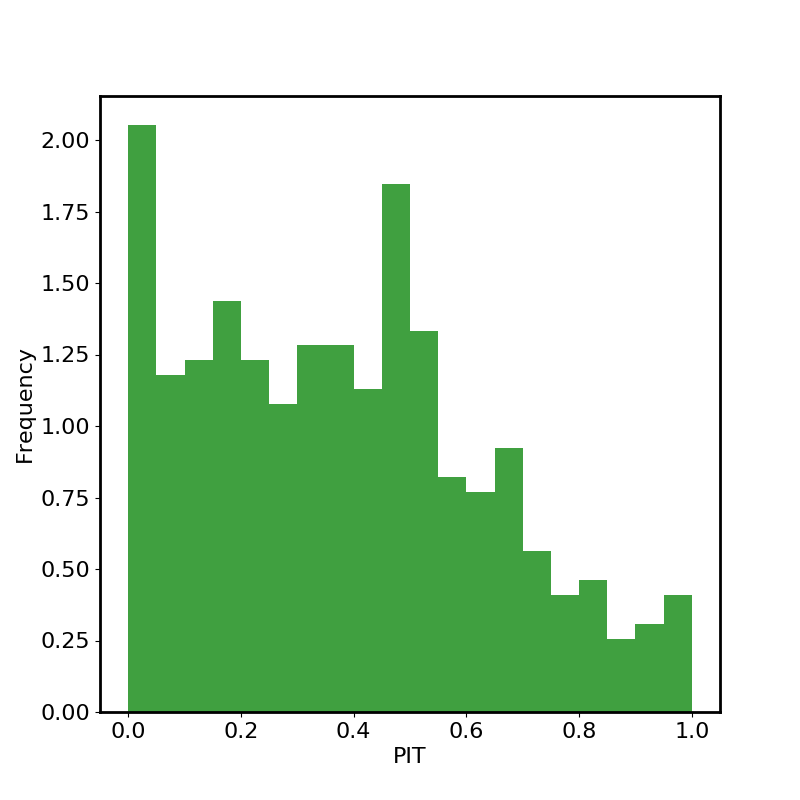}\includegraphics[width=6cm]{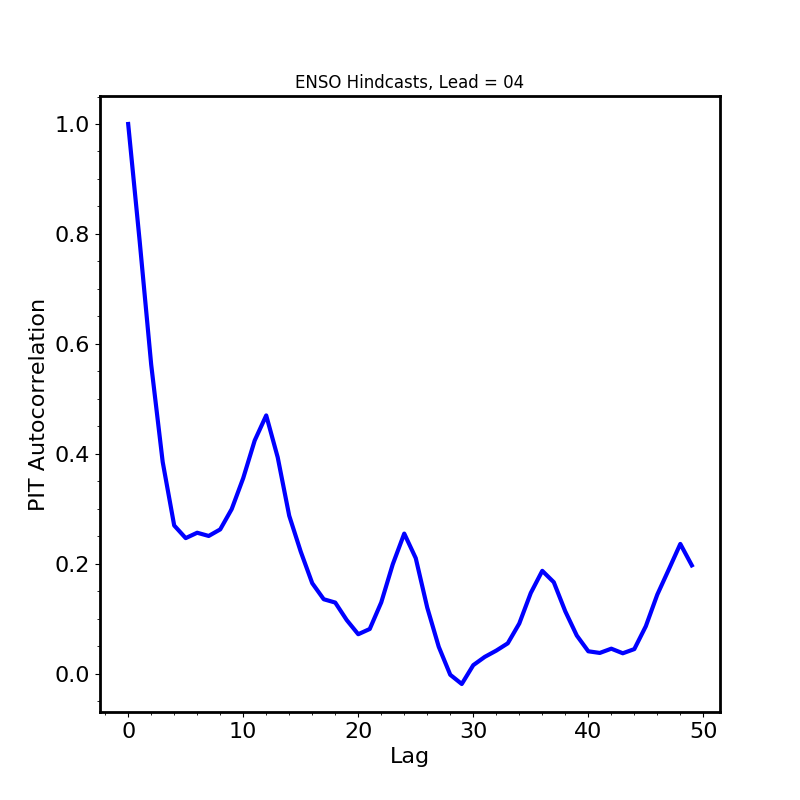}
\par\end{centering}
\begin{centering}
\includegraphics[width=6cm]{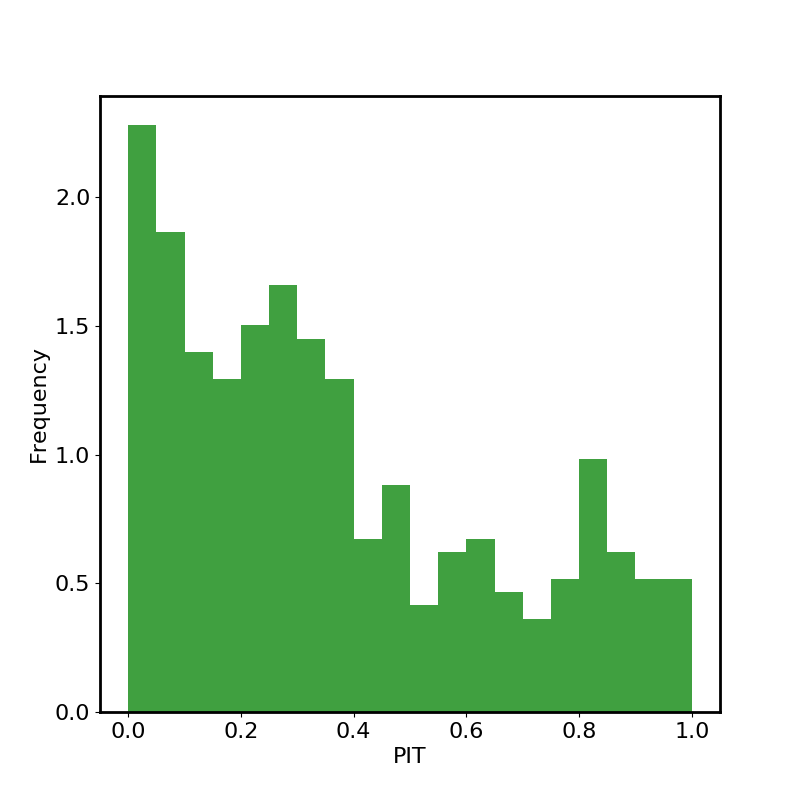}\includegraphics[width=6cm]{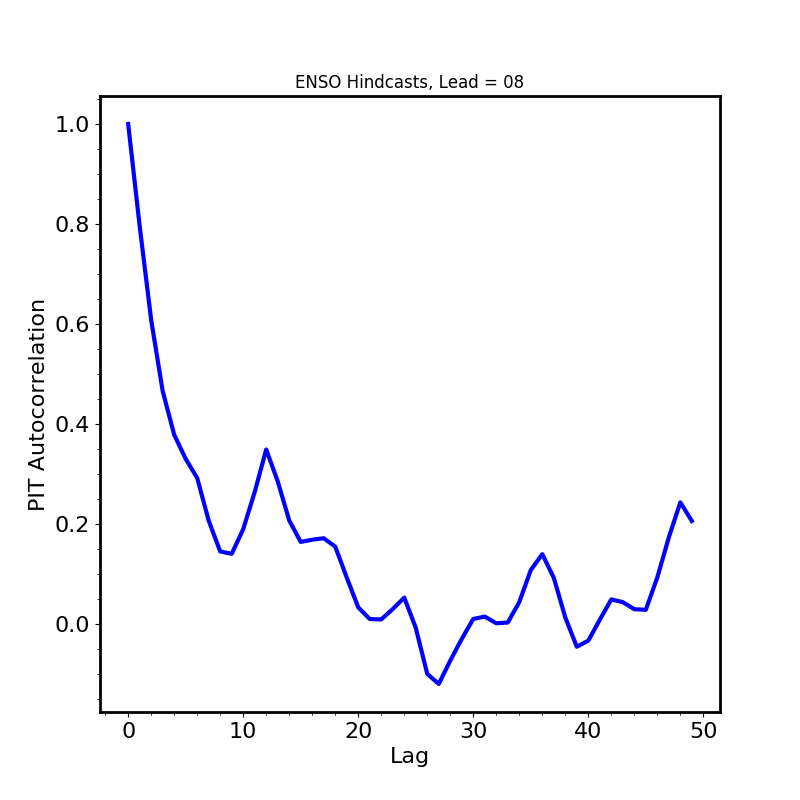}
\par\end{centering}
\centering{}\includegraphics[width=6cm]{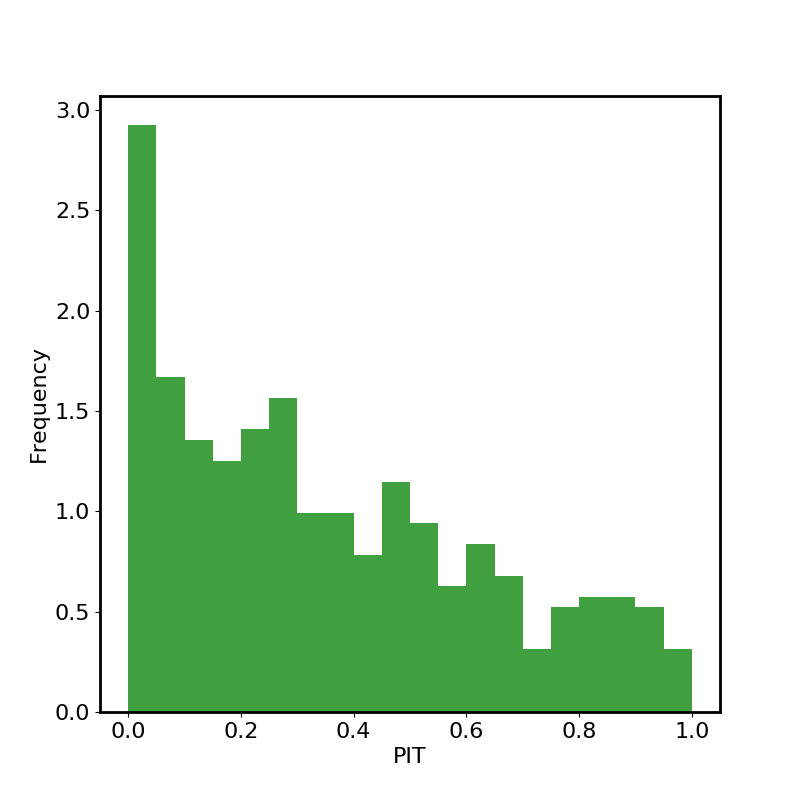}\includegraphics[width=6cm]{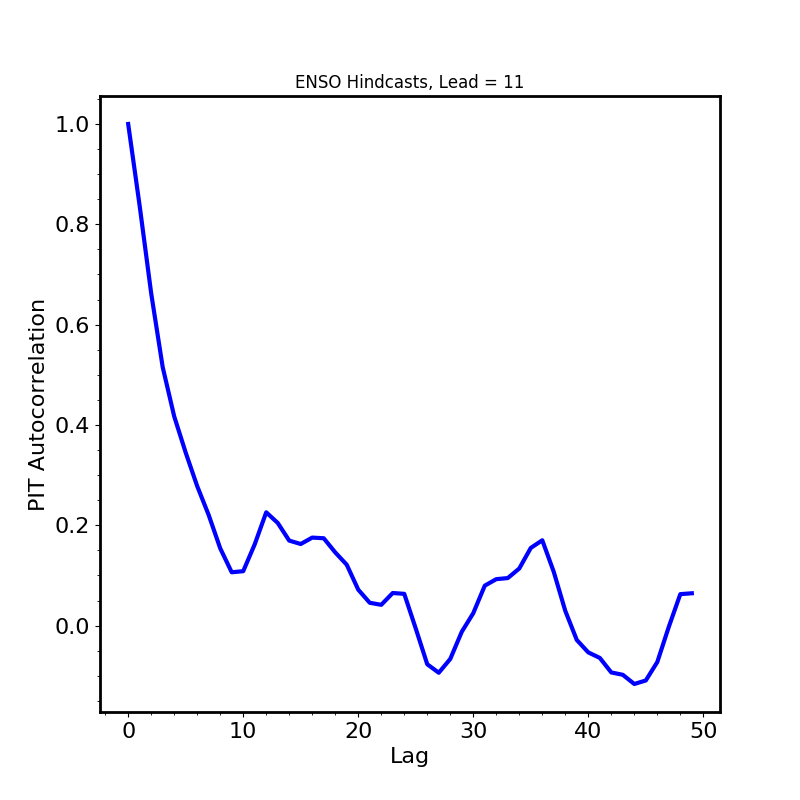}\caption{\label{fig:ENSO_1}BMA forecasts of NINO 3.4 index, based on NMME
hindcast data. Panels from top to bottom correspond to lead times
of 4, 8, and 11 months forecast lead times. Left column: PIT histograms
resulting from comparison of BMA forecasts with observations. The BMA
forecasts are biased to high values of the NINO 3.4 index. Right column:
Time autocorrelation functions of PIT values. Substantial temporal
correlations exist out to and past 15-month lags.}
\end{figure}

We examine the NMME model predictions of the intensity of El Ni\~no-Southern
Oscillation (ENSO) phenomenon. The ENSO state is often characterized
by the NINO 3.4 index, which is the monthly mean sea surface temperature
(SST) averaged over the equatorial Pacific region: 5S to 5N, and 170W
to 120W. Tippet et al. \citep{tippett2017assessing} also use the
NINO 3.4 index to assess the skill of the NMME models; that paper contains
a useful description of all the models and their particular configurations
for the hindcasts, and a discussion of the errors and known problems
in the model forecasts. For this study, no data corrections have been
made, and model climatologies have not been removed---the index
values are based on real temperatures instead of temperature anomalies.

The NMME hindcast dataset is available at \url{http://iridl.ldeo.columbia.edu/SOURCES/.Models/.NMME/}.
The NMME hindcasts are created monthly, have lead times that range
from 1 to 12 months, and are validated with the observed NINO 3.4
index for the period January 1982\textendash October 2017. The observed
index values are derived from NOAA\textquoteright s Optimum Interpolation
Sea Surface Temperature data (OISST, version 2, \citep{reynolds2002improved})
which are available at \url{http://www.cpc.ncep.noaa.gov/data/indices/sstoi.indices}. 

Our object in this study is to work with as long a stretch of data
as possible and for that stretch of data to represent model output
that is temporally as homogeneous as possible, because if the
model composition were to fluctuate during the study, or be substantially
different between training and test data sets, the recalibration procedure
could not be expected to be effective. Of the 15 NMME models, only
6 were run daily during the entire period of the project, while others
were retired at various stages of the project. Of those 6, 5 ran with
the same ensemble size throughout, while the remaining model had an
ensemble size that varied with sufficient frequency to create concern
for the homogeneity of the sample. Consequently, we subsetted
the hindcast data, choosing only the 5 models that were run consistently
monthly for 35 years of hindcasts. These models and their respective
ensemble sizes are displayed in Table \ref{tab:NMME}.

\begin{figure}[tp]
\begin{centering}
\includegraphics[width=6cm]{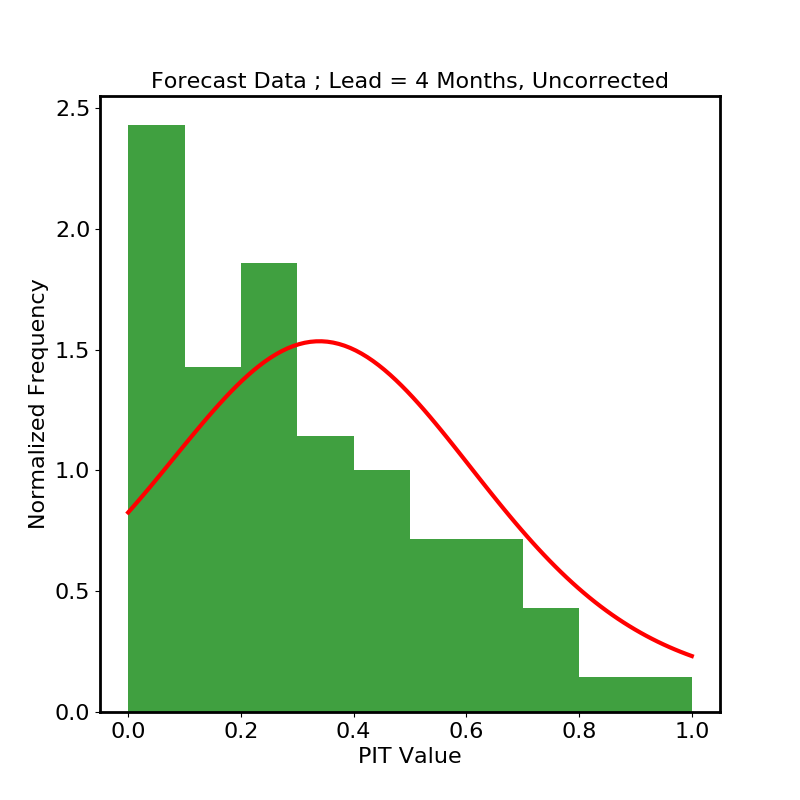}\includegraphics[width=6cm]{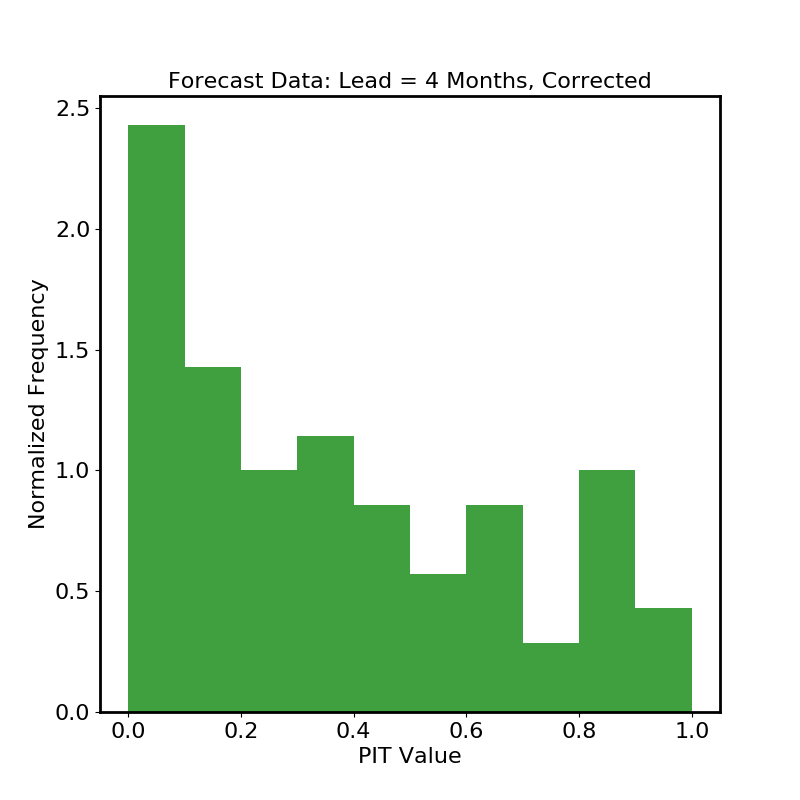}
\par\end{centering}
\begin{centering}
\includegraphics[width=6cm]{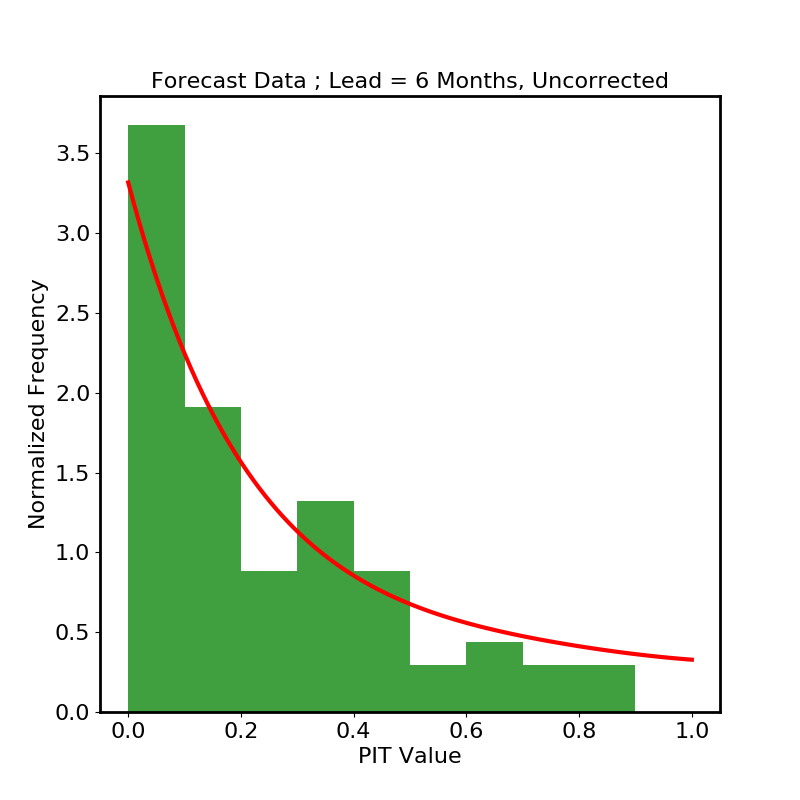}\includegraphics[width=6cm]{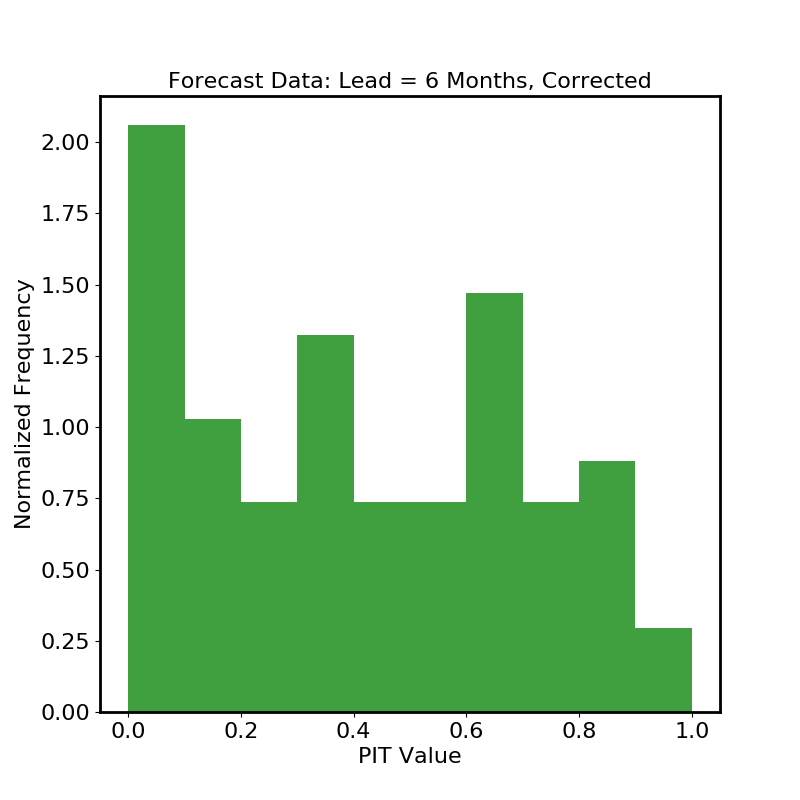}
\par\end{centering}
\begin{centering}
\includegraphics[width=6cm]{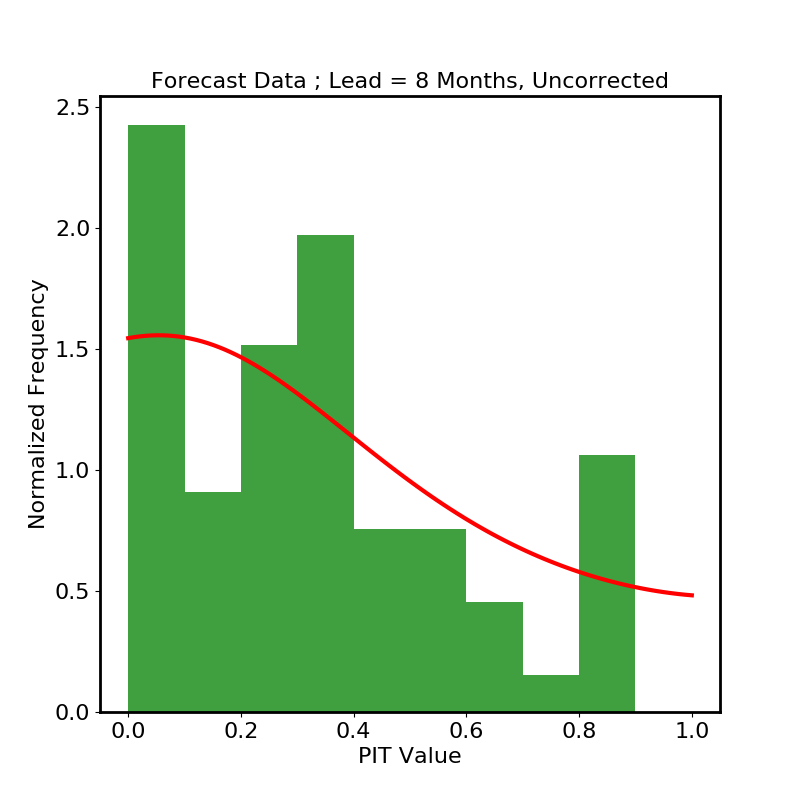}\includegraphics[width=6cm]{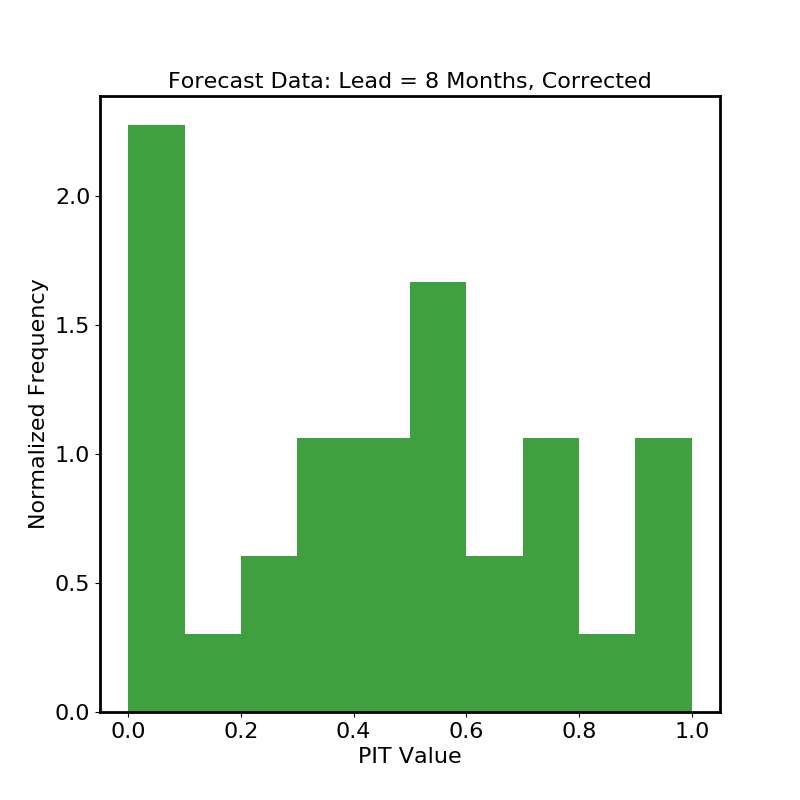}
\par\end{centering}
\begin{centering}
\caption{\label{fig:ENSO_2}PIT histograms. Left column: Uncorrected forecasts.
Solid red line shows fit to training data histogram. Right column:
Recalibrated forecasts. From top to bottom, forecast lead times of
4, 6, and 8 months.}
\par\end{centering}
\end{figure}

To convert the forecast simulation ensembles to
continuous forecast distributions, we chose the method of Bayesian
model averaging (BMA) \citep{raftery2005using}, as adapted to multi-model
ensembles with exchangeable members by Fraley et al. \citep{Fraley_et_al-2010}.
Briefly, BMA models the forecast PDFs consequent from the ensemble
predictions by using a mixture model, with mixture weights ascribed to
different components. 
Ensemble members from a single model are ``exchangeable'' \citep{Fraley_et_al-2010}
in that none of them may be regarded as bearing better or worse information
than their ensemble partners bear. They therefore are all assigned equal
weights under the scheme. Ensemble members from different models are
``nonexchangeable'' and so have different weights. Following \citep{Fraley_et_al-2010,raftery2005using},
we use Gaussians for the mixture component PDFs, with Gaussian widths
that are the same within each model ensemble but may differ
from model to model. We first bias-correct the forecasts model by model
using linear regression of the observations on the forecasts in the
training set. Then we center the Gaussians on the bias-corrected ensemble
forecast values and optimize the likelihood of the training data
iteratively by the EM algorithm \citep{dempster1977maximum,tanner1993tools},
updating weights and Gaussian widths at each EM iteration \citep{raftery2005using,Fraley_et_al-2010}.
The converged weights and widths are used to create forecasts in the
test set.

We have available a total of 430 monthly hindcast simulations. We
consider lead times of 1--11 months for each hindcast. We train the
BMA forecasting machinery on the first 36 hindcasts and use the machinery
to create forecasts from the remaining 394 hindcasts, one for each
lead time.

The ensemble forecast results are summarized in Figure \ref{fig:ENSO_1}.
The three rows of the figure correspond to lead times of 4, 8, and
11 months. The left column depicts the PIT histograms, which show
clear evidence that the BMA forecasts are biased to high values of
NINO 3.4 index, despite the preliminary bias corrections. Clearly
potential leverage exists here for the recalibration procedure
to do its work. However, there is a fly in the ointment: the right
column of Figure \ref{fig:ENSO_1} displays the temporal autocorrelation
functions of the PIT values, which are clearly significantly correlated
out to 15-month lags and beyond. While not entirely surprising, this
is a serious potential restraint on the effectiveness of the method,
since with only 394 forecasts to work with, a thinning by a factor
of 15 leaves hardly enough forecasts to form a training set, to say
nothing of a test set.

We compromise, \emph{faute de mieux}, on a thinning by a factor of
5---below this factor we find correlations unacceptably
compromise the fit of $\pi(F|{\cal F},C)$, above it we have too few
forecasts to work with. We choose a recalibration training set of
64 forecasts, leaving $394-5\times64=74$ forecasts in the test set.
For each lead time, we fit $\pi(F|{\cal F},C)$ to the corresponding
PIT histogram and use it to compute $EI[\pi(F|{\cal F},C)]$, $\overline{\Delta S}$,
$\textrm{Var}(\Delta S)$ and FAM, then to run 74 rounds of Entropy
Game between the BMA forecasts and the recalibrated forecasts.

\begin{figure}[tp]
\begin{centering}
\includegraphics[width=4.6cm]{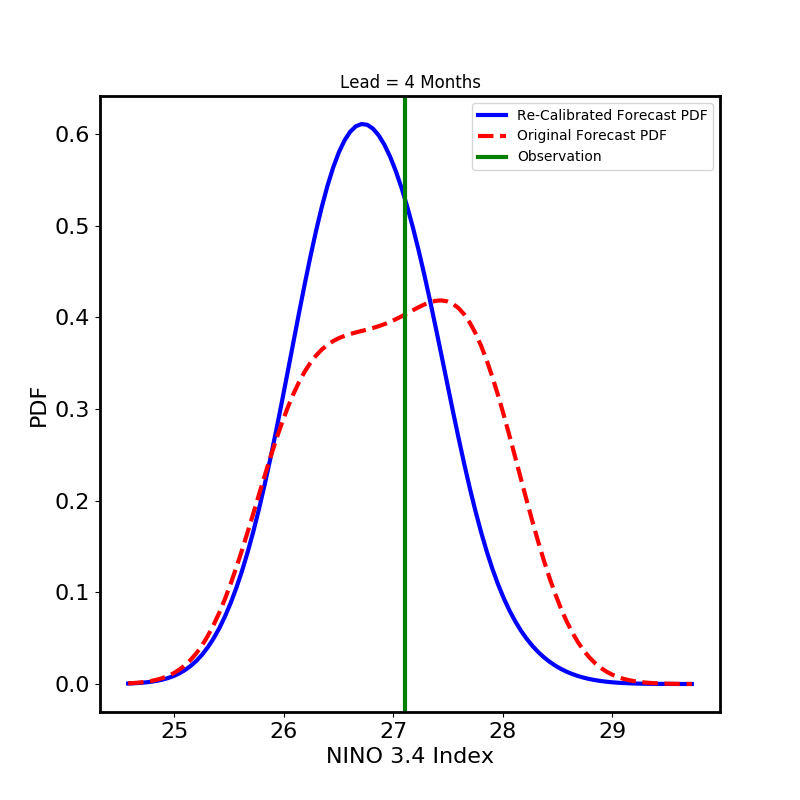}\includegraphics[width=4.6cm]{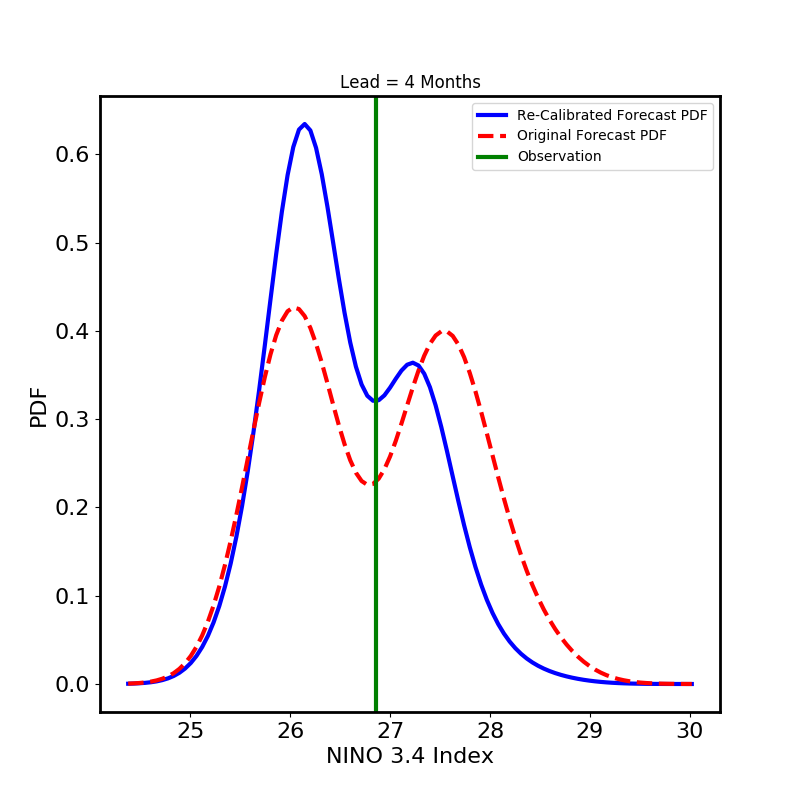}\includegraphics[width=4.6cm]{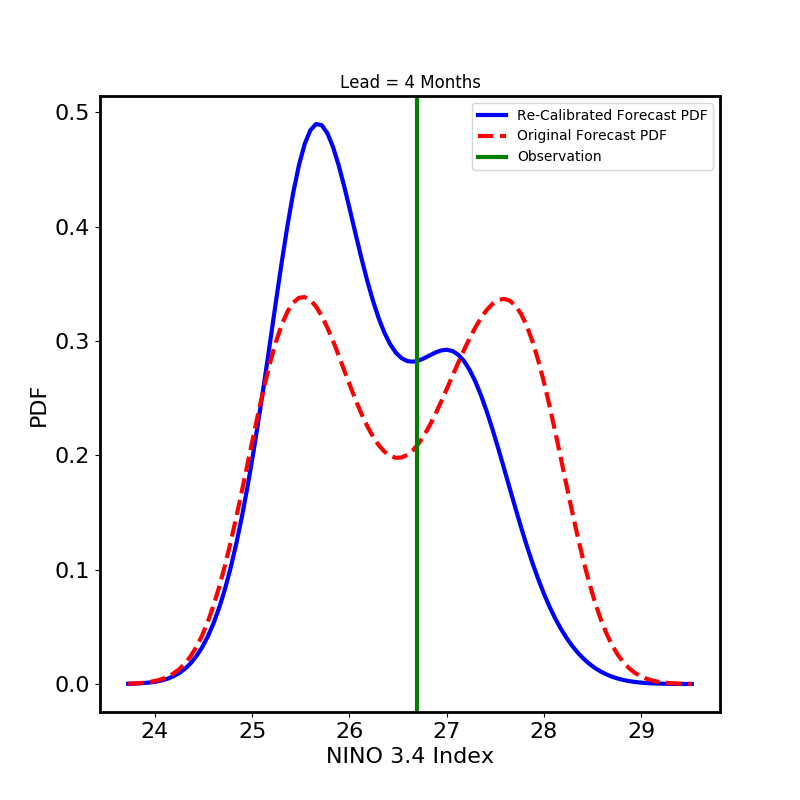}
\par\end{centering}
\begin{centering}
\includegraphics[width=4.6cm]{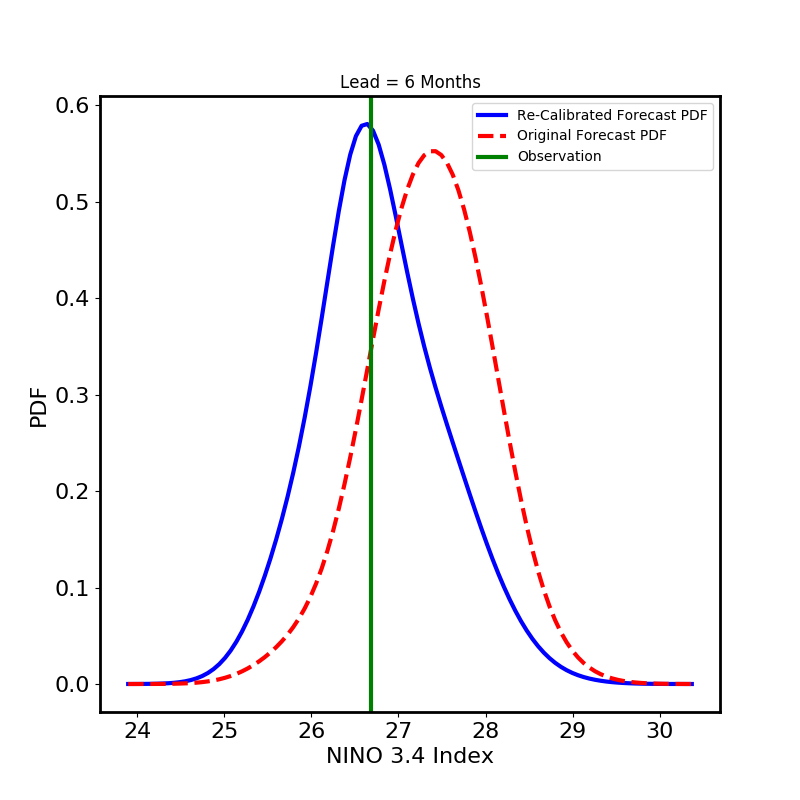}\includegraphics[width=4.6cm]{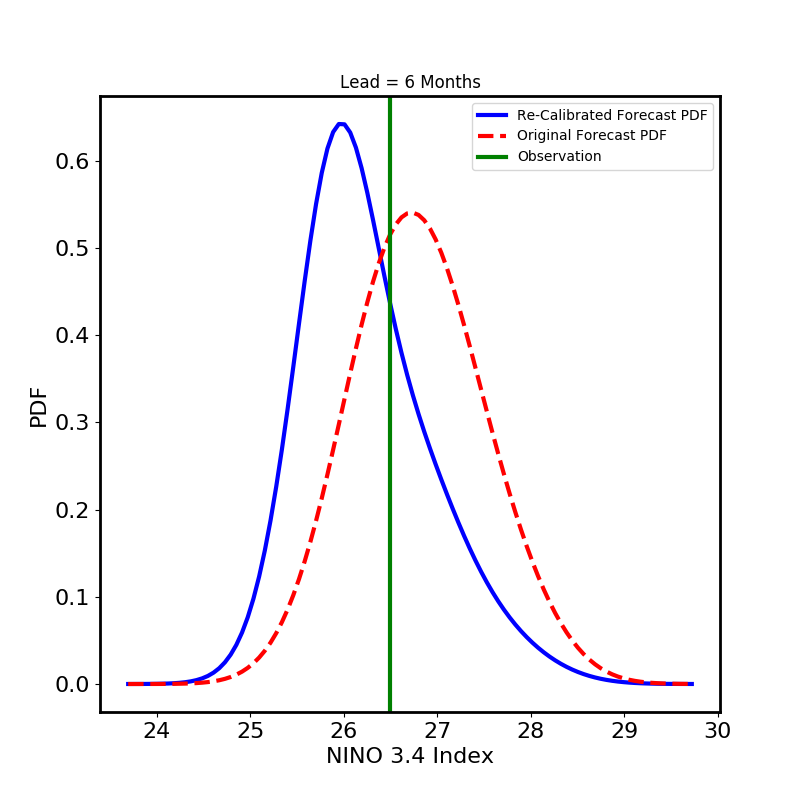}\includegraphics[width=4.6cm]{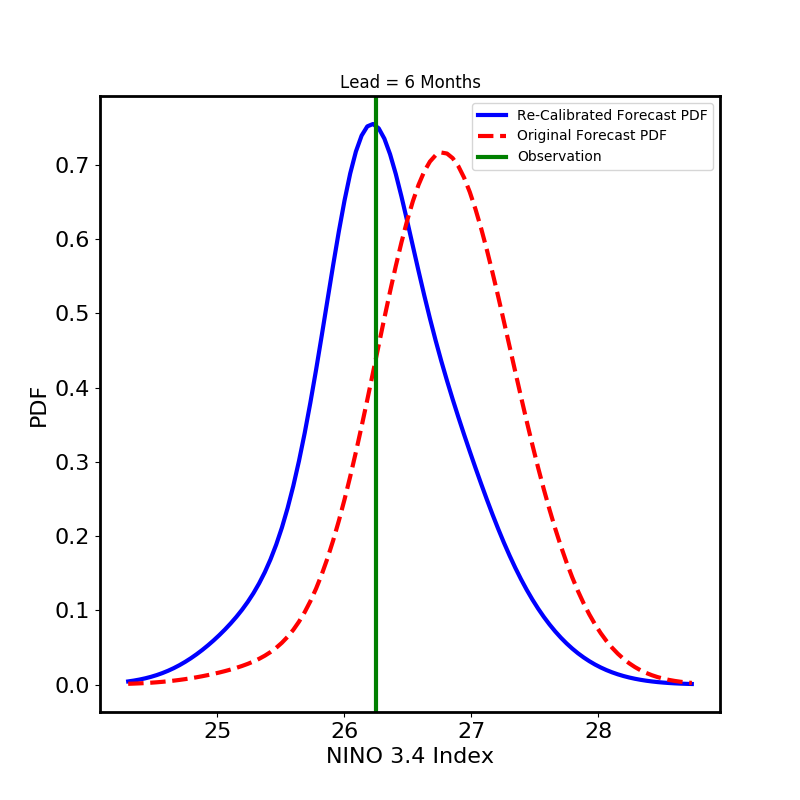}
\par\end{centering}
\begin{centering}
\includegraphics[width=4.6cm]{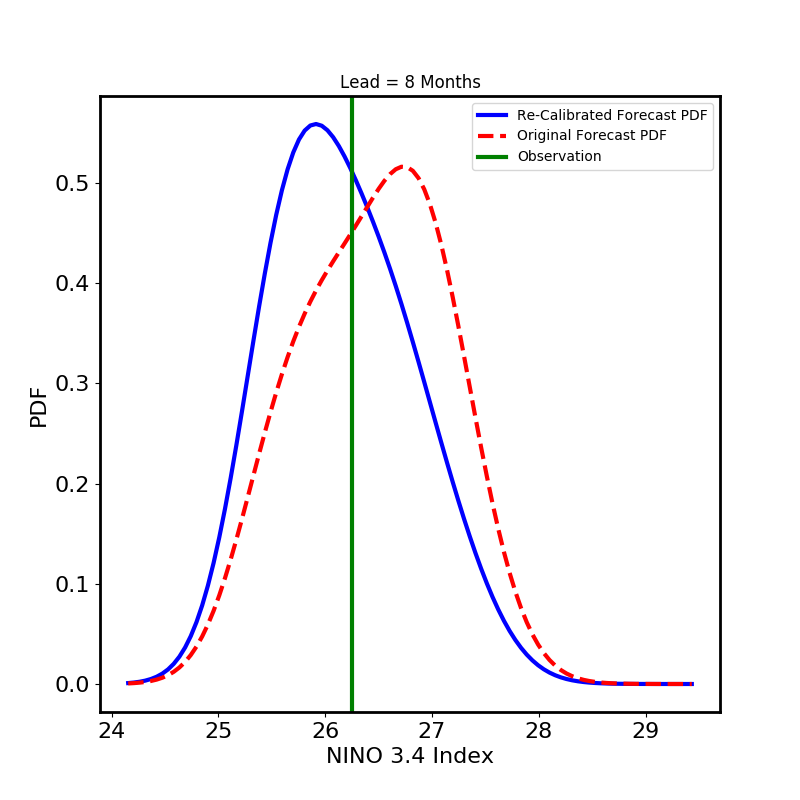}\includegraphics[width=4.6cm]{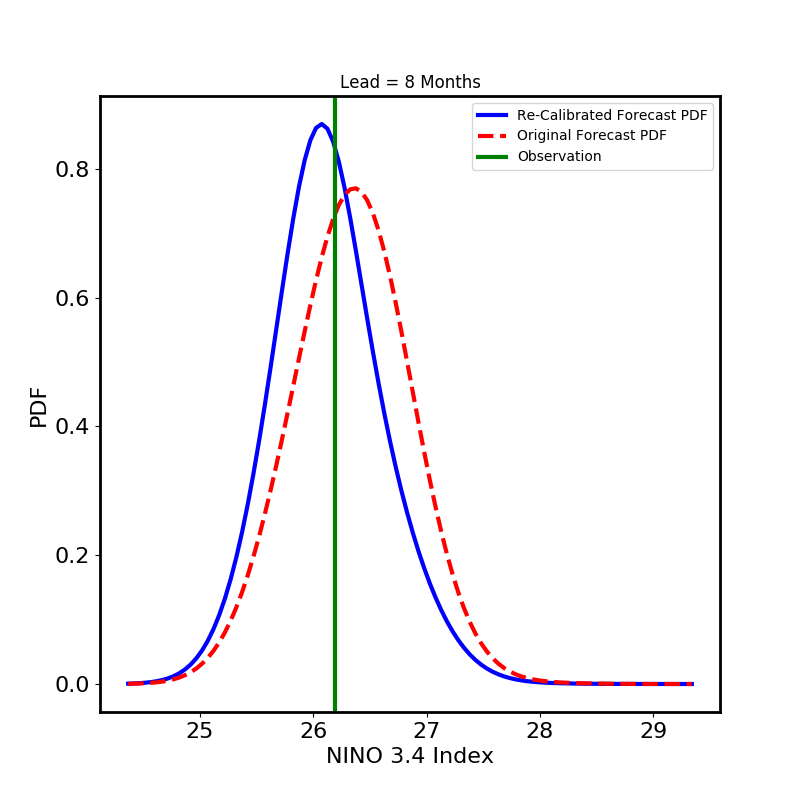}\includegraphics[width=4.6cm]{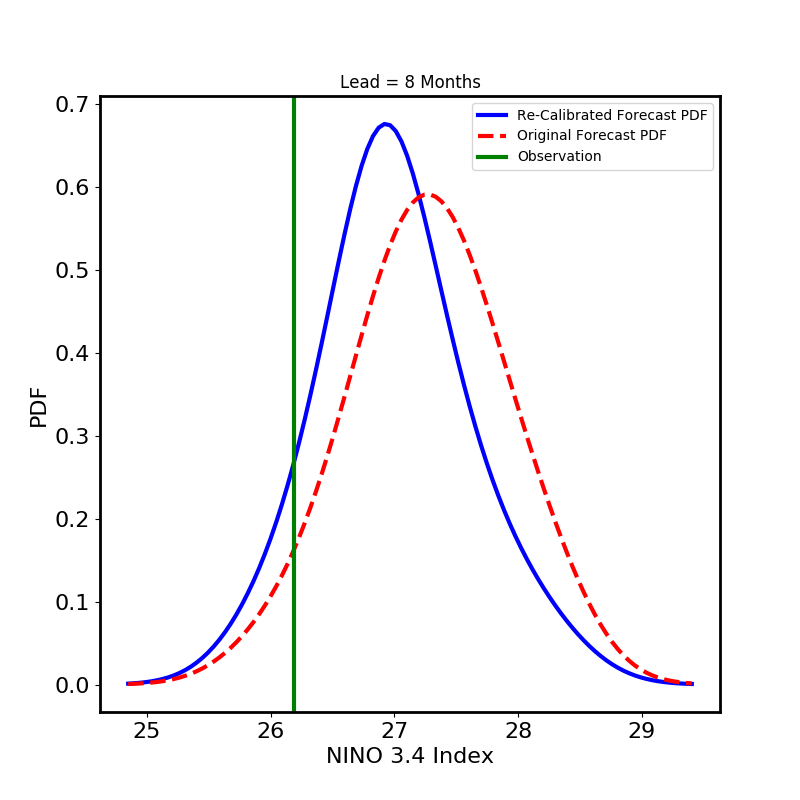}
\par\end{centering}
\centering{}\caption{\label{fig:ENSO_4}ENSO3.4 forecasts. The panels show ensemble forecasts,
recalibrated forecasts, and observation for the first 3 forecasts
in the set of 74 comprising the test set. The dashed red curve is the
ensemble forecast, the solid blue curve is the recalibrated forecast,
and the green vertical line shows the observation. Top row: 4-month forecast
lead. Middle row: 6-month forecast lead. Bottom row: 8-month forecast
lead.}
\end{figure}

\begin{figure*}[tp]
\begin{centering}
\includegraphics[width=7cm]{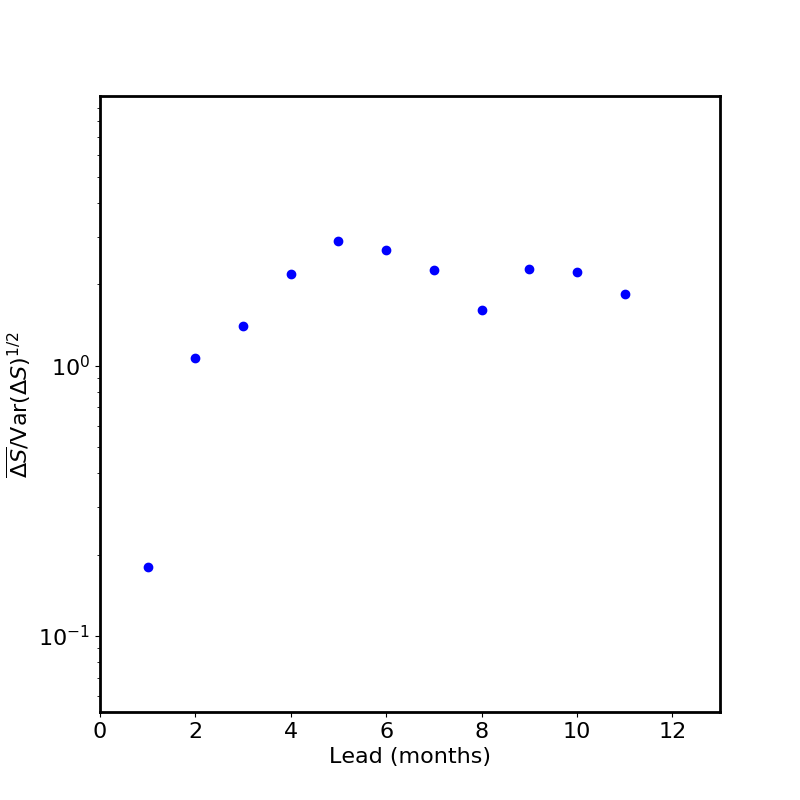}\includegraphics[width=7cm]{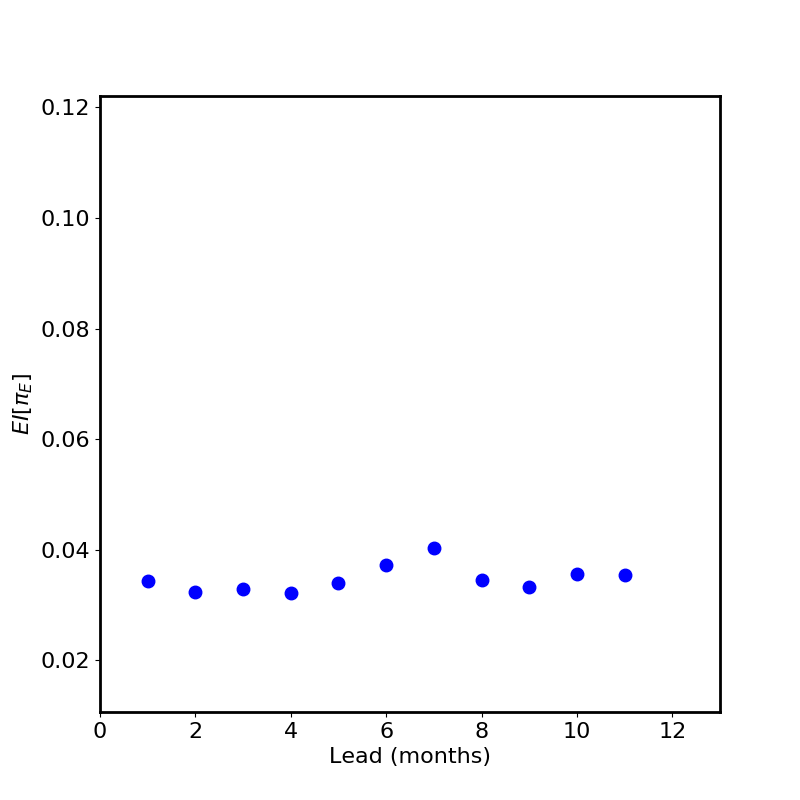}
\par\end{centering}
\centering{}\includegraphics[width=7cm]{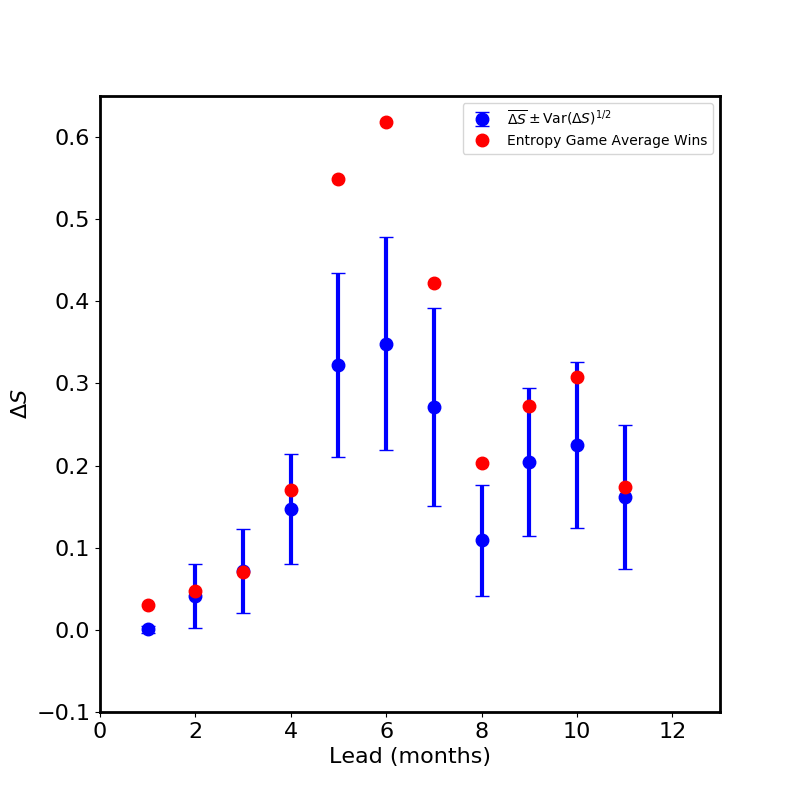}\includegraphics[width=7cm]{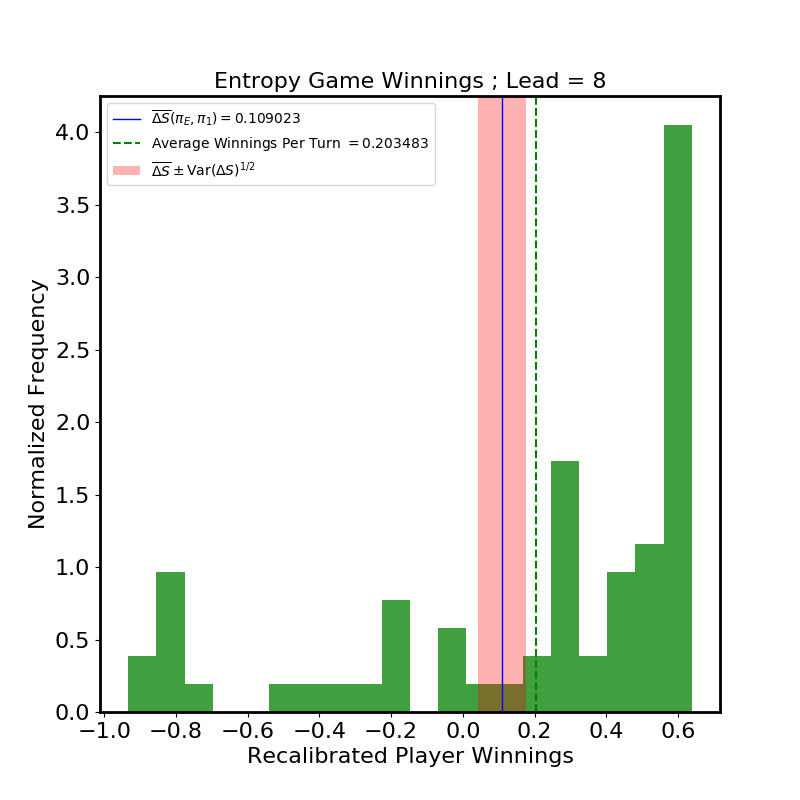}\caption{\label{fig:ENSO_3}Results of recalibration of BMA ensemble forecasts
of NINO 3.4 index. Top left Panel: FAM plot, as a function of forecast
lead time now. Top right Panel: $EI$ plot, again as a function of
lead time. Lower left Panel: Expected entropy game winnings (red dots)
and predictions (blue dots and errorbars) as a function of lead time.
Lower right Panel: Histogram of outcomes over 74 rounds of the Entropy
Game for the 8-month lead time case. The green dashed line is the
empirical mean, the blue solid line is the predicted mean, and the
pink band is the predicted $1-\sigma$ interval.}
\end{figure*}

Figure \ref{fig:ENSO_2} shows the result of the recalibration procedure
on the PIT histograms for forecast leads of 4, 6, and 8 months. The
74 test forecasts are histogrammed into 20 bins. The recalibrated
forecasts (right column) have improved probabilistic calibration over
the published forecasts (left column). The effect is not as dramatic
as for the circuit data of \S\ref{subsec:A-Nonlinear-Circuit}, in
part because of the paucity of data and in part no doubt because
of model inadequacy due to residual correlations in the data. Nonetheless
the improvement in calibration is clear.

Figure \ref{fig:ENSO_4} shows a few sample published and recalibrated
forecasts for leads of 4, 6, and 8 months. Comparison with Figure
\ref{fig:ENSO_2} shows that the recalibration is attempting to correct
the bias by shifting the mass of the probability distributions to
lower values of the NINO 3.4 index.

In Figure \ref{fig:ENSO_3}, the top-left panel shows FAM as a function
of forecast lead. FAM increases dramatically from lead=1 month to
lead=2 months, then settles down between a value of 1 and 2, suggesting
moderate confidence in the performance of the recalibrated forecast
for months 2 and later. The top-right panel shows $EI\left[\pi(F|{\cal F},C)\right]$,
which is fairly steady with a shallow peak near lead=7 months, which
indicates that the quality of the fit of $\pi(F|{\cal F},C)$ to the
PIT training data is fairly uniform across lead times.

The lower left panel shows entropy game average winnings (red dots)
compared with performance measures $\overline{\Delta S}$ (blue dots)
and $\mathrm{Var}(\Delta S)$ (blue errorbars). The performance measures
do modestly well in predicting winnings, given the relatively modest
size of the available training set and the correlations that still
reside therein. The performance advantage of the recalibrated forecasts
is striking, especially beginning around lead=4 months.

The lower-right panel shows a histogram of outcomes over 74 rounds
of the entropy game in the 8-month forecast lead case. Again, the
histogram structure is interpretable in terms of the PIT histogram
fit in the middle-left panel of Figure \ref{fig:ENSO_2}, with the
mode at $\log_{2}1.5\approx0.6$ and a tail extending to $\log_{2}0.5=-1$.

The empirical average entropy game winnings are in the range 0.2\textendash 0.6
bits, corresponding to a range of per turn wealth multipliers of $2^{0.2}$\textendash $2^{0.6}$$=$1.15\textendash 1.52
in a Kelly-style odds-making/betting game. Even despite
the limitations set by the small amount of available training data,
the recalibrated forecaster can expect to bankrupt the BMA forecaster
after relatively few turns, especially if betting on leads of about
6 months or so. 

\section{Discussion\label{sec:Discussion}}

In this work we have gone to great length to emphasize
the importance of weighing information---and of using
quantitative measures of information---in reasoning about
probabilistic forecasts of continuous variables. 
We showed that the information interpretation of calibration is
useful because we may use it to build out of an arbitrary forecast system a
related recalibrated system that is expected to be \emph{much} better calibrated
than the original system
in those cases where the probabilistic calibration of the original
system was noticeably poor. 

We validated the forecast recalibration theory on two very different
examples: (1) a nonlinear circuit whose output is forecast by iterating a radial basis function model constructed in delay space (See \cite{machete2013} for details) and a smoothing using
kernel dressing and climatology blending, and (2) 30 years of monthly
NINO 3.4 index observations, using forecasts generated from NWP ensembles
smoothed by BMA. In each case, the nature of the observational data,
the input elements to the forecast system, and the method of generating
probabilistic forecasts were different. We emphasize
that the recalibration procedure was successful in both cases, producing
objectively superior forecasts (as measured by entropy game outcomes)
without needing to care much about the inner nature of the forecasts
that it improves upon.

In the ENSO study, the recalibrated forecasts were easily able to
outperform the BMA forecasts, despite the modest training set size.
This result is particularly striking in view of the fact that training was
performed on simulations and data spanning about 27 years (after thinning),
during which time some secular evolution of NINO 3.4 index dynamics
certainly occurred due to carbon forcing, so that the test set comprising
the remaining data necessarily represents somewhat different climatology
from the training set. The success of the method suggests that while
the climatology may evolve, the \emph{miscalibration} of the forecast
system may be more stable over time and hence may remain a reliable guide to
recalibration.

We showed that the recalibrated forecasts have better ignorance scores
than do the original published forecasts and can consistently win bets
in the entropy game, a game that, while not fair (because the recalibrated
forecast has more information than the original forecast, and hence
the player that wields it has an edge), is not in any way biased toward
one player or another by its rules. We also pointed out that while
the entropy game is an abstract game, its expected winnings are directly
related to the wealth multiplication factor of the player with the
recalibrated forecast in Kelly-style odds-setting-and-betting games
on outcomes such as percentiles of the original forecast.

For recalibration to work well, much depends on the power and flexibility
of the modeling system used to fit the training set of PIT values;
and for the performance of the recalibrated forecast to be predictable,
the modeling system must give access to entropy measures of the distributions
being estimated. The Gaussian process measure estimation scheme described
in \ref{sec:AppendixA}, by providing a hyperparametric regression
estimate of the PIT measure that yields estimates of Kullback-Leibler
divergences from the true distribution, provides a highly satisfactory
solution for this application.

This is far from saying that further development is unnecessary. In
the first place, the studies presented in this work employed only
the most basic and simple GP kernel---the squared-exponential---in 
modeling PIT distributions. In fact, there was evidence
in \S\ref{subsec:A-Nonlinear-Circuit}, in the top-right panel of
Figure \ref{fig:Circuit_2}, that at the largest training-set sizes
the $EI[\pi(F|{\cal F},C)]$ plot deviates from the expected $N_{t}^{-1}$
behavior (see Equation \ref{eq:ES_Asym}), which could indicate a
model defect that is masked by noise for smaller training sets.
This sort of situation is probably not rare, so it would be worth
investigating the effectiveness of more flexible covariances and possibly
mixtures of such covariances.

Furthermore, the GPME model's reliance on i.i.d. training data to create a regression model
$\pi\left(F|\mathcal{F},C\right)$ qualifies the success of the recalibration
procedure in removing the i.i.d. restrictions of the methods described in the DHT \cite{diebold1999multivariate} and
KFE \cite{pmlr-v80-kuleshov18a} papers. 
Additionally, the necessity of thinning forecasts to create an approximately
uncorrelated PIT training sample can be a daunting prospect in cases
where the
data is not sufficiently abundant to support adequate thinning. In
the ENSO case an acceptable compromise was fortunately found. Nonetheless,
thinning seems an undesirable nuisance imposed by a somewhat simplistic
model---the log-Gaussian Cox process, which leads to simple
closed-form expressions at the cost of requiring that the data be
statistically independent. It would be interesting and useful to develop
the modeling in a way as to account for correlations, instead of ignoring
them, possibly by modeling the PIT distribution in two dimensions,
PIT value and time, by using a two-dimensional
Gaussian process. Other alternatives are certainly worth considering.

In this work, we have also argued from a perspective on forecast quality
assessment that emphasizes the importance of decision support. Forecast skill scores can often be difficult to interpret in terms
of decision support by someone wishing to ascertain the superiority
of some forecast set over another; and since different choices of skill
score do not agree on a unique sort order of forecast excellence,
the process of preferring some forecast sets over others on the basis
of skill has something of a beauty-contest air about it \citep{smith2015towards}.
We have seen that one can rate the performance of
forecast sets concretely, in terms of their ability to consistently
win bets against other forecast sets. It would perhaps be well to
emphasize this concrete interpretation of skill, since it would seem
to translate more directly into actionable decision-making information.

\section*{Acknowledgements}
The authors wish to thank both referees and this journal's associate editor, for
critiques that materially strengthened this article, and L.~A.~Smith for valuable discussions.
This material was based upon work supported by the US Department of Energy, Office of 
Science, Office of Advanced Scientific Computing Research, under Contract DE-AC02-06CH11347.
Jennifer Adams acknowledges support from NSF (1338427), NOAA (NA14OAR4310160)
and NASA (NNX14AM19G).

The submitted manuscript has been created by UChicago Argonne, LLC, Operator of
Argonne National Laboratory (``Argonne''). Argonne, a U.S. Department of Energy
Office of Science laboratory, is operated under Contract No. DE-AC02-06CH11357.
The U.S. Government retains for itself, and others acting on its behalf, a paid-up
nonexclusive, irrevocable worldwide license in said article to reproduce, prepare
derivative works, distribute copies to the public, and perform publicly and display
publicly, by or on behalf of the Government. The Department of Energy will provide
public access to these results of federally sponsored research in accordance with the
DOE Public Access Plan \url{http://energy.gov/downloads/doe-public-access-plan}.

\appendix

\section{Gaussian Process Probability Measure Estimation\label{sec:AppendixA}}

A large literature on probability density estimation exists, and
a number of popular techniques including kernel density estimation
(KDE) \citep[Chapter 6]{scott2015multivariate}, nearest-neighbor
estimation \citep[p. 257]{ivezic2014}, and Gaussian mixture modeling
\citep[p. 259]{ivezic2014}, as well as more sophisticated methods
based on stochastic processes well adapted to density estimation,
such as Dirichlet processes \citep{escobar1995bayesian}.

For this work, we choose a nonparametric approach based on Gaussian
process (GP) modeling. GP modeling is a popular approach for modeling spatial,
time series, and spatiotemporal data \citep{stein2012interpolation,cressie2015statistics}
and has recently received a lucid introductory
treatment in \citep{Rasmussen2006}. Some work applying GP modeling
to density estimates from Poisson-process data has appeared recently
\citep{flaxman2015fast}. We have selected and developed this technique
because it is easy to implement and leads to readily computed closed-form
expressions for the Shannon entropies that we require.

Our scheme builds on the same foundation as described in \citep{flaxman2015fast}:
we model a log-Gaussian Cox process (LGCP), wherein a Gaussian process
model is placed on the log-density of an inhomogeneous Poisson process,
described further below. Our scheme has a few new features compared
to the work described in \citep{flaxman2015fast}: We show how to
approximately normalize the density distributions so that they are,
in fact, approximately probability distributions; we exhibit closed-form
expressions for Shannon entropies associated with fit uncertainties
in the estimated densities; and we point out a curious---and to our 
knowledge, previously unrecognized---feature
of this LGCP, which is that the Laplace method approximation to the
Poisson likelihood yields a much better approximation to a Gaussian
when the log-density is approximated than when the density is approximated
directly.

\subsection{Poisson Number Density Estimation}

Suppose we have some i.i.d. sample points from some space. How do we estimate
the distribution that gave rise to those points?

More precisely: Suppose we have an absolutely continuous finite measure
$\mu$ over a set $\Gamma\subset\mathbf{R}^{D}$, representable by
a density $\rho(\boldsymbol{x})$ so that. $d\mu(\boldsymbol{x})=\rho(\boldsymbol{x})d^{D}\boldsymbol{x}$.
For any subset $b\subset\Gamma$, we interpret $\mu(b)$ as the mean
of a Poisson distribution describing the number of events of some
type that may occur in $b$. Evidently, $\mu(\Gamma)$ is the total
expected number of events in $\Gamma$, and $\mu/\mu(\Gamma)$ is
a probability measure over $\Gamma$, represented by a normalized
probability density $\pi(\boldsymbol{x})=\rho(\boldsymbol{x})/\mu(\Gamma)$.
We are given $N$ samples from this probability measure, denoted $\boldsymbol{x}_{k}$,
$k=1,\ldots,N$. Suppose that $N\gg1$. In this situation, which is
common in many fields, one often would like a sensible way to estimate
$\mu$, or, equivalently, $\rho(\boldsymbol{x})$. 

The density $\rho(\boldsymbol{x})$ is known imperfectly, since it
must be estimated from the data $\boldsymbol{x}_{k}$. Just as statistical
estimation of a real-valued scalar quantity $v$ leads naturally to
consideration of a real-valued scalar random variable $V$ whose possible
realizations are values of $v$, the estimation of a density $\rho(\boldsymbol{x})$
from data leads naturally to consideration of a density-function-valued
random variable $R(\boldsymbol{x})$ whose realizations are possible
density functions $\rho(\boldsymbol{x})$. The simplest nontrivial
theory of such stochastic functions is the theory of Gaussian processes
\citep{stein2012interpolation,Rasmussen2006}, which is what we exploit
here.

We partition $\Gamma$ into bins, which will constitute measurable
training sets. In order to avoid a coarse binning of the space that
smears out spatial structure in $\rho(\boldsymbol{x})$, the bins
should be geometrically small compared to the length scales in the
measure. On the other hand, we would like to write down a likelihood
for the data that somehow leverages the Gaussian nature of the GP
model. But the Poisson likelihood, appropriate for this kind of data,
is acceptably Gaussian in $\mu$ (the Poisson mean) only when $\mu$
is dispiritingly large, $\mu>15$ or so. Hence our requirement for small
bins is, in principle, in tension with our requirement for populous
bins.

Furthermore, a GP model of $\rho(\boldsymbol{x})$ will certainly
not respect the positivity constraint $\rho(\boldsymbol{x})>0$. It
might do so approximately, for certain choices of kernel function,
in regions with abundant sample points, but we would like
our model to apply correctly to sparsely sampled regions as well as
crowded ones.

While these obstacles seem considerable, they are surmountable. The
key element of the estimation procedure described below is log-Gaussian
Cox process (LGCP), which models $\ln\rho$, rather than $\rho$ directly,
so that $\rho$ automatically satisfies the positivity constraint.
It turns out that by a stroke of good luck, choosing to model $\ln\rho$
as a GP (instead of $\rho$ directly) solves two problems at once.
In the first place, such a model automatically satisfies the positivity
constraint $\rho(\boldsymbol{x})>0$. More subtly, as a function of
$\ln\rho$, the Laplace approximation to the Poisson likelihood \emph{approaches
the Gaussian regime much more rapidly} than it does as a function
of $\rho$!

This is not difficult to demonstrate. Consider a single Poisson variate
$n$ with mean $\mu=\exp(l)$. The Poisson likelihood for an observation
of $n$ is $\pi=e^{-\mu}\mu^{n}$. The normal approximation is achieved
by expanding $\ln\pi$ in a Taylor series about its maximum. As a
function of $\mu$, this is to say
\begin{equation}
\ln\pi=-n+n\ln n-\frac{1}{2}\frac{1}{\sigma_{\mu}^{\,2}}(\mu-\bar{\mu})^{2}+R_{1}(n,\mu),\label{eq:Residual_mu}
\end{equation}
where $\bar{\mu}=n$, $\sigma_{\mu}^{\,2}=n$, and where we denote
the approximation residual by $R_{1}(n,\mu)$. As a function of $l$,
we may similarly expand around the maximum, obtaining
\begin{equation}
\ln\pi=-n+n\ln n-\frac{1}{2}\frac{1}{\sigma_{l}^{\,2}}(l-\bar{l})^{2}+R_{2}(n,l),\label{eq:Residual_l}
\end{equation}
with $\bar{l}=\ln n$ and $\sigma_{l}^{\,2}=1/n$. The functions $R_{1}$
and $R_{2}$ are defined implicitly by Equations (\ref{eq:Residual_mu}--\ref{eq:Residual_l})
and may be computed directly from those formulae. The result is plotted
in Figure \ref{Fig:l_mu}.
\begin{center}
\begin{figure}
\begin{centering}
\includegraphics[scale=0.5]{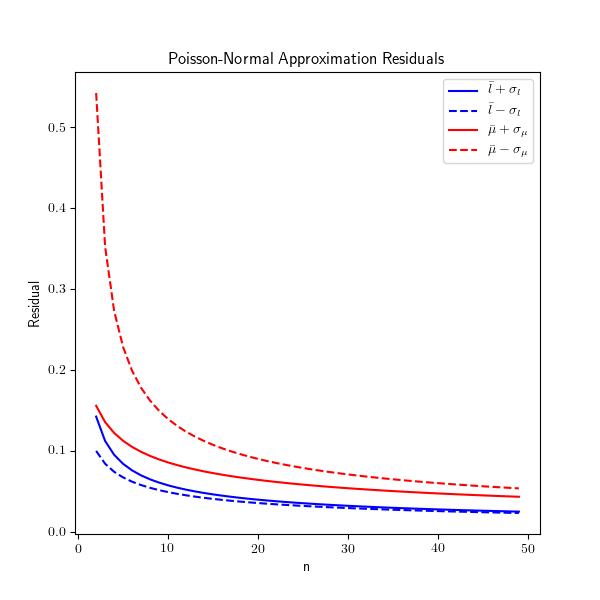}
\par\end{centering}
\caption{
\label{Fig:l_mu}Poisson-normal approximation residuals. The blue
curves show the residual magnitudes $\left|R_{l}(n)\right|$ at a
deviation of 1-$\sigma$ above (solid) and below (dashed) the mode
$\bar{l}$. The red curves display the analogous behavior for $\left|R_{\mu}(n)\right|$.}
\end{figure}
\par\end{center}

The figure shows values of the residuals computed at their respective
$\pm\sigma$ point about their respective means, that
is, $R_{1}(n,\bar{\mu}\pm\sigma_{\mu})$ and $R_{2}(n,\bar{l}\pm\sigma_{l})$.
They show that at these characteristic values of the normal distribution
to be approximated, the residual $R_{2}$ is considerably smaller
than the corresponding $R_{1}$, especially at the $-\sigma$ points.
In fact, the accuracy attained by the Gaussian approximation in $\mu$
at $n=14$ is exceeded at $n=3$ by the accuracy of the approximation
in $l$. By $n=10$ or so, the accuracy of the normal approximation
to $l$ exceeds that attained by the approximation to $\mu$ at $n>40$.
This is reassuring because it suggests that the Laplace approximation
error can be well-controlled even for small bin counts of order 5--10.

Now suppose the space $\Gamma$ has volume $\int_{\Gamma}d^{D}\boldsymbol{x}=\Omega$.
We break up the space into $B$ disjoint bins $b_{\nu},\nu=1,\ldots,B$,
satisfying $\bigcup_{\nu}b_{\nu}\subset\Gamma$, $b_{\nu}\bigcap b_{\beta}=\emptyset$
if $\nu\neq\beta$, $\int_{b_{\nu}}d^{D}\boldsymbol{x}\equiv v_{\nu}$.
We associate each bin $b_{\nu}$ with a coordinate label $\boldsymbol{x}_{\nu}$,
which is usually the location of the center of the bin. In what follows,
we will assume that the bins are small in the sense that the density
$\rho(\boldsymbol{x})$ is approximately constant over each bin.

Since the density is approximately constant over each bin, we may
set $\int_{b_{\nu}}d^{D}\boldsymbol{x}\,\rho(\boldsymbol{x})=\rho(\boldsymbol{x}_{\nu})v_{\nu}.$
We want to model the log of $\rho$ rather than $\rho$ directly.
To do so, we need a reference volume scale to nondimensionalize
$\rho$. The volume $\Omega$ will do nicely. We therefore set
\begin{equation}
l_{\nu}\equiv\ln\left(\rho(\boldsymbol{x}_{\nu})\Omega\right),\label{eq:lnudef}
\end{equation}
from which follows
\begin{eqnarray}
\rho(\boldsymbol{x}_{\nu})v_{\nu} & = & \frac{v_{\nu}}{\Omega}e^{l_{\nu}}\nonumber \\
 & \equiv & \omega_{\nu}e^{l_{\nu}},\label{eq:rho-l}
\end{eqnarray}
defining the dimensionless volume element $\omega_{\nu}\equiv v_{\nu}/\Omega$.

Suppose we observe $n_{\nu}$ samples from the density in bin $b_{\nu}$.
If we are given the density $\rho(\boldsymbol{x})$, we may compute
the Poisson process likelihood $\mathcal{L}\left(\boldsymbol{X}|\boldsymbol{l}\right)$ of the data. Setting $\rho(\boldsymbol{x}_{\nu})\equiv\rho_{\nu}$
and $\boldsymbol{X}=\{\boldsymbol{x}_{k},k=1,\ldots,N\}$ for brevity,
we have

\begin{eqnarray}
\mathcal{L}\left(\boldsymbol{X}|\boldsymbol{l}\right) & = & \left[\prod_{\nu=1}^{B}\exp\left(-\omega_{\nu}e^{l_{\nu}}+n_{\nu}l_{\nu}\right)\times\omega_{\nu}^{\,n_{\nu}}\right]\nonumber \\
 & \approx & \left[\prod_{\nu=1}^{B}\exp\left(-n_{\nu}+n_{\nu}\ln n_{\nu}-\frac{n_{\nu}}{2}\left(l_{\nu}-\ln\left(n_{\nu}/\omega_{\nu}\right)\right)^{2}\right)\right]\nonumber \\
 & = & \mathrm{const.}\times\exp\left[-\frac{1}{2}\left(\boldsymbol{l}-\boldsymbol{l}_{1}\right)^{T}\boldsymbol{D}^{-1}\left(\boldsymbol{l}-\boldsymbol{l}_{1}\right)\right],\label{eq:PP_Lik}
\end{eqnarray}
where we have appealed to the normal approximation discussed above,
and we defined 
\begin{eqnarray}
\boldsymbol{l} & \equiv & \left[l_{1},\ldots l_{B}\right]{}^{T},\label{eq:vecl_def}\\
\boldsymbol{l}_{1} & \equiv & \left[\ln(n_{1}/\omega_{1}),\ldots,\ln(n_{B}/\omega_{B})\right]^{T}\label{eq:vecl1_def}\\
\boldsymbol{D} & \equiv & \mathrm{diag}\left[n_{1}^{\,-1},\ldots,n_{B}^{\,-1}\right]\label{eq:D_Def}
\end{eqnarray}
Clearly at this point that we have committed to having no
empty bins, since any empty bin completely compromises the normal
approximation that we have just introduced. In fact, as discussed above,
we will be happy with bin sample counts in the 5--10 region.

Note that we have implicitly assumed here that the data are well described
by a Poisson point process and, in particular, that the $\boldsymbol{x}_{k}$
are i.i.d. If this assumption is incorrect, then
Equation (\ref{eq:PP_Lik}) is not the correct expression for the
likelihood. In some cases, when the $\boldsymbol{x}_{k}$ arise from
a stationary time-series with nonzero correlations $<\boldsymbol{x}_{k}\boldsymbol{x}_{l}>=f(|k-l|)$,
we may be able to thin out the data by a factor suggested by the
shape of the autocorrelation function, so as to obtain an approximately
uncorrelated data sample that may be correctly modeled by using Equation
(\ref{eq:PP_Lik}).

We will model the function $l(\boldsymbol{x})$ using a constant-mean
Gaussian process,
\begin{equation}
l\sim GP\left(l_{0}(\boldsymbol{x}),K(\boldsymbol{x},\boldsymbol{x}^{\prime};\theta)\right),\label{eq:GPdef}
\end{equation}
where the mean function $l_{0}(\boldsymbol{x})$ is in fact a constant
$l_{0}$ and the covariance $K(\boldsymbol{x},\boldsymbol{x}^{\prime};\theta)$
is a positive-definite (as an integral kernel) function, parametrized
by some hyperparameters denoted by $\theta$. For example, we might
choose a stationary kernel with a scale hyperparameter $\sigma$ and
an amplitude hyperparameter $A$, that is,
\begin{equation}
K(\boldsymbol{x},\boldsymbol{x}^{\prime})=Ak\left(\frac{\boldsymbol{x}-\boldsymbol{x}^{\prime}}{\sigma}\right),\label{eq:Kernel_param}
\end{equation}
in which case $\theta=(A,\sigma)$. The specific form of the covariance
function is not needed here, and it could in general be chosen from
the many known valid covariance forms, as seems suited to the type
of measure being modeled \citep[Chapter 4]{Rasmussen2006}. 

The choice of a parametrized mean level $l_{0}$ is made here because
a zero-mean GP (the more usual choice) effectively makes a choice
of amplitude scale for $\rho$ that is not selected by the data. One commonly 
obviates this kind of issue by mean-subtracting the data
$\boldsymbol{l}_{1}$. However, many weighted means of the
data in $\boldsymbol{l}_{1}$ could be chosen for this purpose,
and it is not a priori clear that the usual unweighted mean $\bar{l}=B^{-1}\sum_{\nu=1}^{B}l_{\nu}$
is optimal among these. Setting the mean level as an adjustable parameter
selects a certain weighted mean as the maximum-likelihood estimate
(MLE) of $l_{0}$ and turns out to be computationally inexpensive,
as we will see below.

The covariance matrix $\boldsymbol{Q}$ arising from the GP model
is just the Gram matrix of the covariance function,
\begin{equation}
\left[\boldsymbol{Q}\right]_{\nu\nu^{\prime}}=K(\boldsymbol{x}_{\nu},\boldsymbol{x}_{\nu^{\prime}};\theta),\label{eq:Gram_matrix}
\end{equation}
where the indices $\nu,\nu^{\prime}$ range over the $B$ bins. The
mean vector arising from the process is $\bar{\boldsymbol{l}}=l_{0}\boldsymbol{u}_{B}$,
where $\boldsymbol{u}_{B}$ is the $B$-dimensional ``one'' vector,
$\left[\boldsymbol{u}_B\right]_{\nu}=1$, $\nu=1,\ldots,B$, and $l_{0}$
is a parameter to be estimated, residing in the mean function rather
than in the covariance kernel.

In terms of $\boldsymbol{Q}$ and $\bar{\boldsymbol{l}}$, the probability
of a density $\rho$ represented by $B$-dimensional vector $\boldsymbol{l}$
is

\begin{equation}
\pi\left(\boldsymbol{l}|I\right)d^{B}\boldsymbol{l}=(2\pi)^{-N/2}\left[\det\boldsymbol{Q}\right]^{-1/2}\exp\left[-\frac{1}{2}\left(\boldsymbol{l}-l_{0}\boldsymbol{u}_{B}\right)^{T}\boldsymbol{Q}^{-1}\left(\boldsymbol{l}-l_{0}\boldsymbol{u}_{B}\right)\right]d^{B}\boldsymbol{l},\label{eq:Prob_pi}
\end{equation}
where we have symbolically collected in $I$ all conditioning information
such as parameters $\theta,l_{0}$ and covariance kernel choices. 

The Poisson process likelihood of Equation (\ref{eq:PP_Lik}) can
be marginalized over the GP distribution for $\rho$ of Equation (\ref{eq:Prob_pi})
to produce the \emph{marginal likelihood}:
\begin{eqnarray}
\mathcal{L}\left(\boldsymbol{X}|I\right) & = & \int d^{B}\boldsymbol{l}\,\mathcal{L}\left(\boldsymbol{X}|\boldsymbol{l}\right)\pi(\boldsymbol{l}|I)\nonumber \\
 & = & (2\pi)^{-N/2}\left[\det\boldsymbol{Q}\right]^{-1/2}\times\int d^{B}\boldsymbol{l}\,\exp\Biggl[-\frac{1}{2}\left(\boldsymbol{l}-l_{0}\boldsymbol{u}_{B}\right)^{T}\boldsymbol{Q}^{-1}\left(\boldsymbol{l}-l_{0}\boldsymbol{u}_{B}\right)\nonumber \\
 &  & \hspace{2.2cm}-\frac{`1}{2}\left(\boldsymbol{l}-\boldsymbol{l}_{1}\right)^{T}\boldsymbol{D}^{-1}\left(\boldsymbol{l}-\boldsymbol{l_{1}}\right)\Biggr].\label{eq:marglik_1}
\end{eqnarray}
This is, unsurprisingly, the form for the marginal likelihood of a
GP trained on noisy data $\boldsymbol{l}_{1}$ with noise covariance
$\boldsymbol{D}$. We may therefore take over the standard result
for the marginal likelihood \citep[Eq. 2.30]{Rasmussen2006}, adapted
for the case of non-constant mean and heteroskedastic noise:

\begin{equation}
\mathcal{L}\left(\boldsymbol{X}|I\right)=\mathrm{const.}\times\left[\det\left(\boldsymbol{Q}+\boldsymbol{D}\right)\right]^{-1/2}\exp\left[-\frac{1}{2}w\right],\label{eq:marglik_2}
\end{equation}
where
\begin{equation}
w\equiv\left(\boldsymbol{l}_{1}-l_{0}\boldsymbol{u}_{B}\right)\left(\boldsymbol{Q}+\boldsymbol{D}\right)^{-1}\left(\boldsymbol{l}_{1}-l_{0}\boldsymbol{u}_{B}\right).\label{eq:w_1}
\end{equation}

We may use Equation (\ref{eq:marglik_2}) to obtain an MLE of $l_{0}$ as a function of the kernel parameters
$\theta$. We define the function $\tilde{l}_{0}(\theta)$ by
\begin{equation}
\tilde{l}_{0}(\theta)=\frac{\boldsymbol{l}_{1}^{T}\left(\boldsymbol{Q}+\boldsymbol{D}\right)^{-1}\boldsymbol{u}_{B}}{\boldsymbol{u}_{B}^{T}\left(\boldsymbol{Q}+\boldsymbol{D}\right)^{-1}\boldsymbol{u}_{B}}.\label{eq:l0_mle}
\end{equation}
Then $l_{0}=\tilde{l}_{0}(\theta)$ is the conditional MLE of $l_{0}$
given $\theta$. Defining $\boldsymbol{g\equiv}\left(\boldsymbol{Q}+\boldsymbol{D}\right)^{-1}\boldsymbol{u}_{B}$,
we may write 
\begin{equation}
\tilde{l}_{0}(\theta)=\frac{\sum_{\nu=1}^{B}l_{1\nu}g_{\nu}}{\sum_{\nu-1}^{B}g_{\nu}},\label{eq:l_wgt_avg}
\end{equation}
which shows that the MLE of $l_{0}$ is in fact a weighted average
of $\boldsymbol{l}_{1}$, as asserted previously.

Combining Equations (\ref{eq:l0_mle}) and (\ref{eq:w_1}), we obtain
\begin{equation}
w\biggl|_{l_{0}=\tilde{l}_{0}(\theta)}=\boldsymbol{l}_{1}^{T}\left(\boldsymbol{Q}+\boldsymbol{D}\right)^{-1}\boldsymbol{l}_{1}-\frac{\left[\boldsymbol{l}_{1}^{T}\left(\boldsymbol{Q}+\boldsymbol{D}\right)^{-1}\boldsymbol{u}_{B}\right]^{2}}{\boldsymbol{u}_{B}^{T}\left(\boldsymbol{Q}+\boldsymbol{D}\right)^{-1}\boldsymbol{u}_{B}},\label{eq:w_mle}
\end{equation}
which, by the Schwartz inequality, is non-negative definite and can
attain a zero value only when $\boldsymbol{l}_{1}=\alpha\boldsymbol{u}_{B}$
for some scalar $\alpha$. 

Combining Equation(\ref{eq:w_mle}) with the negative log of Equation
(\ref{eq:marglik_2}), we conclude that the MLE of $\theta,l_{0}$
are obtained by minimizing the objective function $S(\theta)$:
\begin{equation}
S(\theta)\equiv\ln\det\left(\boldsymbol{Q}+\boldsymbol{D}\right)+\boldsymbol{l}_{1}^{T}\left(\boldsymbol{Q}+\boldsymbol{D}\right)^{-1}\boldsymbol{l}_{1}-\frac{\left[\boldsymbol{l}_{1}^{T}\left(\boldsymbol{Q}+\boldsymbol{D}\right)^{-1}\boldsymbol{u}_{B}\right]^{2}}{\boldsymbol{u}_{B}^{T}\left(\boldsymbol{Q}+\boldsymbol{D}\right)^{-1}\boldsymbol{u}_{B}},\label{eq:Obj_Fn}
\end{equation}
\begin{eqnarray}
\theta^{(MLE)} & = & \argmin_{\theta}S(\theta),\label{eq:theta_mle}\\
l_{0}^{(MLE)} & = & \tilde{l}_{0}\left(\theta^{(MLE)}\right).\label{eq:l0_mle_2}
\end{eqnarray}

Suppose that we would like to estimate the density $\rho(\boldsymbol{y}_{a})$---or 
more to the point, the log-density $l_{a}^{(pred)}\equiv\ln\left(\rho(\boldsymbol{y}_{a})\Omega\right)$---at 
a set of points $\boldsymbol{y}_{a},$ $a=1,\ldots,P$. We may do this 
by following the standard methodological path of GP regression.
We first extend the covariance matrix to an $(B+P)\times(B+P)$ matrix
$\hat{\boldsymbol{Q}}$, writing
\begin{equation}
\hat{\boldsymbol{Q}}\equiv\left[\begin{array}{cc}
\boldsymbol{Q}^{(pred)} & \boldsymbol{k}^{T}\\
\boldsymbol{k} & \boldsymbol{Q}
\end{array}\right],\label{eq:Qhat}
\end{equation}
with 
\begin{eqnarray}
\left[\boldsymbol{Q}^{(pred)}\right]_{ab} & \equiv & K(\boldsymbol{y}_{a},\boldsymbol{y}_{b};\theta),\label{eq:Qpred}\\
\left[\boldsymbol{k}\right]_{\nu a} & \equiv & K(\boldsymbol{x}_{\nu},\boldsymbol{y}_{a};\theta).\label{eq:kvec}
\end{eqnarray}
We further define $\boldsymbol{l}^{(pred)}=[l_{1}^{(pred)},\ldots,l_{P}^{(pred)}]^{T}$,
and $P$-dimensional ``one'' vector $[\boldsymbol{u}_{P}]_{a}=1$,
$a=1,\ldots,P$. Then we may take over the standard formula for predictions
by a GP trained with noisy data \citep[pp. 16--18]{Rasmussen2006},
again adapted for nonzero mean and heteroskedastic noise. That is,
$\boldsymbol{l}^{(pred)}\sim\mathcal{N}(\boldsymbol{\lambda}^{(pred)},\boldsymbol{C}^{(pred)})$,
with
\begin{equation}
\boldsymbol{\lambda}^{(pred)}=l_{0}\boldsymbol{u}_{P}+\boldsymbol{k}_{y}^{T}\left(\boldsymbol{Q}+\boldsymbol{D}\right)^{-1}(\boldsymbol{l}_{1}-l_{0}\boldsymbol{u}_{B})\label{eq:lambda_pred}
\end{equation}
and
\begin{eqnarray}
\boldsymbol{C}^{(pred)} & = & \boldsymbol{Q}^{(pred)}-\boldsymbol{k}^{T}\left(\boldsymbol{Q}+\boldsymbol{D}\right)^{-1}\boldsymbol{k}.\label{eq:C_pred}
\end{eqnarray}
This is essentially a new, updated Gaussian process, with ``trained''
mean function
\begin{equation}
\lambda(\boldsymbol{x})=l_{0}+\sum_{\nu=1}^{B}K(\boldsymbol{x}_{\nu},\boldsymbol{x};\theta)\left[\left(\boldsymbol{Q}+\boldsymbol{D}\right)^{-1}(\boldsymbol{l}_{1}-l_{0}\boldsymbol{u}_{B})\right]_{\nu},\label{eq:lambda_pred_fn}
\end{equation}
and ``trained'' covariance function
\begin{equation}
C(\boldsymbol{x},\boldsymbol{y})=K(\boldsymbol{x},\boldsymbol{y};\theta)-\sum_{\nu,\mu=1}^{B}K(\boldsymbol{x},\boldsymbol{x}_{\mu};\theta)K(\boldsymbol{y},\boldsymbol{x}_{\nu};\theta)\left[\left(\boldsymbol{Q}+\boldsymbol{D}\right)^{-1}\right]_{\nu\mu}.\label{eq:C_pred_fn}
\end{equation}

Returning to the higher-level ``random function'' view, we may summarize
the story so far as follows. There is an unknown unnormalized density
$\rho(\boldsymbol{x})$, from which a set of points $\boldsymbol{X}=\{\boldsymbol{x}_{k},k=1,\ldots,N\}$
is iid sampled. Since $\rho(\boldsymbol{x})$ is imperfectly known,
we represent it by a function-valued random variable $R(\boldsymbol{x})$
and its scaled logarithm $l(\boldsymbol{x})=\ln(\Omega\rho(\boldsymbol{x}))$
by a function-valued random variable $L(\boldsymbol{x})$ to be estimated
by GPME. Realizations of $L(\cdot)$ are possible log-density functions
$l(\cdot)$. Similarly, realizations of the function-valued random
variable $R(\cdot)=\Omega^{-1}\exp(L(\cdot))$ are possible density
functions $\rho(\cdot)$.

The prior distribution over $L(\cdot)$ is a hierarchical model featuring
a Gaussian process with constant mean function $l_{0}$ and a covariance
function $K(\boldsymbol{x},\boldsymbol{y};\theta)$, as well as some
prior distribution over $l_{0},\theta$ that we will not need to specify
since we will proceed by maximizing the likelihood with respect to
these parameters (when the parameter priors change slowly compared
with the likelihood, this is approximately MAP estimation). We represent
this prior distribution by the notation $L(\cdot)|I\sim GP[l_{0},K(\cdot,\cdot)]$.
Training with the data $\boldsymbol{X}$ yields an updated posterior
distribution for $L(\cdot)\,|\,(\boldsymbol{X},I)$, which is a Gaussian
process with mean function $\lambda(\boldsymbol{x})$ and covariance
function $C(\boldsymbol{x},\boldsymbol{y})$. Notationally, $L(\cdot)\,|\,(\boldsymbol{X},I)\sim GP\left[\lambda(\cdot),C(\cdot,\cdot)\right]$.

So far, we have modeled the imperfectly known log-density $L(\boldsymbol{x})$,
rather than $R(\boldsymbol{x})$. This choice has consequences for
the inferred Poisson process density, which is not, as one might naively
assume, simply a constant times $\exp(\lambda(\boldsymbol{x}))$.
In a small volume $v\equiv\Omega\omega$ about a location $\boldsymbol{x}$,
the expected number of events $n$\emph{ given} the imperfectly known
log-density $L(\boldsymbol{x})$ is
\begin{equation}
vR(\boldsymbol{x})=\omega\exp\left[L(\boldsymbol{x})\right].\label{eq:Exp_1}
\end{equation}
Given the GP posterior predictive distribution $L(\cdot)\,|\,\boldsymbol{X}\sim GP[\lambda(\cdot),C(\cdot,\cdot)]$
, the effective number density $\rho_{E}(\boldsymbol{x})$ at $\boldsymbol{x}$
is given by
\begin{eqnarray}
\rho_{E}(\boldsymbol{x}) & = & E_{L(\cdot)\,|\,(\boldsymbol{X},I)}\left\{ R(\boldsymbol{x})\right\} \nonumber \\
 & = & \Omega^{-1}\left(2\pi C(\boldsymbol{x},\boldsymbol{x})\right)^{-1/2}\int dl\ \exp\left\{ -\frac{1}{2}\frac{\left[l-\lambda(\boldsymbol{x})\right]^{2}}{C(\boldsymbol{x},\boldsymbol{x})}\right\} \times\exp(l)\nonumber \\
 & = & \Omega^{-1}\exp\left[\lambda(\boldsymbol{x})\right]\times\exp\left[\frac{1}{2}C(\boldsymbol{x},\boldsymbol{x})\right].\label{eq:eta}
\end{eqnarray}
We see that the log expected number of events is shifted with respect
to the log-density $l(\boldsymbol{x})$ by the nonconstant factor
$C(\boldsymbol{x},\boldsymbol{x})/2$.

\subsection{Probability Density Estimation\label{subsec:Effective-Probability-Density}}

The posterior predictive probability density $\pi(\boldsymbol{x}|\boldsymbol{X},I)$
that a future event should occur within a differential volume $d^{D}\boldsymbol{x}$
of $\boldsymbol{x}$ is the normalized version of $\rho_{E}(\boldsymbol{x})$:
\begin{equation}
\pi(\boldsymbol{x}|\boldsymbol{X},I)=\rho_{E}(\boldsymbol{x})/\int d^{D}\boldsymbol{x}\,\rho_{E}(\boldsymbol{x}),\label{eq:Prob1}
\end{equation}
where the notation $\pi(\boldsymbol{x}|\boldsymbol{X},I)$ will be
justified below. In the asymptotic limit $N\rightarrow\infty$, the
normalization constant may be directly estimated from the data. We
replace the integral by a sum over the training bins, in effect selecting
the same prediction points as training bin centers:
\begin{equation}
A\equiv\int d^{D}\boldsymbol{x}\,\rho_{E}(\boldsymbol{x})\approx\sum_{\nu=1}^{B}\omega_{\nu}\exp\left\{ \lambda_{\nu}\right\} \times\exp{\frac{1}{2}C_{\nu\nu}},\label{eq:Norm1}
\end{equation}
where
\begin{eqnarray}
\boldsymbol{\lambda} & = & l_{0}\boldsymbol{u}_{B}+\boldsymbol{Q}\left(\boldsymbol{Q}+\boldsymbol{D}\right)^{-1}\left(\boldsymbol{l}_{1}-l_{0}\boldsymbol{u}_{B}\right)\nonumber \\
 & = & \boldsymbol{l}_{1}-\boldsymbol{D}(\boldsymbol{Q}+\boldsymbol{D})^{-1}\left(\boldsymbol{l}_{1}-l_{0}\boldsymbol{u}_{B}\right)\label{eq:lambda1}
\end{eqnarray}
and
\begin{eqnarray}
\boldsymbol{C} & = & \boldsymbol{Q}-\boldsymbol{Q}\left(\boldsymbol{Q}+\boldsymbol{D}\right)^{-1}\boldsymbol{Q}\nonumber \\
 & = & \boldsymbol{D}-\boldsymbol{D}\left(\boldsymbol{Q}+\boldsymbol{D}\right)^{-1}\boldsymbol{D}.\label{eq:C1}
\end{eqnarray}

In the asymptotic limit, $\boldsymbol{D}\rightarrow0$, and we have
$\boldsymbol{\lambda}\approx\boldsymbol{l}_{1}$, $\boldsymbol{C}\approx\boldsymbol{0}$.
Since $\left[\boldsymbol{l}_{1}\right]_{\nu}=\ln\frac{n_{\nu}}{\omega_{\nu}}$,
we have
\begin{eqnarray}
A & \approx & \sum_{\nu=1}^{B}\omega_{\nu}\times\left(\frac{n_{\nu}}{\omega_{\nu}}\right)\nonumber \\
 & = & N,\label{eq:Norm2}
\end{eqnarray}
which is an entirely unsurprising result.

Combining Equations (\ref{eq:eta}), (\ref{eq:Prob1}), and (\ref{eq:Norm1}),
we may write
\begin{equation}
\pi(\boldsymbol{x}|\boldsymbol{X},I)=(A\Omega)^{-1}\exp\left(\lambda(\boldsymbol{x})+\frac{1}{2}C(\boldsymbol{x},\boldsymbol{x})\right).\label{eq:pi_E}
\end{equation}

We have seen that GPME furnishes a tractable posterior distribution
over $R(\boldsymbol{x})$, the imperfectly known Poisson number density
that estimates $\rho(\boldsymbol{x})$. Now consider the normalized
probability density $\pi(\boldsymbol{x})=\rho(\boldsymbol{x})/J[\rho]$,
where $J[\rho]\equiv\int d^{D}\boldsymbol{x}\,\rho(\boldsymbol{x})$.
We would like to use GPME to obtain a posterior distribution over
$\Pi(\boldsymbol{x})$, the imperfectly known probability density
(a probability-density-valued random variable) that estimates $\pi(\boldsymbol{x})$.

An unfortunate property of the the GPME scheme is that while it
yields a tractable distribution over log number density $L(\boldsymbol{x})$,
it does not yield a tractable distribution over probability density
$\ln\Pi(\boldsymbol{x})$. The reason is that the transformation from
$L(\boldsymbol{x})$ to $\ln\Pi(\boldsymbol{x})$ is nonlinear and
does not transform the Gaussian distribution in $L$ into another
tractable distribution over either $\Pi$ or $\ln\Pi$. The normalization
factor $A$ discussed above pertains to the effective number density
$\rho_{E}(\boldsymbol{x})$, which, according to the first line of
Equation (\ref{eq:eta}), is the expectation of the density $R(\boldsymbol{x})$
over the posterior distribution of the GP. This factor allows us to
transition from the expected density to the posterior predictive probability
density $\pi(\boldsymbol{x}|\boldsymbol{X},I)$. The factor $A$ is
not, in general, the normalization appropriate to the imperfectly known
density $R(\boldsymbol{x})$.

If we are satisfied with approximate normalization, however, then
the factor $A$ \emph{is} an appropriate normalization. The reason
is that, as we now show, in the asymptotic limit $E_{L(\cdot)\,|\,\boldsymbol{X}}\left\{ J\right\} =A\approx N$,
$\mathrm{Var}_{L(\cdot)\,|\,\boldsymbol{X}}\left\{ J\right\} \approx N$
so that $E_{L(\cdot)\,|\,\boldsymbol{X}}\left\{ J\right\} /\sqrt{\mathrm{Var}_{L(\cdot)\,|\,\boldsymbol{X}}\left\{ J\right\} }\approx N^{-1/2}$.
Consequently, for large $N$, only a small error is committed by replacing
$J[\rho]$ by $A$, and we may set
\begin{equation}
\Pi(\boldsymbol{x})\approx A^{-1}R(\boldsymbol{x})=(A\Omega)^{-1}e^{L(\boldsymbol{x})}.\label{eq:pi_approx}
\end{equation}
This amounts to a constant offset of $\ln\Pi$ from $L$, so that
the Gaussian distribution over $L(\boldsymbol{x})$ is simply mean-shifted
by $-\ln(A\Omega)$ to produce the Gaussian distribution over $\ln\Pi$.

To show the required expectations, we again approximate the integral $J$, a random variable, by the sum over observed bins,
\begin{eqnarray}
J & = & \int d^{D}\boldsymbol{x}\,R(\boldsymbol{x})\nonumber \\
 & \approx & \sum_{\nu=1}^{B}\omega_{\nu}e^{L(\boldsymbol{x}_{\nu})},\label{eq:J1}
\end{eqnarray}
so that
\begin{eqnarray}
E_{L(\cdot)\,|\,\boldsymbol{X}}\left\{ J\right\}  & \approx & \sum_{\nu=1}^{B}\omega_{\nu}E_{L(\cdot)\,|\,\boldsymbol{X}}\left\{ e^{L(\boldsymbol{x}_{\nu})}\right\} \nonumber \\
 & = & \sum_{\nu=1}^{B}\omega_{\nu}\exp\left[\lambda(\boldsymbol{x})\right]\times\exp\left[\frac{1}{2}C(\boldsymbol{x},\boldsymbol{x})\right]\nonumber \\
 & = & A.\label{eq:EJ}
\end{eqnarray}

Furthermore,
\begin{equation}
E_{L(\cdot)\,|\,\boldsymbol{X}}\left\{ J^{2}\right\} \approx\sum_{\nu=1}^{B}\sum_{\mu=1}^{B}\omega_{\nu}\omega_{\mu}E_{L(\cdot)\,|\,\boldsymbol{X}}\left\{ e^{L(\boldsymbol{x}_{\mu})+L(\boldsymbol{x}_{\nu})}\right\} .\label{eq:EJ2_1}
\end{equation}

Defining the $B$-dimensional vector $\boldsymbol{m}$ by
\begin{equation}
[\boldsymbol{m}(\mu,\nu)]_{\sigma}=\delta_{\nu\sigma}+\delta_{\mu\sigma},\label{eq:mvec}
\end{equation}
we may write this as
\begin{eqnarray}
E_{L(\cdot)\,|\,\boldsymbol{X}}\left\{ J^{2}\right\}  & \approx & \sum_{\nu=1}^{B}\sum_{\mu=1}^{B}\omega_{\nu}\omega_{\mu}(2\pi)^{-B/2}\left(\det\boldsymbol{C}\right)^{-1/2}\nonumber \\
 &  & \times\int d^{B}\boldsymbol{l}\,\exp\left\{ -\frac{1}{2}\left(\boldsymbol{l}-\boldsymbol{\lambda}\right)^{T}\boldsymbol{C}^{-1}(\boldsymbol{l}-\boldsymbol{\lambda})+\boldsymbol{m}(\mu,\nu)^{T}\boldsymbol{l}\right\} \nonumber \\
 & = & \sum_{\nu=1}^{B}\sum_{\mu=1}^{B}\omega_{\nu}\omega_{\mu}\exp\left[\boldsymbol{m}(\mu,\nu)^{T}\boldsymbol{\lambda}\right]\times\exp\left[\frac{1}{2}\boldsymbol{m}(\mu,\nu)^{T}\boldsymbol{C}\boldsymbol{m}(\mu,\nu)\right]\nonumber \\
 & = & \sum_{\nu=1}^{B}\sum_{\mu=1}^{B}\omega_{\nu}\omega_{\mu}\exp\left[\lambda_{\nu}+\lambda_{\mu}\right]\times\exp\left[\frac{1}{2}C(\boldsymbol{x}_{\nu},\boldsymbol{x}_{\nu})+\frac{1}{2}C(\boldsymbol{x}_{\mu},\boldsymbol{x}_{\mu})+C(\boldsymbol{x}_{\nu},\boldsymbol{x}_{\mu})\right].\nonumber \\
\label{eq:EJ2_2}
\end{eqnarray}
We then have
\begin{eqnarray}
\mathrm{Var}_{L(\cdot)\,|\,\boldsymbol{X}}\left\{ J\right\}  & = & E_{L(\cdot)\,|\,\boldsymbol{X}}\left\{ J^{2}\right\} -\left[E_{L(\cdot)\,|\,\boldsymbol{X}}\left\{ J\right\} \right]^{2}\nonumber \\
 & \approx & \sum_{\nu=1}^{B}\sum_{\mu=1}^{B}\omega_{\nu}\omega_{\mu}\exp\left[\lambda_{\nu}+\lambda_{\mu}\right]\nonumber \\
 &  & \times\exp\left[\frac{1}{2}C(\boldsymbol{x}_{\nu},\boldsymbol{x}_{\nu})+\frac{1}{2}C(\boldsymbol{x}_{\mu},\boldsymbol{x}_{\mu})\right]\left\{ \exp\left[C(\boldsymbol{x}_{\nu},\boldsymbol{x}_{\mu})\right]-1\right\} .\nonumber \\
\label{eq:VarJ_1}
\end{eqnarray}

In the asymptotic limit, by Equation (\ref{eq:C1}) $\boldsymbol{C}\rightarrow\boldsymbol{D}$,
which is diagonal and has small matrix elements $1/n_{\nu}$, so that
\begin{eqnarray}
\mathrm{Var}_{L(\cdot)\,|\,\boldsymbol{X}}\left\{ J\right\}  & \approx & \sum_{\nu=1}^{B}\sum_{\mu=1}^{B}\omega_{\nu}\omega_{\mu}\exp\left[\lambda_{\nu}+\lambda_{\mu}\right]\times\exp\left[\frac{1}{2}C(\boldsymbol{x}_{\nu},\boldsymbol{x}_{\nu})+\frac{1}{2}C(\boldsymbol{x}_{\mu},\boldsymbol{x}_{\mu})\right]\left[\boldsymbol{D}\right]_{\nu\mu}\nonumber \\
 & \approx & \sum_{\nu=1}^{B}\omega_{\nu}^{2}\times\left(\frac{n_{\nu}}{\omega_{\nu}}\right)^{2}\times n_{\nu}^{-1}\nonumber \\
 & = & N.\label{eq:VarJ_2}
\end{eqnarray}

Equations (\ref{eq:EJ}) and (\ref{eq:VarJ_2}) are the required relations
that allow us to approximate $\Pi(\boldsymbol{x})\approx A^{-1}L(\boldsymbol{x})$
in the asymptotic regime and hence approximate the posterior distribution
over $\ln\Pi$ by a Gaussian process. In this light, we may cast the
posterior predictive distribution $\pi(\boldsymbol{x}|\boldsymbol{X},I)$,
defined in Equation (\ref{eq:Prob1}), as
\begin{eqnarray}
\pi(\boldsymbol{x}|\boldsymbol{X},I) & = & \rho_{E}(\boldsymbol{x})/A\nonumber \\
 & = & E_{\Pi\,|\,(\boldsymbol{X},I)}\left\{ \Pi(\boldsymbol{x})\right\} .\label{eq:Prob3}
\end{eqnarray}
This equation provides the justification for attaching the notation
$\pi(\boldsymbol{x}|\boldsymbol{X},I)$ to the posterior predictive
distribution.

\subsection{Entropy Estimation}

The practical output of the probability density estimation procedure
is the posterior predictive probability density $\pi(\boldsymbol{x}|\boldsymbol{X},I)$.
One might ask how far this is from the imperfectly known true probability
density $\Pi(\boldsymbol{x})$. The Kullback-Leibler divergence between
the two distributions is
\begin{equation}
KL\left[\Pi\,||\,\pi(\boldsymbol{x}|\boldsymbol{X},I)\right]=\int d^{D}\boldsymbol{x}\,\Pi(\boldsymbol{x})\ln\frac{\Pi(\boldsymbol{x})}{\pi(\boldsymbol{x}|\boldsymbol{X},I)},\label{eq:KL}
\end{equation}
which is a random variable that measures departure of $\Pi(\boldsymbol{x})$
from $\pi(\boldsymbol{x}|\boldsymbol{X},I)$. We may calculate the
expected divergence
\begin{eqnarray}
ES\left[\pi\left(\boldsymbol{x}|\boldsymbol{X}\right)\right] & \equiv & E_{\Pi\,|\,\boldsymbol{X}}\left\{ KL\left[\Pi\,||\,\pi(\boldsymbol{x}|\boldsymbol{X},I)\right]\right\} \nonumber \\
 & = & \Omega^{-1}\int d^{D}\boldsymbol{x}\,E_{\Pi\,|\,\boldsymbol{X}}\left\{ A^{-1}e^{L(\boldsymbol{x})}\left(L(\boldsymbol{x})-\ln A\right)\right\} \nonumber \\
 &  & -\int d^{D}\boldsymbol{x}\,\pi(\boldsymbol{x}|\boldsymbol{X},I)\ln\left(\Omega\pi(\boldsymbol{x}|\boldsymbol{X},I)\right).\label{eq:ES_1}
\end{eqnarray}
We have
\begin{eqnarray}
E_{\Pi\,|\,\boldsymbol{X}}\left\{ A^{-1}e^{L(\boldsymbol{x})}\right\}  & = & A^{-1}e^{\lambda(\boldsymbol{x})+\frac{1}{2}C(\boldsymbol{x},\boldsymbol{x})}\nonumber \\
 & = & \Omega\pi(\boldsymbol{x}|\boldsymbol{X},I)\label{eq:Eel}
\end{eqnarray}
and
\begin{eqnarray}
E_{\Pi\,|\,\boldsymbol{X}}\left\{ A^{-1}L(\boldsymbol{x})e^{L(\boldsymbol{x})}\right\}  & = & A^{-1}\left(2\pi C(\boldsymbol{x},\boldsymbol{x})\right)^{-1/2}\int dl\,\exp\left[-\frac{1}{2}\frac{(l-\lambda(\boldsymbol{x}))^{2}}{C(\boldsymbol{x},\boldsymbol{x})}\right]\times l\,e^{l}\nonumber \\
 & = & \Omega\pi(\boldsymbol{x}|\boldsymbol{X},I)\left[\ln\left(\Omega\pi(\boldsymbol{x}|\boldsymbol{X},I)\right)+\ln A+\frac{1}{2}C(\boldsymbol{x},\boldsymbol{x})\right].\nonumber \\
\label{eq:Elel}
\end{eqnarray}
Putting all this together, we find
\begin{equation}
ES\left[\pi(\boldsymbol{x}|\boldsymbol{X},I)\right]=\int d^{D}\boldsymbol{x}\,\pi(\boldsymbol{x}|\boldsymbol{X},I)\times\frac{1}{2}C(\boldsymbol{x},\boldsymbol{x}).\label{eq:ES_2}
\end{equation}
This is an intuitively reasonable result: the expected divergence between the
estimated probability density and the true density is proportional
the average of the posterior variance weighted by the posterior predictive
density. As the quality of the fit improves, the variance decreases
and takes $ES\left[\pi(\boldsymbol{x}|\boldsymbol{X},I)\right]$ down
with it.

We may obtain an asymptotic estimate of $ES\left[\pi(\boldsymbol{x}|\boldsymbol{X},I)\right]$
by using the training bins as prediction points and approximating
the integral by its finite Riemann sum, as we did above. Then, 
\begin{eqnarray}
ES\left[\pi(\boldsymbol{x}|\boldsymbol{X},I)\right] & \approx & \frac{1}{A}\sum_{\nu=1}^{B}\omega_{\nu}e^{\lambda_{\mu}+C_{\nu\nu}}\times\frac{1}{2}C_{\nu\nu}\nonumber \\
 & \approx & \frac{1}{2N}\sum_{\nu=1}^{B}\omega_{\nu}\left(\frac{n_{\nu}}{\omega_{\nu}}\right)\times\frac{1}{n_{\nu}}\nonumber \\
 & = & B/2N.\label{eq:ES_Asym}
\end{eqnarray}

This asymptotic behavior is reassuring since its simple dependence on
the average number of events per bin is in accordance with intution.
Of course, how rapidly the asymptotic result becomes a reasonable
approximation depends on how quickly the first term in Equation (\ref{eq:C1})
eclipses the second term as $N\rightarrow\infty$, which is to say,
on the choice of covariance kernel function $K(\boldsymbol{x}_{1},\boldsymbol{x}_{2})$,
on the best-fit hyperparameters and, therefore, ultimately on the
data and the distribution that gave rise to it.

Suppose someone has proposed a different probability density $p(\boldsymbol{x})$
as the source of the data. Can we tell whether $\pi(\boldsymbol{x}|\boldsymbol{X},I)$
is an improvement on $p(\boldsymbol{x})$?

Define
\begin{eqnarray}
\Delta S\left[\pi(\boldsymbol{x}|\boldsymbol{X},I),p(\boldsymbol{x})\right] & \equiv & KL\left[\Pi(\boldsymbol{x})||p(\boldsymbol{x})\right]-KL\left[\Pi(\boldsymbol{x})||\pi(\boldsymbol{x}|\boldsymbol{X},I)\right].\nonumber \\
 & = & \int d^{D}\boldsymbol{x}\,\Pi(\boldsymbol{x})\,\ln\frac{\pi(\boldsymbol{x}|\boldsymbol{X},I)}{p(\boldsymbol{x})}.\label{eq:DeltaS}
\end{eqnarray}
The quantity $\Delta S$ is a random variable, in consequence of the
uncertainty in the imperfectly known distribution $\Pi(\boldsymbol{x})$.
Taking the expectation value of Equation (\ref{eq:DeltaS}), we find
\begin{equation}
\overline{\Delta S}=E_{\Pi\,|\,(\boldsymbol{X},I)}\left\{ \Delta S\right\} =KL\left[\pi(\boldsymbol{x}|\boldsymbol{X},I)||p(\boldsymbol{x})\right].\label{eq:EDeltaS}
\end{equation}

By the properties of the entropy, we have $\overline{\Delta S}\ge0$,
with equality holding only if $p(\boldsymbol{x})=\pi(\boldsymbol{x}|\boldsymbol{X},I)$
almost everywhere. This is not to say that $\pi(\boldsymbol{x}|\boldsymbol{X},I)$
is always superior to any other distribution $p(\boldsymbol{x})$,
however (what if $p$ were, in fact, the ideal distribution $\pi(\boldsymbol{x}|I)$?).
The quantity $\Delta S$ is uncertain and may in fact be negative; 
$\overline{\Delta S}$ is merely its expected value.
The distribution for $\Delta S$ is too difficult to compute, but
we may compute its variance straightforwardly:
\begin{equation}
E_{\Pi\,|\,(\boldsymbol{X},I)}\left\{ \left(\Delta S\right)^{2}\right\} =\int d^{D}\boldsymbol{x}_{1}d^{D}\boldsymbol{x}_{2}\,\ln\frac{\pi(\boldsymbol{x}_{1}|\boldsymbol{X})}{p(\boldsymbol{x}_{1})}\,\ln\frac{\pi(\boldsymbol{x}_{2}|\boldsymbol{X})}{p(\boldsymbol{x}_{2})}\,E_{\Pi\,|\,\boldsymbol{X}}\left\{ \Pi(\boldsymbol{x}_{1})\Pi(\boldsymbol{x}_{2})\right\} .\label{eq:EDeltaS2}
\end{equation}
But
\begin{eqnarray}
E_{\Pi\,|\,(\boldsymbol{X},I)}\left\{ \Pi(\boldsymbol{x}_{1})\Pi(\boldsymbol{x}_{2})\right\}  & = & (A\Omega)^{-2}(2\pi)^{-1}\left(\det\boldsymbol{C}_{2}\right)^{-1/2}\nonumber \\
 &  & \times\int d^{2}\boldsymbol{l}\,\exp\left\{ -\frac{1}{2}\left(\boldsymbol{l}-\boldsymbol{\lambda}_{2}\right)^{T}\boldsymbol{C}_{2}^{-1}\left(\boldsymbol{l}-\boldsymbol{\lambda}_{2}\right)+\boldsymbol{u}_{2}^{T}\boldsymbol{l}\right\} ,\nonumber \\
\label{eq:Epipi}
\end{eqnarray}
where $\boldsymbol{\lambda}_{2}^{T}=[\lambda(\boldsymbol{x}_{1}),\lambda(\boldsymbol{x}_{2})]$,
$\left[\boldsymbol{C}_{2}\right]_{ij}=C(\boldsymbol{x}_{i},\boldsymbol{x}_{j})$
with $i,j=1,2$, and $\boldsymbol{u}_{2}$ is the two-dimensional ``one''
vector. In other words,
\begin{eqnarray}
E_{\Pi\,|\,(\boldsymbol{X},I)}\left\{ \Pi(\boldsymbol{x}_{1})\Pi(\boldsymbol{x}_{2})\right\}  & = & (A\Omega)^{-2}\exp\left\{ \lambda(\boldsymbol{x}_{1})+\lambda(\boldsymbol{x}_{2})+\frac{1}{2}C(\boldsymbol{x}_{1},\boldsymbol{x}_{1})+\frac{1}{2}C(\boldsymbol{x}_{2},\boldsymbol{x}_{2})+C(\boldsymbol{x}_{1},\boldsymbol{x}_{2})\right\} \nonumber \\
 & = & \pi(\boldsymbol{x}_{1}|\boldsymbol{X},I)\pi(\boldsymbol{x}_{2}|\boldsymbol{X},I)\exp\left\{ C(\boldsymbol{x}_{1},\boldsymbol{x}_{2})\right\} .\label{eq:Epipi_2}
\end{eqnarray}

Substituting in Equation (\ref{eq:EDeltaS2}), we have
\begin{eqnarray}
E_{\Pi\,|\,(\boldsymbol{X},I)}\left\{ \left(\Delta S\right)^{2}\right\}  & = & \int d^{D}\boldsymbol{x}_{1}d^{D}\boldsymbol{x}_{2}\,\left(\pi(\boldsymbol{x}_{1}|\boldsymbol{X},I)\ln\frac{\pi(\boldsymbol{x}_{1}|\boldsymbol{X},I)}{p(\boldsymbol{x}_{1})}\right)\nonumber \\
 &  & \times\left(\pi(\boldsymbol{x}_{2}|\boldsymbol{X},I)\ln\frac{\pi(\boldsymbol{x}_{2}|\boldsymbol{X},I)}{p(\boldsymbol{x}_{2})}\right)\times\exp\left\{ C(\boldsymbol{x}_{1},\boldsymbol{x}_{2})\right\} .\nonumber \\
\label{eq:EDeltaS3}
\end{eqnarray}
It follows immediately that 
\begin{eqnarray}
\mathrm{Var}_{\Pi\,|\,(\boldsymbol{X},I)}\left\{ \Delta S\right\}  & = & \int d^{D}\boldsymbol{x}_{1}d^{D}\boldsymbol{x}_{2}\,\left(\pi(\boldsymbol{x}_{1}|\boldsymbol{X},I)\ln\frac{\pi(\boldsymbol{x}_{1}|\boldsymbol{X},I)}{p(\boldsymbol{x}_{1})}\right)\nonumber \\
 &  & \hspace{2cm}\times\left(\pi(\boldsymbol{x}_{2}|\boldsymbol{X},I)\ln\frac{\pi(\boldsymbol{x}_{2}|\boldsymbol{X},I)}{p(\boldsymbol{x}_{2})}\right)\nonumber \\
 &  & \hspace{2cm}\times\left(\exp\left\{ C(\boldsymbol{x}_{1},\boldsymbol{x}_{2})\right\} -1\right).\label{eq:VarDeltaS}
\end{eqnarray}

The asymptotic approximations for $\overline{\Delta S}$ and $\mathrm{Var}(\Delta S)$
are
\begin{eqnarray}
\lim_{N\rightarrow\infty}\overline{\Delta S} & = & \sum_{\nu=1}^{B}\left(\frac{n_{\nu}}{N}\right)\left[\ln\left(\frac{n_{\nu}}{Nv_{\nu}}\right)-\ln p(\boldsymbol{x}_{\nu})\right],\label{eq:EDeltaS_Asym}\\
\lim_{N\rightarrow\infty}\mathrm{Var}(\Delta S) & = & \frac{1}{N}\sum_{\nu=1}^{B}\left(\frac{n_{\nu}}{N}\right)\left[\ln\left(\frac{n_{\nu}}{Nv_{\nu}}\right)-\ln\pi_{1}(\boldsymbol{x}_{\nu})\right]^{2}.\label{eq:VarDeltaS_Asym}
\end{eqnarray}
Since in general $n_{\nu}/N$ tends to a finite value in the limit,
we see that $\mathrm{Var}(\Delta S)\sim\mathcal{O}(N^{-1})$
and tends to zero in the limit. On the other hand, if $p(\boldsymbol{x})$
is misspecified, one expects $\lim_{N\rightarrow\infty}\overline{\Delta S}$
to be a finite positive value. It follows that the GP measure estimate
$\pi(\boldsymbol{x}|\boldsymbol{X},I)$ can achieve significant performance
improvement over a misspecified distribution $p(\boldsymbol{x})$
in the limit of large data and that we would expect to be able to
exploit this superior performance (in betting against the owner of
$p(\boldsymbol{x})$, say), once $N$ is large enough that $\overline{\Delta S}/\sqrt{\mathrm{Var}(S)}\gg1$.

If $p(\boldsymbol{x})$ is \emph{not} misspecified---if it happens to 
be the true distribution $\pi(\boldsymbol{x})$---then we can set 
$n_{\nu}/N=\pi(\boldsymbol{x}_{\nu})(1+\epsilon)$,
where $\epsilon\sim n_{\nu}^{-1/2}$ , in the limit of large $N$.
We therefore have $\overline{\Delta S}\sim\mathcal{O}(N^{1/2})$ in
this limit, so that asymptotically $\overline{\Delta S}/\sqrt{\mathrm{Var}(S)}$
tends to a constant.

\section*{References}
\bibliographystyle{elsarticle-num}
\bibliography{refs}

\end{document}